\documentclass[sn-nature]{sn-jnl}

\usepackage{graphicx}%
\usepackage{multirow}%
\usepackage{amsmath,amssymb,amsfonts}%
\usepackage{amsthm}%
\usepackage{mathrsfs}%
\usepackage[title]{appendix}%
\usepackage{xcolor}%
\usepackage{textcomp}%
\usepackage{manyfoot}%
\usepackage{booktabs}%
\usepackage{algorithm}%
\usepackage{algorithmicx}%
\usepackage{algpseudocode}%
\usepackage{listings}%
\usepackage{aas_macros}
\usepackage{siunitx}

\usepackage{soul}

\newcommand{\del}[1]{{\color{red}{\st{#1}}}}

\begin{document}

\setlength{\baselineskip}{18pt}

%
\title[an Extremely Eccentric Double Neutron Star System]
{PSR J0653+0443: An Eclipsing Millisecond Pulsar in an Extremely Eccentric Double Neutron Star System} 
%


\author[1]{\fnm{Z.~L.} \sur{Yang}}

\author*[1,2,3]{\fnm{J.~L.} \sur{Han}}\email{hjl@nao.cas.cn}%

\author[1,3]{\fnm{L.~G.} \sur{Hou}}

\author[1,2]{\fnm{Z.~W.} \sur{Chen}}

\author[4]{\fnm{Y.~G.} \sur{Gong}}

\author[5]{\fnm{Y.~L.} \sur{Guo}}

\author[2,1]{\fnm{Y.~J.} \sur{Lu}}

\author[4]{\fnm{C.} \sur{Zhang}}

\author[1,2]{\fnm{K.~Y.} \sur{Song}}

\author[7]{\fnm{B.} \sur{Liu}}

\author[8,12]{\fnm{D.} \sur{Lai}}

\author[1,2,3]{\fnm{C.} \sur{Wang}}

\author[1,3]{\fnm{W.~Q.} \sur{Su}}

\author[1,2,3]{\fnm{P.~F.} \sur{Wang}}

\author[9]{\fnm{X.~P.} \sur{You}}

\author[1,2]{\fnm{N.~N.} \sur{Cai}}

\author[9]{\fnm{Q.} \sur{Gao}}

\author[1,2]{\fnm{W.~C.} \sur{Jing}}

\author[6]{\fnm{B.} \sur{Wang}}

\author[10,11]{\fnm{H.~G.} \sur{Wang}}

\author[1,2]{\fnm{T.} \sur{Wang}}

\author[1,2]{\fnm{J.} \sur{Xu}}

\author[1,2]{\fnm{Y.} \sur{Yan}}

\author[1,2]{\fnm{D.~J.} \sur{Zhou}}




\affil[1]{National Astronomical Observatories, Chinese Academy of Sciences, A20 Datun Road, Chaoyang District, Beijing 100101, China}

\affil[2]{School of Astronomy and Space Sciences, University of Chinese Academy of Sciences, Beijing 100049, China}

\affil[3]{State Key Laboratory of Radio Astronomy and Technology, Beijing 100101, China }

\affil[4]{Institute of Fundamental Physics and Quantum Technology, Department of Physics, School of Physical Science and Technology, Ningbo University, Zhejiang 315211, China}

\affil[5]{School of Astronomy and Space Science, Nanjing University, Nanjing 210023, China}

\affil[6]{Yunnan Observatories, Chinese Academy of Sciences, Kunming 650216, China}

\affil[7]{Institute for Astronomy, School of Physics, Zhejiang University, 310058 Hangzhou, China}

\affil[8]{Tsung-Dao Lee Institute, Shanghai Jiaotong University, Shanghai 201210, China} 

\affil[9]{School of Physical Science and Technology, Southwest University, Chongqing 400715, China}

\affil[10]{Department of Astronomy, Guangzhou University, Guangzhou 510006, China}

\affil[11]{National Astronomical Data Center, Great Bay Area, Guangzhou 510006, China}

\affil[12]{Department of Astronomy, Center for Astrophysics and Planetary Science, Cornell University, Ithaca, NY 14853, USA}


\abstract{
Double neutron star (DNS) systems emit gravitational waves (GW) as they orbit each other \cite{Taylor+1982ApJ...253..908T}, with waveforms dependent on their orbital properties \cite{Poisson+2014grav.book.....P}. The number of DNS systems surviving binary star evolution in the Milky Way and the distribution of their orbital parameters are fundamental questions for GW astronomy and binary evolution studies. 
Here, we report the discovery of a new Galactic field DNS system, PSR J0653+0443, comprising two ~1.4 \(M_\odot\) components in an extremely eccentric orbit (\(e=0.992387\), \(P_{\rm orb}=79.084\) d). Its unique orbital dynamics include a 46-minute transit from true anomaly \( \theta=-90^\circ \) to \(90^\circ\) and striking periastron relative motion (maximum acceleration 1.11~km~s\(^{-2}\), peak speed 1.14×10\(^6\)~m~s\(^{-1}\)). The pulsar undergoes an 85-second conjunction eclipse, indicating a nearly edge-on binary orbit.
Short-term variations in the periastron advance induce significant pulse arrival time modulation spanning \(-2.91\) to \(+2.49\) ms --- at least an order of magnitude larger than that observed in any other DNS system. From this modulation, the non-relativistic periastron advance rate is constrained to \( (-1.5^{+0.9}_{-0.8})\times10^{-4}\) deg yr\(^{-1}\).
The extreme orbital eccentricity can only be attributed to a large kick velocity imparted during the iron-core collapse supernova of the progenitor of the second component. The predicted detection probability of such a PSR J0653+0443-like DNS system is \( \lesssim 5 \times 10^{-3} \), implying that our discovery of this system challenges current understanding of binary evolution \cite{Tauris+2023pbse.book.....T}.
This extraordinary DNS system emits periodic gravitational wave bursts, with the GW energy flux peaking at 0.357 mHz. This makes it an ideal target for the future LISAmax mission \cite{2023CQGra..40s5022M}.
}
\keywords{Double Neutron Stars, Eclipse, Periastron Advance}


\maketitle

{\bf One sentence summary:} {\it  We discovered an outstanding DNS system with an extremely high eccentricity, a source for periodical gravitational wave bursts, and it is the second DNS system with an observed eclipse.} \\[3mm]

Double neutron star (DNS) systems are evolutionary products of massive binary stars \cite{Tauris+2017ApJ...846..170T}. Since the discovery of the first DNS system (PSR B1913+16) in 1974 \cite{Hulse+1975ApJ...195L..51H}, over 30 
such systems have been identified 
(Table~\ref{tab:DNSlist}). Measuring the orbital parameters and their variations enables precise tests of various gravitational theories \cite{Kramer+2021PhRvX..11d1050K}. 
For instance, orbital damping caused by gravitational wave (GW) emission is manifested by the orbital period derivative ($\Dot{P}_{\rm orb}$), providing indirect evidence for the existence of GW \cite{Hulse+1975ApJ...195L..51H, Taylor+1982ApJ...253..908T}.
Some DNS systems will eventually merge within a Hubble time, appearing as GW events such as  GW170817 \cite{Abbott+2017PhRvL.119p1101A} and GW190425 \cite{Abbott+2020ApJ...892L...3A} detected by LIGO. 
DNS systems can be well approximated as a two-point-masses system.   
When the two NSs orbit in an ellipse, the periastron advance rate is not constant in time near periastron (where the separation is smallest), general relativistic effects are strongest, and the periastron advance rate is expected to peak. 

\begin{figure}[b]
    \centering
    \includegraphics[width=0.65\columnwidth]{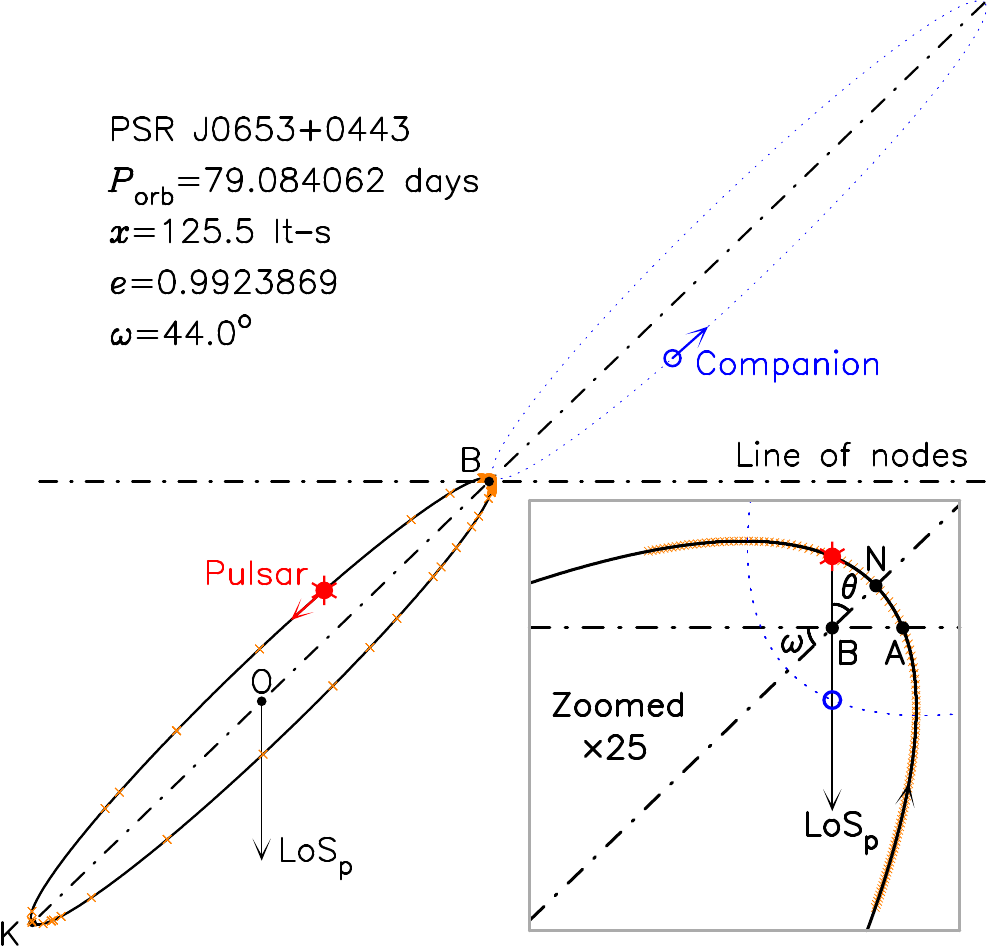}
    \caption{\textbf{Schematic of the face-on orbit of PSR J0653+0443.} 
    Two neutron stars orbit each other with an orbital period of $P_{\rm orb}=$ 79.0840618~days and an extreme high orbital eccentricity of $e=0.9923869$. 
    The line of nodes (the intersection of the orbital plane and the plane of the sky) is indicated by a dot-dashed line. The projected line of sight onto the orbital plane is marked by the arrow (LoS$_{\rm p}$). 
    Key orbital points are defined as follows: {\bf B} ( barycenter of the binary), {\bf O} (orbital center of the pulsar), {\bf K} (apastron), {\bf N} (periastron),  {\bf A} (ascending node), and {\bf D} (descending node, not shown). 
    The projected semi-major axis of the pulsar's orbit ({\bf OK} or {\bf ON}) has a length of $a_{\rm p} = x/\sin i$, where $x$ denotes the projected semi-major axis of the pulsar's orbit, and $i$ is the inclination angle between the orbital plane and the line of sight. The longitude of periastron $\omega$ is the angle of $\angle${\bf ABN} in the enlarged inset. FAST observations are marked by orange crosses at their respective orbital phases (see Table~\ref{tab:obslist}), with a dense coverage near periastron by one point per minute. The companion's orbit is outlined by a dotted line, calculated using a mass ratio of 1.04 (the best-fit value from pulsar timing analysis). Eclipses occur when the pulsar, its companion and the barycenter ({\bf B}) are aligned along the line of sight around the true anomaly angle $\theta = 90^\circ - \omega 
    = 46^\circ$. The key orbital parameters of PSR J0653+0443 are labeled in the top-left corner.
    }
    \label{fig:geo}
\end{figure}

\begin{table*}[t]
	\centering
	\caption{\textbf{Timing parameters and derived properties for PSR~J0653+0443.} The flux density and rotation measure are obtained from the 2-hr observation on 2025 February 19. The uncertainties are given by 1$\sigma$ at a confidence level of 68.3\%. The $m_{\rm c}$ and $m_{\rm t}$ parameters are present in both fitted and derived sections in order to show the refined uncertainties obtained through bayesian chi2 analysis (see Method~\ref{sec:pulsartiming} for details). 
	}
	\label{tab:timingPara}
    \scriptsize
    \renewcommand\arraystretch{0.9}
	\setlength{\tabcolsep}{2mm}
	\begin{tabular}{lc} 
	\hline
		\hline \\[-2.5mm]
        \multicolumn{2}{c}{General information} \\
        \hline \\[-2.5mm]
		Number of TOAs      &     342 \\
		MJD range           &    59902 to 60804 \\
        MJD of period determination & 60300 \\
        UNITS       &   TCB \\
        Solar System ephemeris   &  DE440 \\ [1mm]
        \hline \\[-2.5mm]
        \multicolumn{2}{c}{Fitted parameters}\\
        \hline \\[-2.5mm]
        Right ascension, $\alpha$ (J2000) & $06^{\rm h}53^{\rm m}04^{\rm s}.57459\pm0^{\rm s}.00024$\\
		Declination, $\delta$ (J2000)  & $+04^\circ43'16''.349\pm0''.013$\\
		Proper motion in R.A. (mas yr$^{-1}$) & -10$\pm$5\\
		Proper motion in DEC. (mas yr$^{-1}$)  & -4$\pm$12\\
		Spin frequency, $\nu$ (Hz) & 21.036552768034$\pm$0.000000000014\\ 
		Spin frequency derivative, $\Dot{\nu}$ (Hz $\rm s^{-1}$)  & (-5.09$\pm$0.06)$\times$10$^{-17}$\\ 
		Dispersion measure, DM (pc $\rm cm^{-3}$)  &  27.419(4)\\
        Rotation measure, RM (rad\,m$^{-2}$) & 21.4(6) \\
        Flux density, S$_{\rm 1250 MHz}$ (\textmu Jy) & 102.2(4) \\
		Orbital period, $P_{\rm orb}$ (day) & 79.08406184(12)\\
		Epoch of periastron, $T_0$ (MJD) & 60725.59202195(7) \\
	    $\zeta_1\equiv x\sin\omega/e$ (lt-s)  & 87.905453(6) \\
	    $\zeta_2\equiv x\cos\omega\sqrt{1-e^2}$ (lt-s) & 11.118(4) \\
	    $\delta_{\rm e}\equiv\sqrt{1-e^2}$  & 0.1231593(4) \\
	    Companion mass, $m_{\rm c}$ ($M_\odot$) & 1.44(10) \\
	    Total mass, $m_{\rm t}$ ($M_\odot$) & 2.814(20) \\
	    Periastron advance rate $\dot{\omega}$ 
	    (deg\,yr$^{-1}$) & 0.017791(10) \\[1mm]
	    \hline\\[-2.5mm]
	    \multicolumn{2}{c}{Model derived parameters} \\ 
	    \hline\\[-2.5mm]
		Spin period, $P$ (s)  & 0.04753630554525(3)\\
		Spin period derivative, $\dot{P}$ (s\,s$^{-1}$) & 1.150(14)$\times$10$^{-19}$\\
		Characteristic age, $\tau_{\rm c}$ (Gyr)  &  6.6\\
		Surface magnetic field, $B_{\rm s}$ ($10^{9}$G)  &  2.4 \\
		Spin-down luminosity, $\dot{E}$ (erg\,s$^{-1}$) & 4.2$\times10^{31}$ \\
		DM Distance, $d_{\rm NE2001}$ (kpc)  &   0.2 \\
		DM Distance, $d_{\rm YMW16}$ (kpc)  &   1.3 \\
		Projected semi-major axis, $x$ (lt-s) & 125.539(26)\\
        Longitude of periastron, $\omega$ (deg) & 44.018(11)\\
        Orbital eccentricity, $e$  &  0.99238692(5) \\
		Mass function ($M_\odot$)  &  0.33966(21) \\
		Coalescence timescale (Gyr) & 2.5\\
		Einstein delay, $\gamma$ (ms) & 45(4)\\
		$\Dot{\omega}_{\rm EX}$ (deg\,yr$^{-1}$)$^\dag$ & (-1.5$^{+0.9}_{-0.8})\times10^{-4}$ \\
        Pulsar mass, $m_{\rm p}$ ($\rm M_\odot$)$^\dag$  &  {1.38$^{+0.06}_{-0.10}$}\\
        Companion mass, $m_{\rm c}$ ($\rm M_\odot$)$^\dag$  &  {1.44$^{+0.08}_{-0.06}$}\\
        Total mass, $m_{\rm t}$ ($\rm M_\odot$)$^\dag$  &  {2.814$^{+0.018}_{-0.021}$}\\
        Orbital inclination, $i$ ($^\circ$)$^\dag$  &  {76$^{+8}_{-15}$}\\
		\hline \hline\\
		$^\dag$best-fit value and 1$\sigma$ uncertainty derived from $\chi^2$ analysis.
    \end{tabular}
\end{table*}

\begin{figure}[t]
    \centering
    \includegraphics[width=0.6\columnwidth]{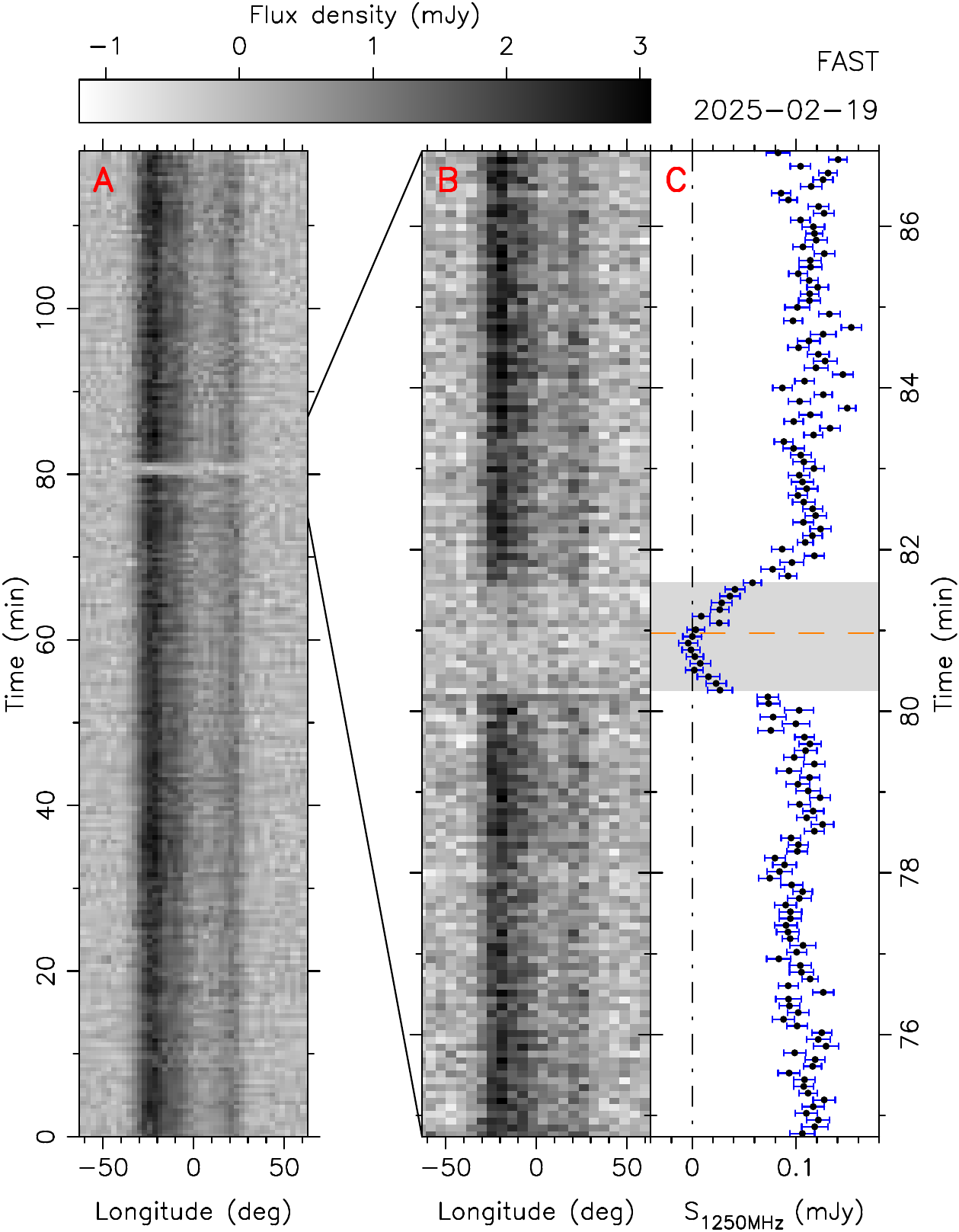}
    \caption{\textbf{Time-phase diagram showing eclipse near the periastron passage.} 
    The FAST observations were carried out on February 19, 2025, during which an 85-second eclipse was detected near the conjunction phase of $\theta + \omega = 90^\circ$, indicated by the orange dashed line. Panel {\bf A} presents the time-phase diagram for sub-integrated profiles, with 128 bins and a 30-second sub-integration time. Panel {\bf B} shows a zoomed-in view of this eclipse region, using profiles with 64 bins and a 5-second sub-integration time. The flux densities and their associated uncertainties plotted in Panel {\bf C} were derived by comparing the observed profiles to a standard profile, via the \texttt{psrflux} routine within the \texttt{PSRCHIVE} package \cite{Hotan+2004PASA...21..302H}. 
    }
    \label{fig:eclipse}
\end{figure}

\section*{Pulsar~J0653+0443 and its extraordinary orbit} 
\label{sec2}

\begin{figure}[t]
    \centering
    \includegraphics[width=0.52\columnwidth]{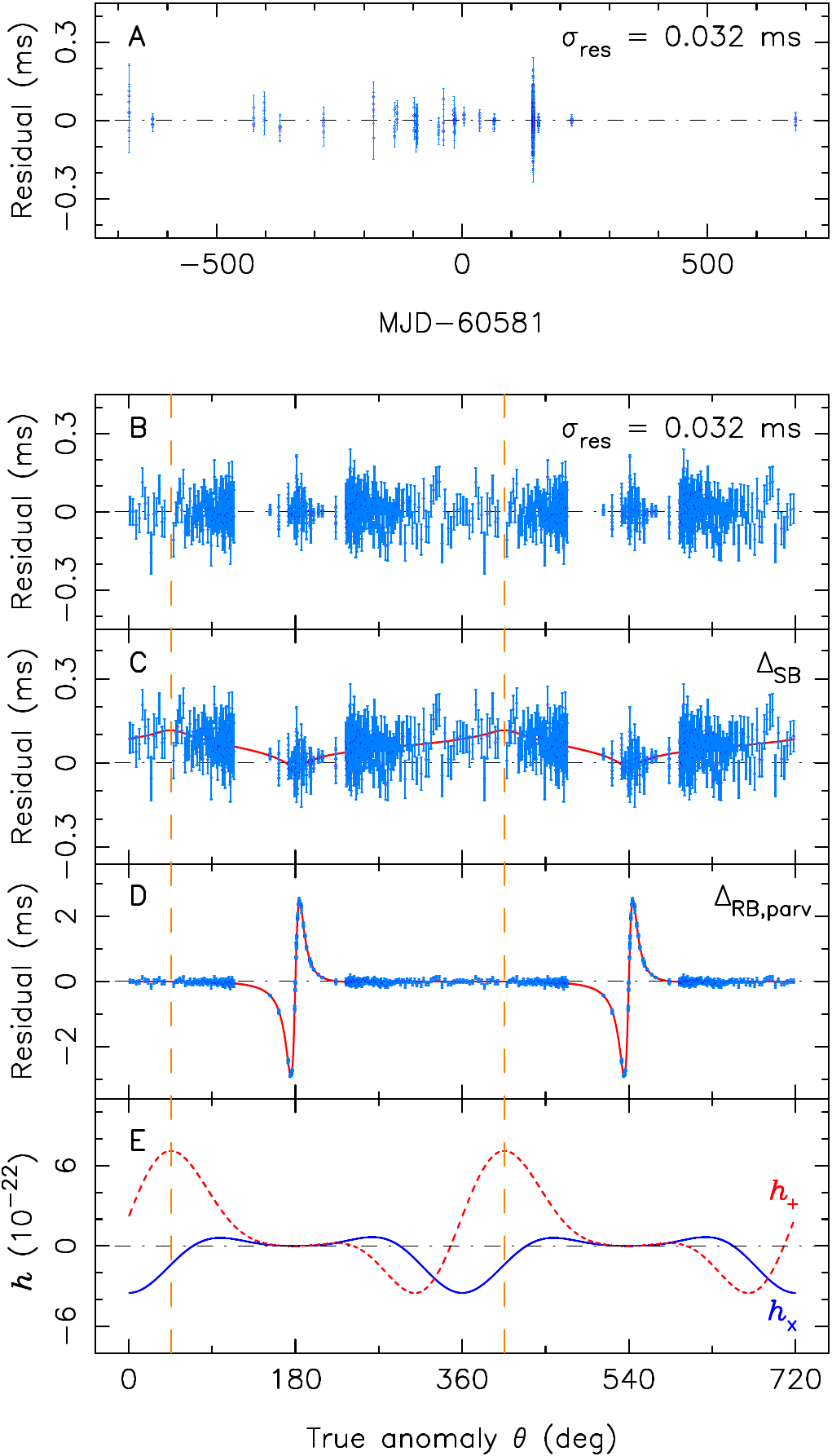}
    \caption{\textbf{Timing residuals of PSR~J0653+0443 and the strain amplitude of GWs.}
   Panels ({\bf A}) and ({\bf B}) show timing residuals as a function of modified Julian date (MJD) and orbital phase (expressed by the true anomaly $\theta$, covering two orbital periods), respectively, after fitting data with parameters in Table~\ref{tab:timingPara}. The pulsar undergoes an 85-second  eclipse near $\theta = 90^\circ-\omega = 46^\circ$ (orange  dashed line, see Fig.~\ref{fig:eclipse}), corresponding to the orbital phase where the pulsar lies behind its companion along the line of sight. The weighted root-mean-square of the residual of TOAs  ($\sigma_{\rm res}$)  is labeled in the panels. Panel ({\bf C}) shows the TOA residual term induced by the Shapiro delay (Eq.~\ref{Delta_S} for $\Delta_{\rm SB}$), and panel ({\bf D}) presents the term induced by the orbital-phase-dependent periastron advance (Eq.~\ref{delta_T} for $\Delta_{\rm RB, parv}$). The red curves are theoretically derived by using the best-fit parameters in Table~\ref{tab:timingPara}. Error bars of data points represent the $1\sigma$ uncertainties of TOAs. Panel ({\bf E}) shows the strain amplitude of GWs \cite{Poisson+2014grav.book.....P} of the plus and cross polarizations ($h_+$ , $h_{\times}$) of the PSR~J0653\del{3}+0443 DNS system (see Fig.~\ref{fig:hplusvstime}), assuming a distance of 1.0~kpc.}
    \label{timing_res}
\end{figure}

PSR~J0653+0443, a millisecond pulsar, was first detected by the Five-hundred-meter Aperture Spherical radio Telescope (FAST) \cite{Nan+2006ScChG..49..129N} on November 20, 2022 \cite{Han+2025RAA....25a4001H}. It is a bright pulsar with a spin period of $P = 47.5$~ms. The detection was confirmed via a 15-minute FAST tracking observation on January 07, 2023  (Table~\ref{tab:obslist}). 
Using spin period variations induced by the orbital motion, we initially derived the approximate orbital parameters from the first seven FAST observations. We extracted the barycenter period and epoch of each observation using {\sc Presto} \cite{Ransom+2001PhDT.......123R}, and then performed a four-dimensional search over orbital period $P_{\rm orb}$, time of periastron passage $T_0$, orbital eccentricity $e$ and longitude of periastron $\omega$ to fit the observed barycenter periods across different epochs. This fitting employed the linear least-squares method described in \cite{Bhattacharyya+2008MNRAS.387..273B}, revealing that PSR J0653+0443 resides in a binary system with a wide, highly eccentric orbit ($P_{\rm orb}\simeq79$~days, $e\gtrsim 0.9$). Subsequently, we conducted a series of targeted FAST observations  ---  particularly those covering the periastron passages  --- to refine the orbital parameters  (Table~\ref{tab:obslist}). Using pulse times of arrival (TOAs) measured by FAST over a span of more than 2 years, we obtained a phase-connected timing solution for the pulsar  (Table~\ref{tab:timingPara}); the timing residuals are plotted against observation time and orbital phase (expressed in the true anomaly angle $\theta$) in Fig.~\ref{timing_res}.  

In pulsar timing, the R\"{o}mer delay term describes the light travel time variation caused by the pulsar’s orbital motion, and can be expressed as \cite{Blandford+1976ApJ...205..580B}
\begin{equation}
    \Delta_{\rm RB}=\frac{x}{c}[\sin\omega(\cos u-e)+\sqrt{1-e^2}\cos\omega\sin u],
    \label{Romer_delay}
\end{equation}
where $u$ is the eccentric anomaly, $x$ is the projected semi-major axis,
and $c$ is the speed of light. 
For this highly eccentric DNS system,  we replaced orbital parameters $x$, $\omega$, and $e$ with $\zeta_1=x\sin\omega/e$, $\zeta_2=x\cos\omega\sqrt{1-e^2}$ and $\delta_{\rm e}=\sqrt{1-e^2}$ when fitting the R\"{o}mer delay. This substitution avoids the high covariance between the three original parameters (see Method~\ref{sec:KandPK}). Based on the R\"{o}mer delay measurement, we determined the pulsar's orbital period and projected semi-major axis as $P_{\rm orb}=79.0840618 \pm 0.0000001$~days and $x=125.54 \pm 0.03$~lt-s, respectively. The mass function of the PSR~J0653+0443 binary system is thus derived as  
\begin{equation}
f(m_{\rm p},m_{\rm c},i)=(4\pi^2/G)x^3/P_{\rm orb}^2=(m_{\rm c}\sin i)^3/(m_{\rm p}+m_{\rm c})^2=0.33966(21)~M_\odot,
\label{mf}
\end{equation}
where $m_{\rm p}$ and $m_{\rm c}$ are the masses of the pulsar and its companion, respectively, $i$ is the inclination angle of the orbit, and $G$ is the gravitational constant. 
Using the total mass derived from periastron advance ($m_{\rm t} \simeq 2.81~{M}_{\odot}$; see below) and $\sin i \le 1$, the companion mass is constrained to be $m_{\rm c} \ge 1.38 M_\odot$, consistent with a neutron star nature. 
Assuming that the recycled pulsar has mass comparable to or greater than the second-born counterpart \cite{Farrow+2019ApJ...876...18F}, as observed in most known DNS systems, we further constrain the orbital inclination angle to $\sin i \gtrsim 0.98$, indicating that the PSR~J0653+0443 system has a nearly edge-on orbit. The projected orbit schematic of this DNS system is shown in Fig.~\ref{fig:geo}. 
Despite the 79-day orbital period, the pulsar takes only 46 minutes to move from $\theta = -90^\circ \ (270^\circ) $ to $ 90^\circ $ (see Method~\ref{sec:KandPK}), where $\theta $ is the true anomaly angle. 

Periastron passage occurs every 79-day orbit, but it is not always observable with FAST (Table~\ref{tab:periastron_passage}).  On February 19, 2025, this DNS system passed through periastron and was observed by FAST, as it fell within FAST's zenith angle of $<28.5^{\circ}$ (see Fig.~\ref{fig:motion_trajectory}). We conducted a 2-hour tracking observation of PSR~J0653+0443 covering the periastron passage (see zoomed part of Fig.~\ref{fig:geo}), and  precisely measured the orbital eccentricity ($e= 0.99238692\pm0.00000005$), the highest ever measured for stellar binary systems. Prior to this, the highest recorded eccentricity for a binary system was $e= 0.985\pm0.002$ (WDS~05354$-$3316 in \cite{Tokovinin+2020AJ....160....7T}). 
We also detected an 85-second eclipse of the radio emission, occurring near $\theta = 90^\circ - \omega = 46^\circ$ (grey-shaded region in Fig.~\ref{fig:eclipse}),  when the pulsar, its companion, and the barycenter were aligned along the line of sight (Fig.~\ref{fig:geo}). The eclipse spanned a true anomaly range of $\Delta\theta=6.84^\circ$, corresponding to the pulsar’s passage behind the effective eclipsing magnetosphere of the companion neutron star ---  confirming that PSR J0653+0443 resides in an edge-on orbit. Prior to this, DNS eclipses had only been detected in the double pulsar system, PSR J0737$-$3039A \cite{Lyne+2004Sci...303.1153L}. The ``light curve'' (intensity variation) of the gradual eclipse of PSR~J0653+0443 is similar to that of PSR~J0737$-$3039A at low time resolution \cite{Lower+2024MNRAS.534.3936L}. The relatively weak radio signals from PSR~J0653+0443 precluded the detection of the periodical variations induced by the rotation of companion neutron star --- a phenomenon observed for PSR~J0737$-$3039A at high time resolution \cite{Breton+2008Sci...321..104B, Lower+2024A&A...682A..26L}. 
At about $\theta  = 46^\circ$, the separation between the two neutron stars is $2.28$ light-seconds (lt-s), and the effective size of the companion’s eclipsing magnetosphere can be constrained by the eclipse duration to  $4\times10^4$ km. 
%
%

Using polarization signals recorded during the 2-hour FAST tracking observation on 19 February 2025, we obtained the polarization profile of PSR J0653+0443 after polarization calibration (Fig.~\ref{pol}).



\section*{The relativistic effects of PSR~J0653+0443}
\label{sec3}

Among all known DNS systems (Table~\ref{tab:DNSlist}), PSR~J0653+0443 has both the longest orbital period and the highest orbital eccentricity (Fig.~\ref{fig:pe2}). With such a high eccentricity, the relative velocity of the two neutron stars and orbital acceleration between PSR~J0653+0443 and its companion vary strongly with the orbital phase, reaching the maximum ($v_{\rm pc,max} = 1.14\times10^3~{\rm km~s}^{-1}$, $Acc_{\rm pc,max} = 1.11\times10^3~{\rm m~s}^{-2}$) when the two objects are closest with the smallest separation $S_{\rm pc,min}=0.83R_\odot$ at the true anomaly of $\theta=0^{\circ}$ (Fig.~\ref{fig:geo}). Compared to other DNS systems, such as PSR J1946+2052 that has the shortest orbital period \cite{Stovall+2018ApJ...854L..22S} or PSR J1757$-$1854 with the largest orbital acceleration \cite{Cameron+2018MNRAS.475L..57C}, PSR~J0653+0443 exhibits an even higher relative velocity, though its orbital acceleration falls between these two extremes.

Given that PSR J0653+0443 and its companion run on a nearly edge-on orbit, the Shapiro delay \cite{Damour+1986AIHPA..44..263D, lori2012}
\begin{equation}
    \Delta_{\rm SB} = -\frac{2G m_{\rm c}}{c^3} \ln\left\{1 - e \cos u - \sin i \left[ \sin \omega (\cos u - e) + \sqrt{1 - e^2} \cos \omega \sin u \right] \right\},
    \label{Delta_S}
\end{equation}
can be constrained from the timing residual. 
%
%
The TOA residual induced by the Shapiro delay, together with the theoretical curve calculated using the best-fit values of $m_{\rm c}=1.44~M_\odot$ and $\sin i=0.970$ (see Table~\ref{tab:timingPara}), are presented in Fig.~\ref{timing_res}C.

%
General relativity predicts that the longitude of periastron $\omega$ advances at the rate of \cite{Damour+1992PhRvD..45.1840D,lori2012}
\begin{equation}
    \Dot{\omega}_{\rm GR}=3(\frac{P_{\rm orb}}{2\pi})^{-5/3}G^{2/3}m_{\rm t}^{2/3}c^{-2}(1-e^2)^{-1}.
    \label{omdot}
\end{equation}
Note that this is the secular (orbital averaged) precession rate, and is often measured well for pulsar binaries in eccentric orbit by timing analysis. We measured the periastron advance rate for PSR~J0653+0443 as $\dot{\omega} = 0.017791\pm0.000010$~degree per year, which is consistent with the predicated $\Dot{\omega}_{\rm GR}$ from general relativity.

The periastron advancing rate has a short-term variation along the orbit, 
with $\omega$ varying as a function of $\theta$ by \cite{Damour+1985AIHPA..43..107D}
\begin{equation}
    \omega-\omega_0=k\theta.
    \label{dom_1}
\end{equation}
Here $\omega_0$ is the longitude of periastron at the time of periastron passage $T_0$. In general relativity, $k = \dot{\omega}_{\rm GR}P_{\rm orb}/(2\pi)$.
This orbital-phase-dependent periastron advance causes a periodical modulation on TOA, which can be expressed by a change in the R\"{o}mer delay of (see Method~\ref{sec:KandPK}) 
\begin{equation}
    \Delta_{\rm RB,parv} 
    =   \dot{\omega}_{\rm GR}\left(\frac{P_{\rm orb}}{2\pi}\right)(\theta-M)\frac{\partial\Delta_{\rm RB}}{\partial\omega},
    \label{delta_T}
\end{equation}
here $M=2\pi(t-T_0)/P_{\rm orb}$ is the mean anomaly, and $t$ is the time measured in the proper reference frame of the pulsar. The nearly parabolic orbit of PSR~J0653+0443 makes it an ideal system suitable to demonstrate the varying periastron advance. 
As demonstrated in Fig.~\ref{timing_res}D, $\Delta_{\rm RB,parv}$ varies in the range of $-2.91$~ms to +2.49~ms. Compared to the 
other DNS systems (see Table~\ref{tab:DNSlist}), it is more than one order of magnitude greater than the next largest one ($-159$~{\textmu}s to 128~{\textmu}s for PSR J2140$-$2311B \cite{Balakrishnan+2023ApJ...942L..35B}). 

Using two years of timing data for PSR J0653+0443, we have determined the masses of the pulsar and its companion, with results presented in Table~\ref{tab:timingPara}. Employing the DDeGR+ model, following  \cite{Splaver+2002ApJ...581..509S} we derived the best-fit values 
for the total binary mass, orbital inclination angle, pulsar mass, and companion mass 
as $m_{\rm t} = 2.814^{+0.018}_{-0.021}~M_\odot$, $i = 76^{+8}_{-15}$~deg, $m_{\rm p} = 1.38^{+0.06}_{-0.10}~M_\odot$, and $m_{\rm c} = 1.44^{+0.08}_{-0.06}~M_\odot$ (see Fig.~\ref{mtot-cosi}). The $\pm1\sigma$ uncertainties for these parameters are obtained via $\chi^2$ analysis.

Beside the Shapiro delay and the periastron advance, another relativistic effect --- the Einstein delay --- also induces a significant TOA delay per orbit \cite{Damour+1986AIHPA..44..263D, lori2012}, as described by
%
 $   \Delta_{\rm EB} = \gamma \sin u. $
%
In general relativity, $\gamma$ is given by \cite{Damour+1986AIHPA..44..263D}
\begin{equation}
    \gamma=T_\odot^{2/3} e \left(\frac{P_{\rm orb}}{2\pi}\right)^{1/3}\frac{m_{\rm c}(m_{\rm p}+2m_{\rm c})}{(m_{\rm p}+m_{\rm c})^{4/3}},
\end{equation}
with $T_\odot^{2/3}\equiv G{\rm M}_\odot/c^3\approx 4.92549~\mu$s.  However, $\gamma$ is highly covariant with $\zeta_2=x\cos\omega\sqrt{1-e^2}$ in the absence of significant periastron precession, making a direct measurement of $\gamma$ currently unfeasible (see Method~\ref{sec:PK}). Using the best-fit values for the PSR J0653+0443 binary 
given in  Table~\ref{tab:timingPara}, we estimated $\gamma$ to be 45~ms, with a $1\sigma$ uncertainty of 4~ms (as that of $\zeta_2/c$).

In addition to the relativistic effect, a non-relativistic periastron advance contribution $\dot{\omega}_{\rm EX}$ (such as that due tothe proper motion of the binary) may exist. Using Eq.~(\ref{dom_3}), we introduced $\dot{\omega} \equiv \dot{\omega}_{\rm GR} + \dot{\omega}_{\rm EX}$ into our timing model, enabling the $\dot{\omega}_{\rm EX}$ measurement via comparing the long-term $\omega$ evolution and the short-term advance rate variation (Method~\ref{sec:pulsartiming}). Our 
$\chi^2$ analysis to the timing data yielded $\dot{\omega}_{\rm EX} = (-1.5^{+0.9}_{-0.8}) \times 10^{-4}\ \text{deg}\,\text{yr}^{-1}$ (Fig.~\ref{omdot1-omdot2}), which is more precise than for the double pulsar PSR J0737$-$3039A/B system ($\dot{\omega}_{\rm EX} = 3(4) \times 10^{-4}\ \text{deg}\,\text{yr}^{-1}$) \cite{Kramer+2021PhRvX..11d1050K}. The result means that $\dot{\omega}_{\rm EX}$ is non-zero at the 68\% confidence level, but consistent with zero at the 95\% confidence level.

\section*{Evolution path and fate of PSR~J0653+0443}
\label{sec4}

DNS systems are the evolutionary products of binary systems containing two massive stars  \cite{Tauris+2017ApJ...846..170T, Tauris+2023pbse.book.....T}. The more massive star in the binary evolves first, forming a neutron star in a supernova (SN) explosion. If the binary system survives in this event, the less massive star will continue to evolve and eventually expand and transfer mass to the first-born neutron star. Due to a large mass ratio, this mass transfer process is unstable, leading to the formation of a common envelope. 
The outcome of this common envelope evolution 
is either a merger of the two objects (producing a massive neutron star or black hole) or a compact neutron star-helium star binary after the common envelope is ejected \cite{Paczynski+1976IAUS...73...75P, Iben+1993PASP..105.1373I, Ivanova+2013A&ARv..21...59I}.
This helium star continues its nuclear evolution: when its core helium is exhausted, the star expands again and initiates 
the so-called Case BB Roche-lobe overflow \cite{Dewi+2002MNRAS.331.1027D}. After the second SN explosion, a new DNS system forms if the binary system survives. Similar to other DNS systems, the extremely large orbital eccentricity of PSR~J0653+0443 should arise from the birth kick imparted to the second-born neutron star \cite{Tauris+2017ApJ...846..170T} during iron-core collapse supernova (Fe CC-SN, see Method~\ref{sec:sn}) 
\cite{mth+19}. We present an evolutionary example of PSR J0653+0443's progenitor neutron star-helium star system using the stellar evolution code \textsc{Mesa} \cite{Paxton+2011ApJS..192....3P, Paxton+2019ApJS..243...10P} in Fig.~\ref{casebb}. 

We explored the probability of forming a PSR J0653+0443-like DNS system during the second SN explosion using population synthesis (see Method~\ref{sec:pop}), considering various 
pre-SN orbital periods and kick velocities, as illustrated in  Fig.~\ref{fig:space}. The 
probability of forming DNS systems for an orbital period in the range between 50 to 110 days and an orbital eccentricity larger than 0.95 is less that $\lesssim5\times10^{-3}$. 
The fact that we detected such a binary with such an extreme eccentricity among 28 Galactic field DNS systems (Table~\ref{tab:DNSlist}) suggesting that either we are very lucky or the current binary evolution theory and current understanding of core-collapse SNe needs modifications.

Due to GW emission, the orbit of PSR~J0653+0443 is shrinking and its eccentricity is decreasing, while the minimum separation between the pulsar and its companion remains nearly constant after the system's formation.
The GW emission of this extreme eccentric binary is characterized by ``periodical bursts'' (Fig.~\ref{fig:hplusvstime}) \cite{Poisson+2014grav.book.....P, Xuan+2025PhRvD.111d3018X}. Assuming a distance of 1.0~kpc, the strain amplitude of the plus and cross polarizations ($h_+$ , $h_{\times}$) of the GW bursts are shown in Fig.~\ref{timing_res}E.
Currently, the GW emission has a peak frequency of 0.357 mHz (about the $2440$th harmonics of the binary orbital frequency), and the characteristic strain amplitude is below the noise levels of the space-based GW detectors such as LISA \cite{LISA:2017pwj}, Taiji \cite{Hu:2017mde}, and TianQin \cite{TianQin:2015yph, Gong:2021gvw}, but significantly above those of Folkner \cite{Mueller+2019BAAS}
and LISAmax \cite{2023CQGra..40s5022M} (see Fig.~\ref{fig:sensitivity}),
suggesting it can serve as an archetypal Galactic laboratory for testing the GW waveform with extremely large harmonics predicted by general relativity. 
It has a coalescence timescale of 2.54~Gyr --- less than the Hubble time --- and will eventually produce a high-frequency GW event similar to the merger events GW170817 \cite{Abbott+2017PhRvL.119p1101A} and GW190425 \cite{Abbott+2020ApJ...892L...3A}.

\newpage 
\bibliography{J0653}

@ARTICLE{2023CQGra..40s5022M,
       author = {{Martens}, W. and {Khan}, M. and {Bayle}, J.-B.},
        title = "{LISAmax: improving the low-frequency gravitational-wave sensitivity by two orders of magnitude}",
      journal = {Classical and Quantum Gravity},
         year = 2023,
        month = oct,
       volume = {40},
       number = {19},
          eid = {195022},
        pages = {195022},
          doi = {10.1088/1361-6382/acf3c7},
archivePrefix = {arXiv},
       eprint = {2304.08287},
 primaryClass = {gr-qc},
       adsurl = {https://ui.adsabs.harvard.edu/abs/2023CQGra..40s5022M}
}

@ARTICLE{mth+19,
       author = {{M{\"u}ller}, Bernhard and {Tauris}, Thomas M. and {Heger}, Alexander and {Banerjee}, Projjwal and {Qian}, Yong-Zhong and {Powell}, Jade and {Chan}, Conrad and {Gay}, Daniel W. and {Langer}, Norbert},
        title = "{Three-dimensional simulations of neutrino-driven core-collapse supernovae from low-mass single and binary star progenitors}",
      journal = {\mnras},
         year = 2019,
        month = apr,
       volume = {484},
       number = {3},
        pages = {3307-3324},
          doi = {10.1093/mnras/stz216},
archivePrefix = {arXiv},
       eprint = {1811.05483},
 primaryClass = {astro-ph.SR},
       adsurl = {https://ui.adsabs.harvard.edu/abs/2019MNRAS.484.3307M}
}

@article{Mueller+2019BAAS,
	author = {Mueller, Guido and Baker, John and Barke, Simon and Bender, Peter L. and Berti, Emanuele and Caldwell, Robert and Conklin, John W. and Cornish, Neil and Ferrara, Elizabeth C. and Holley-Bockelmann, Kelly and Kamai, Brittany and Larson, Shane L. and Livas, Jeff and McWilliams, Sean T. and Mueller, Guido and Natarajan, Priyamvada and Rioux, Norman and Sankar, Shannon S. and Schnittman, Jeremy and Shoemaker, Deirdre and Slutsky, Jacob and Stebbins, Robin and Thorpe, Ira and Ziemer, John},
	journal = {Bulletin of the AAS},
	number = {7},
	year = {2019},
	month = {sep 30},
	note = {https://baas.aas.org/pub/2020n7i243},
	publisher = {},
	title = {Space based gravitational wave astronomy beyond {LISA}},
	volume = {51},
}

@ARTICLE{2003MNRAS.344..629D,
       author = {{Dewi}, J.~D.~M. and {Pols}, O.~R.},
        title = "{The late stages of evolution of helium star-neutron star binaries and the formation of double neutron star systems}",
      journal = {\mnras},
         year = 2003,
        month = sep,
       volume = {344},
       number = {2},
        pages = {629-643},
          doi = {10.1046/j.1365-8711.2003.06844.x},
archivePrefix = {arXiv},
       eprint = {astro-ph/0306066},
 primaryClass = {astro-ph},
       adsurl = {https://ui.adsabs.harvard.edu/abs/2003MNRAS.344..629D}
}

@INPROCEEDINGS{Taylor+1987grg..conf..209T,
       author = {{Taylor}, J.~H.},
        title = "{Astronomical and space experiments to test relativity.}",
    booktitle = {General Relativity and Gravitation},
         year = 1987,
       editor = {{MacCallum}, M.~A.~H.},
        month = jan,
        pages = {209-222},
       adsurl = {https://ui.adsabs.harvard.edu/abs/1987grg..conf..209T},
      publisher={Cambridge University Press},
      address={Cambridge}
}

@ARTICLE{Taylor+1989ApJ...345..434T,
       author = {{Taylor}, J.~H. and {Weisberg}, J.~M.},
        title = "{Further Experimental Tests of Relativistic Gravity Using the Binary Pulsar PSR 1913+16}",
      journal = {\apj},
         year = 1989,
        month = oct,
       volume = {345},
        pages = {434},
          doi = {10.1086/167917},
       adsurl = {https://ui.adsabs.harvard.edu/abs/1989ApJ...345..434T}
}

@ARTICLE{Lyne+2004Sci...303.1153L,
       author = {{Lyne}, A.~G. and {Burgay}, M. and {Kramer}, M. and {Possenti}, A. and {Manchester}, R.~N. and {Camilo}, F. and {McLaughlin}, M.~A. and {Lorimer}, D.~R. and {D'Amico}, N. and {Joshi}, B.~C. and {Reynolds}, J. and {Freire}, P.~C.~C.},
        title = "{A Double-Pulsar System: A Rare Laboratory for Relativistic Gravity and Plasma Physics}",
      journal = {Science},
         year = 2004,
        month = feb,
       volume = {303},
       number = {5661},
        pages = {1153-1157},
          doi = {10.1126/science.1094645},
archivePrefix = {arXiv},
       eprint = {astro-ph/0401086},
 primaryClass = {astro-ph},
       adsurl = {https://ui.adsabs.harvard.edu/abs/2004Sci...303.1153L}
}

@ARTICLE{Splaver+2002ApJ...581..509S,
       author = {{Splaver}, Eric M. and {Nice}, David J. and {Arzoumanian}, Zaven and {Camilo}, Fernando and {Lyne}, Andrew G. and {Stairs}, Ingrid H.},
        title = "{Probing the Masses of the PSR J0621+1002 Binary System through Relativistic Apsidal Motion}",
      journal = {\apj},
         year = 2002,
        month = dec,
       volume = {581},
       number = {1},
        pages = {509-518},
          doi = {10.1086/344202},
archivePrefix = {arXiv},
       eprint = {astro-ph/0208281},
 primaryClass = {astro-ph},
       adsurl = {https://ui.adsabs.harvard.edu/abs/2002ApJ...581..509S}
}

@ARTICLE{wang2021MNRAS.506.4654W,
       author = {{Wang}, Bo and {Chen}, Wen-Cong and {Liu}, Dong-Dong and {Chen}, Hai-Liang and {Wu}, Cheng-Yuan and {Tang}, Wen-Shi and {Guo}, Yun-Lang and {Han}, Zhan-Wen},
        title = "{Ultracompact X-ray binaries with He star companions}",
      journal = {\mnras},
         year = 2021,
        month = sep,
       volume = {506},
       number = {3},
        pages = {4654-4666},
          doi = {10.1093/mnras/stab2032},
archivePrefix = {arXiv},
       eprint = {2106.01369},
 primaryClass = {astro-ph.HE},
       adsurl = {https://ui.adsabs.harvard.edu/abs/2021MNRAS.506.4654W}
}

@ARTICLE{1983ApJ...267..322H,
       author = {{Hills}, J.~G.},
        title = "{The effects of sudden mass loss and a random kick velocity produced in a supernova explosion on the dynamics of a binary star of arbitrary orbital eccentricity. Applications to X-ray binaries and to the binary pulsars.}",
      journal = {\apj},
         year = 1983,
        month = apr,
       volume = {267},
        pages = {322-333},
          doi = {10.1086/160871},
       adsurl = {https://ui.adsabs.harvard.edu/abs/1983ApJ...267..322H}
}

@ARTICLE{Lower+2024A&A...682A..26L,
       author = {{Lower}, M.~E. and {Kramer}, M. and {Shannon}, R.~M. and {Breton}, R.~P. and {Wex}, N. and {Johnston}, S. and {Bailes}, M. and {Buchner}, S. and {Hu}, H. and {Venkatraman Krishnan}, V. and {Blackmon}, V.~A. and {Camilo}, F. and {Champion}, D.~J. and {Freire}, P.~C.~C. and {Geyer}, M. and {Karastergiou}, A. and {van Leeuwen}, J. and {McLaughlin}, M.~A. and {Reardon}, D.~J. and {Stairs}, I.~H.},
        title = "{A MeerKAT view of the double pulsar eclipses. Geodetic precession of pulsar B and system geometry}",
      journal = {\aap},
         year = 2024,
        month = feb,
       volume = {682},
          eid = {A26},
        pages = {A26},
          doi = {10.1051/0004-6361/202347857},
archivePrefix = {arXiv},
       eprint = {2311.06445},
 primaryClass = {astro-ph.HE},
       adsurl = {https://ui.adsabs.harvard.edu/abs/2024A&A...682A..26L}
}

@ARTICLE{Breton+2008Sci...321..104B,
       author = {{Breton}, Rene P. and {Kaspi}, Victoria M. and {Kramer}, Michael and {McLaughlin}, Maura A. and {Lyutikov}, Maxim and {Ransom}, Scott M. and {Stairs}, Ingrid H. and {Ferdman}, Robert D. and {Camilo}, Fernando and {Possenti}, Andrea},
        title = "{Relativistic Spin Precession in the Double Pulsar}",
      journal = {Science},
         year = 2008,
        month = jul,
       volume = {321},
       number = {5885},
        pages = {104},
          doi = {10.1126/science.1159295},
archivePrefix = {arXiv},
       eprint = {0807.2644},
 primaryClass = {astro-ph},
       adsurl = {https://ui.adsabs.harvard.edu/abs/2008Sci...321..104B}
}

@ARTICLE{Stovall+2018ApJ...854L..22S,
       author = {{Stovall}, K. and {Freire}, P.~C.~C. and {Chatterjee}, S. and {Demorest}, P.~B. and {Lorimer}, D.~R. and {McLaughlin}, M.~A. and {Pol}, N. and {van Leeuwen}, J. and {Wharton}, R.~S. and {Allen}, B. and {Boyce}, M. and {Brazier}, A. and {Caballero}, K. and {Camilo}, F. and {Camuccio}, R. and {Cordes}, J.~M. and {Crawford}, F. and {Deneva}, J.~S. and {Ferdman}, R.~D. and {Hessels}, J.~W.~T. and {Jenet}, F.~A. and {Kaspi}, V.~M. and {Knispel}, B. and {Lazarus}, P. and {Lynch}, R. and {Parent}, E. and {Patel}, C. and {Pleunis}, Z. and {Ransom}, S.~M. and {Scholz}, P. and {Seymour}, A. and {Siemens}, X. and {Stairs}, I.~H. and {Swiggum}, J. and {Zhu}, W.~W.},
        title = "{PALFA Discovery of a Highly Relativistic Double Neutron Star Binary}",
      journal = {ApJL},
         year = 2018,
        month = feb,
       volume = {854},
       number = {2},
          eid = {L22},
        pages = {L22},
          doi = {10.3847/2041-8213/aaad06},
archivePrefix = {arXiv},
       eprint = {1802.01707},
 primaryClass = {astro-ph.HE},
       adsurl = {https://ui.adsabs.harvard.edu/abs/2018ApJ...854L..22S}
}

@ARTICLE{Keith+2009MNRAS.393..623K,
       author = {{Keith}, M.~J. and {Kramer}, M. and {Lyne}, A.~G. and {Eatough}, R.~P. and {Stairs}, I.~H. and {Possenti}, A. and {Camilo}, F. and {Manchester}, R.~N.},
        title = "{PSR J1753-2240: a mildly recycled pulsar in an eccentric binary system}",
      journal = {\mnras},
         year = 2009,
        month = feb,
       volume = {393},
       number = {2},
        pages = {623-627},
          doi = {10.1111/j.1365-2966.2008.14234.x},
archivePrefix = {arXiv},
       eprint = {0811.2027},
 primaryClass = {astro-ph},
       adsurl = {https://ui.adsabs.harvard.edu/abs/2009MNRAS.393..623K}
}

@ARTICLE{2018MNRAS.476.4315N,
       author = {{Ng}, C. and {Kruckow}, M.~U. and {Tauris}, T.~M. and {Lyne}, A.~G. and {Freire}, P.~C.~C. and {Ridolfi}, A. and {Caiazzo}, I. and {Heyl}, J. and {Kramer}, M. and {Cameron}, A.~D. and {Champion}, D.~J. and {Stappers}, B.},
        title = "{PSR J1755-2550: a young radio pulsar with a massive, compact companion}",
      journal = {\mnras},
         year = 2018,
        month = jun,
       volume = {476},
       number = {4},
        pages = {4315-4326},
          doi = {10.1093/mnras/sty482},
archivePrefix = {arXiv},
       eprint = {1802.08248},
 primaryClass = {astro-ph.HE},
       adsurl = {https://ui.adsabs.harvard.edu/abs/2018MNRAS.476.4315N}
}

@ARTICLE{Su+2024MNRAS.530.1506S,
       author = {{Su}, W.~Q. and {Han}, J.~L. and {Yang}, Z.~L. and {Wang}, P.~F. and {Yuan}, J.~P. and {Wang}, C. and {Zhou}, D.~J. and {Wang}, T. and {Yan}, Y. and {Jing}, W.~C. and {Cai}, N.~N. and {Xie}, L. and {Xu}, J. and {Wang}, H.~G. and {Xu}, R.~X. and {You}, X.~P.},
        title = "{The FAST Galactic Plane Pulsar Snapshot Survey - V. PSR J1901+0658 in a double neutron star system}",
      journal = {\mnras},
         year = 2024,
        month = may,
       volume = {530},
       number = {2},
        pages = {1506-1511},
          doi = {10.1093/mnras/stae888},
archivePrefix = {arXiv},
       eprint = {2403.11635},
 primaryClass = {astro-ph.HE},
       adsurl = {https://ui.adsabs.harvard.edu/abs/2024MNRAS.530.1506S}
}

@ARTICLE{Tokovinin+2020AJ....160....7T,
       author = {{Tokovinin}, Andrei and {Mason}, Brian D. and {Mendez}, Rene A. and {Costa}, Edgardo and {Horch}, Elliott P.},
        title = "{Speckle Interferometry at SOAR in 2019}",
      journal = {\aj},
         year = 2020,
        month = jul,
       volume = {160},
       number = {1},
          eid = {7},
        pages = {7},
          doi = {10.3847/1538-3881/ab91c1},
archivePrefix = {arXiv},
       eprint = {2005.05305},
 primaryClass = {astro-ph.SR},
       adsurl = {https://ui.adsabs.harvard.edu/abs/2020AJ....160....7T}
}

@ARTICLE{Swiggum+2023ApJ...944..154S,
       author = {{Swiggum}, J.~K. and {Pleunis}, Z. and {Parent}, E. and {Kaplan}, D.~L. and {McLaughlin}, M.~A. and {Stairs}, I.~H. and {Spiewak}, R. and {Agazie}, G.~Y. and {Chawla}, P. and {DeCesar}, M.~E. and {Dolch}, T. and {Fiore}, W. and {Fonseca}, E. and {Istrate}, A.~G. and {Kaspi}, V.~M. and {Kondratiev}, V.~I. and {van Leeuwen}, J. and {Levin}, L. and {Lewis}, E.~F. and {Lynch}, R.~S. and {McEwen}, A.~E. and {Al Noori}, H. and {Ransom}, S.~M. and {Siemens}, X. and {Surnis}, M.},
        title = "{The Green Bank North Celestial Cap Survey. VII. 12 New Pulsar Timing Solutions}",
      journal = {\apj},
         year = 2023,
        month = feb,
       volume = {944},
       number = {2},
          eid = {154},
        pages = {154},
          doi = {10.3847/1538-4357/acb43f},
archivePrefix = {arXiv},
       eprint = {2212.03926},
 primaryClass = {astro-ph.HE},
       adsurl = {https://ui.adsabs.harvard.edu/abs/2023ApJ...944..154S}
}

@ARTICLE{Ridolfi+2019MNRAS.490.3860R,
       author = {{Ridolfi}, A. and {Freire}, P.~C.~C. and {Gupta}, Y. and {Ransom}, S.~M.},
        title = "{Upgraded Giant Metrewave Radio Telescope timing of NGC 1851A: a possible millisecond pulsar - neutron star system}",
      journal = {\mnras},
         year = 2019,
        month = dec,
       volume = {490},
       number = {3},
        pages = {3860-3874},
          doi = {10.1093/mnras/stz2645},
archivePrefix = {arXiv},
       eprint = {1909.06163},
 primaryClass = {astro-ph.HE},
       adsurl = {https://ui.adsabs.harvard.edu/abs/2019MNRAS.490.3860R}
}

@ARTICLE{Wu+2023ApJ...958L..17W,
       author = {{Wu}, Q.~D. and {Wang}, N. and {Yuan}, J.~P. and {Li}, D. and {Wang}, P. and {Xue}, M.~Y. and {Zhu}, W.~W. and {Miao}, C.~C. and {Yan}, W.~M. and {Wang}, J.~B. and {Yao}, J.~M. and {Wang}, S.~Q. and {Sun}, S.~N. and {Kou}, F.~F. and {Zhao}, D. and {Chen}, Y.~T. and {Dang}, S.~J. and {Feng}, Y. and {Liu}, Z.~J. and {Miao}, X.~L. and {Meng}, L.~Q. and {Yuan}, M. and {Niu}, C.~H. and {Niu}, J.~R. and {Qian}, L. and {Wang}, S. and {Xie}, X.~Y. and {Xiao}, Y.~F. and {Yue}, Y.~L. and {You}, S.~P. and {Yu}, X.~H. and {Zhao}, R.~S. and {Yuen}, R. and {Zhou}, X. and {Zhang}, L.},
        title = "{PSR J2150+3427: A Possible Double Neutron Star System}",
      journal = {\apjl},
         year = 2023,
        month = nov,
       volume = {958},
       number = {1},
          eid = {L17},
        pages = {L17},
          doi = {10.3847/2041-8213/ad0887},
       adsurl = {https://ui.adsabs.harvard.edu/abs/2023ApJ...958L..17W}
}

@ARTICLE{Jacoby+2006ApJ...644L.113J,
       author = {{Jacoby}, B.~A. and {Cameron}, P.~B. and {Jenet}, F.~A. and {Anderson}, S.~B. and {Murty}, R.~N. and {Kulkarni}, S.~R.},
        title = "{Measurement of Orbital Decay in the Double Neutron Star Binary PSR B2127+11C}",
      journal = {\apjl},
         year = 2006,
        month = jun,
       volume = {644},
       number = {2},
        pages = {L113-L116},
          doi = {10.1086/505742},
archivePrefix = {arXiv},
       eprint = {astro-ph/0605375},
 primaryClass = {astro-ph},
       adsurl = {https://ui.adsabs.harvard.edu/abs/2006ApJ...644L.113J}
}

@ARTICLE{Cameron+2023MNRAS.523.5064C,
       author = {{Cameron}, A.~D. and {Bailes}, M. and {Champion}, D.~J. and {Freire}, P.~C.~C. and {Kramer}, M. and {McLaughlin}, M.~A. and {Ng}, C. and {Possenti}, A. and {Ridolfi}, A. and {Tauris}, T.~M. and {Wahl}, H.~M. and {Wex}, N.},
        title = "{New constraints on the kinematic, relativistic, and evolutionary properties of the PSR J1757-1854 double neutron star system}",
      journal = {\mnras},
         year = 2023,
        month = aug,
       volume = {523},
       number = {4},
        pages = {5064-5085},
          doi = {10.1093/mnras/stad1712},
archivePrefix = {arXiv},
       eprint = {2305.14733},
 primaryClass = {astro-ph.HE},
       adsurl = {https://ui.adsabs.harvard.edu/abs/2023MNRAS.523.5064C}
}

@ARTICLE{Paxton+2019ApJS..243...10P,
       author = {{Paxton}, Bill and {Smolec}, R. and {Schwab}, Josiah and {Gautschy}, A. and {Bildsten}, Lars and {Cantiello}, Matteo and {Dotter}, Aaron and {Farmer}, R. and {Goldberg}, Jared A. and {Jermyn}, Adam S. and {Kanbur}, S.~M. and {Marchant}, Pablo and {Thoul}, Anne and {Townsend}, Richard H.~D. and {Wolf}, William M. and {Zhang}, Michael and {Timmes}, F.~X.},
        title = "{Modules for Experiments in Stellar Astrophysics (MESA): Pulsating Variable Stars, Rotation, Convective Boundaries, and Energy Conservation}",
      journal = {\apjs},
         year = 2019,
        month = jul,
       volume = {243},
       number = {1},
          eid = {10},
        pages = {10},
          doi = {10.3847/1538-4365/ab2241},
archivePrefix = {arXiv},
       eprint = {1903.01426},
 primaryClass = {astro-ph.SR},
       adsurl = {https://ui.adsabs.harvard.edu/abs/2019ApJS..243...10P}
}

@ARTICLE{Paxton+2011ApJS..192....3P,
       author = {{Paxton}, Bill and {Bildsten}, Lars and {Dotter}, Aaron and {Herwig}, Falk and {Lesaffre}, Pierre and {Timmes}, Frank},
        title = "{Modules for Experiments in Stellar Astrophysics (MESA)}",
      journal = {\apjs},
         year = 2011,
        month = jan,
       volume = {192},
       number = {1},
          eid = {3},
        pages = {3},
          doi = {10.1088/0067-0049/192/1/3},
archivePrefix = {arXiv},
       eprint = {1009.1622},
 primaryClass = {astro-ph.SR},
       adsurl = {https://ui.adsabs.harvard.edu/abs/2011ApJS..192....3P}
}

@ARTICLE{Farrow+2019ApJ...876...18F,
       author = {{Farrow}, Nicholas and {Zhu}, Xing-Jiang and {Thrane}, Eric},
        title = "{The Mass Distribution of Galactic Double Neutron Stars}",
      journal = {\apj},
         year = 2019,
        month = may,
       volume = {876},
       number = {1},
          eid = {18},
        pages = {18},
          doi = {10.3847/1538-4357/ab12e3},
archivePrefix = {arXiv},
       eprint = {1902.03300},
 primaryClass = {astro-ph.HE},
       adsurl = {https://ui.adsabs.harvard.edu/abs/2019ApJ...876...18F}
}

@ARTICLE{Hobbs+2006MNRAS.369..655H,
       author = {{Hobbs}, G.~B. and {Edwards}, R.~T. and {Manchester}, R.~N.},
        title = "{TEMPO2, a new pulsar-timing package - I. An overview}",
      journal = {MNRAS},
         year = 2006,
        month = jun,
       volume = {369},
       number = {2},
        pages = {655-672},
          doi = {10.1111/j.1365-2966.2006.10302.x},
archivePrefix = {arXiv},
       eprint = {astro-ph/0603381},
 primaryClass = {astro-ph},
       adsurl = {https://ui.adsabs.harvard.edu/abs/2006MNRAS.369..655H}
}

@ARTICLE{Cameron+2018MNRAS.475L..57C,
       author = {{Cameron}, A.~D. and {Champion}, D.~J. and {Kramer}, M. and {Bailes}, M. and {Barr}, E.~D. and {Bassa}, C.~G. and {Bhandari}, S. and {Bhat}, N.~D.~R. and {Burgay}, M. and {Burke-Spolaor}, S. and {Eatough}, R.~P. and {Flynn}, C.~M.~L. and {Freire}, P.~C.~C. and {Jameson}, A. and {Johnston}, S. and {Karuppusamy}, R. and {Keith}, M.~J. and {Levin}, L. and {Lorimer}, D.~R. and {Lyne}, A.~G. and {McLaughlin}, M.~A. and {Ng}, C. and {Petroff}, E. and {Possenti}, A. and {Ridolfi}, A. and {Stappers}, B.~W. and {van Straten}, W. and {Tauris}, T.~M. and {Tiburzi}, C. and {Wex}, N.},
        title = "{The High Time Resolution Universe Pulsar Survey - XIII. PSR J1757-1854, the most accelerated binary pulsar}",
      journal = {\mnras},
         year = 2018,
        month = mar,
       volume = {475},
       number = {1},
        pages = {L57-L61},
          doi = {10.1093/mnrasl/sly003},
archivePrefix = {arXiv},
       eprint = {1711.07697},
 primaryClass = {astro-ph.HE},
       adsurl = {https://ui.adsabs.harvard.edu/abs/2018MNRAS.475L..57C}
}

@ARTICLE{Nan+2006ScChG..49..129N,
       author = {{Nan}, Rendong},
        title = "{Five hundred meter aperture spherical radio telescope (FAST)}",
      journal = {Science in China: Physics, Mechanics and Astronomy},
         year = 2006,
        month = mar,
       volume = {49},
       number = {2},
        pages = {129-148},
          doi = {10.1007/s11433-006-0129-9},
       adsurl = {https://ui.adsabs.harvard.edu/abs/2006ScChG..49..129N}
}

@ARTICLE{Lange+2001MNRAS.326..274L,
       author = {{Lange}, Ch. and {Camilo}, F. and {Wex}, N. and {Kramer}, M. and {Backer}, D.~C. and {Lyne}, A.~G. and {Doroshenko}, O.},
        title = "{Precision timing measurements of PSR J1012+5307}",
      journal = {\mnras},
         year = 2001,
        month = sep,
       volume = {326},
       number = {1},
        pages = {274-282},
          doi = {10.1046/j.1365-8711.2001.04606.x},
archivePrefix = {arXiv},
       eprint = {astro-ph/0102309},
 primaryClass = {astro-ph},
       adsurl = {https://ui.adsabs.harvard.edu/abs/2001MNRAS.326..274L}
}

@ARTICLE{Padmanabh+2023MNRAS.524.1291P,
       author = {{Padmanabh}, P.~V. and {Barr}, E.~D. and {Sridhar}, S.~S. and {Rugel}, M.~R. and {Damas-Segovia}, A. and {Jacob}, A.~M. and {Balakrishnan}, V. and {Berezina}, M. and {Bernadich}, M.~C. and {Brunthaler}, A. and {Champion}, D.~J. and {Freire}, P.~C.~C. and {Khan}, S. and {Kl{\"o}ckner}, H.-R. and {Kramer}, M. and {Ma}, Y.~K. and {Mao}, S.~A. and {Men}, Y.~P. and {Menten}, K.~M. and {Sengupta}, S. and {Venkatraman Krishnan}, V. and {Wucknitz}, O. and {Wyrowski}, F. and {Bezuidenhout}, M.~C. and {Buchner}, S. and {Burgay}, M. and {Chen}, W. and {Clark}, C.~J. and {K{\"u}nkel}, L. and {Nieder}, L. and {Stappers}, B. and {Legodi}, L.~S. and {Nyamai}, M.~M.},
        title = "{The MPIfR-MeerKAT Galactic Plane Survey - I. System set-up and early results}",
      journal = {\mnras},
         year = 2023,
        month = sep,
       volume = {524},
       number = {1},
        pages = {1291-1315},
          doi = {10.1093/mnras/stad1900},
archivePrefix = {arXiv},
       eprint = {2303.09231},
 primaryClass = {astro-ph.HE},
       adsurl = {https://ui.adsabs.harvard.edu/abs/2023MNRAS.524.1291P}
}

@ARTICLE{Han+2021RAA....21..107H,
       author = {{Han}, J.~L. and {Wang}, Chen and {Wang}, P.~F. and {Wang}, Tao and {Zhou}, D.~J. and {Sun}, Jing-Hai and {Yan}, Yi and {Su}, Wei-Qi and {Jing}, Wei-Cong and {Chen}, Xue and {Gao}, X.~Y. and {Hou}, Li-Gang and {Xu}, Jun and {Lee}, K.~J. and {Wang}, Na and {Jiang}, Peng and {Xu}, Ren-Xin and {Yan}, Jun and {Gan}, Heng-Qian and {Guan}, Xin and {Huang}, Wen-Jun and {Jiang}, Jin-Chen and {Li}, Hui and {Men}, Yun-Peng and {Sun}, Chun and {Wang}, Bo-Jun and {Wang}, H.~G. and {Wang}, Shuang-Qiang and {Xie}, Jin-Tao and {Xu}, Heng and {Yao}, Rui and {You}, Xiao-Peng and {Yu}, D.~J. and {Yuan}, Jian-Ping and {Yuen}, Rai and {Zhang}, Chun-Feng and {Zhu}, Yan},
        title = "{The FAST Galactic Plane Pulsar Snapshot survey: I. Project design and pulsar discoveries}",
      journal = {Research in Astronomy and Astrophysics},
         year = 2021,
        month = jun,
       volume = {21},
       number = {5},
          eid = {107},
        pages = {107},
          doi = {10.1088/1674-4527/21/5/107},
archivePrefix = {arXiv},
       eprint = {2105.08460},
 primaryClass = {astro-ph.HE},
       adsurl = {https://ui.adsabs.harvard.edu/abs/2021RAA....21..107H}
}

@ARTICLE{Damour+1985AIHPA..43..107D,
       author = {{Damour}, T. and {Deruelle}, N.},
        title = "{General relativistic celestial mechanics of binary systems. I. The post-Newtonian motion.}",
      journal = {Annales de L'Institut Henri Poincare Section (A) Physique Theorique},
         year = 1985,
        month = jan,
       volume = {43},
       number = {1},
        pages = {107-132},
       adsurl = {https://ui.adsabs.harvard.edu/abs/1985AIHPA..43..107D}
}

@ARTICLE{Abbott+2017PhRvL.119p1101A,
       author = {{Abbott}, B.~P. and {Abbott}, R. and {Abbott}, T.~D. and {Acernese}, F. and {Ackley}, K. and {Adams}, C. and {Adams}, T. and {Addesso}, P. and {Adhikari}, R.~X. and {Adya}, V.~B. and et al.},
        title = "{GW170817: Observation of Gravitational Waves from a Binary Neutron Star Inspiral}",
      journal = {\prl},
         year = 2017,
        month = oct,
       volume = {119},
       number = {16},
          eid = {161101},
        pages = {161101},
          doi = {10.1103/PhysRevLett.119.161101},
archivePrefix = {arXiv},
       eprint = {1710.05832},
 primaryClass = {gr-qc},
       adsurl = {https://ui.adsabs.harvard.edu/abs/2017PhRvL.119p1101A}
}

@ARTICLE{Abbott+2020ApJ...892L...3A,
       author = {{Abbott}, B.~P. and {Abbott}, R. and {Abbott}, T.~D. and {Abraham}, S. and {Acernese}, F. and {Ackley}, K. and {Adams}, C. and {Adhikari}, R.~X. and {Adya}, V.~B. and {Affeldt}, C. and et al.},
        title = "{GW190425: Observation of a Compact Binary Coalescence with Total Mass {\ensuremath{\sim}} 3.4 M$_{{\ensuremath{\odot}}}$}",
      journal = {\apj},
         year = 2020,
        month = mar,
       volume = {892},
       number = {1},
          eid = {L3},
        pages = {L3},
          doi = {10.3847/2041-8213/ab75f5},
archivePrefix = {arXiv},
       eprint = {2001.01761},
 primaryClass = {astro-ph.HE},
       adsurl = {https://ui.adsabs.harvard.edu/abs/2020ApJ...892L...3A}
}

@PHDTHESIS{Ransom+2001PhDT.......123R,
       author = {{Ransom}, Scott Mitchell},
        title = "{New search techniques for binary pulsars}",
       school = {Harvard University, Massachusetts},
         year = 2001,
        month = jan,
       adsurl = {https://ui.adsabs.harvard.edu/abs/2001PhDT.......123R}
}

@BOOK{lori2012,
       author = {{Lorimer}, D.~R. and {Kramer}, M.},
        title = "Handbook of Pulsar Astronomy",
         year = 2012,
       adsurl = {https://ui.adsabs.harvard.edu/abs/2012hpa..book.....L},
      address={Cambridge},
      publisher={Cambridge University Press}
}

@book{PulsarAstronomy2022, 
     edition={5}, 
     series={Cambridge Astrophysics}, 
     title={Pulsar Astronomy}, 
     publisher={Cambridge University Press}, 
     author={Lyne, Andrew and Graham-Smith, Francis and Stappers, Benjamin}, year={2022}, 
     collection={Cambridge Astrophysics},
     address={Cambridge}
}

@ARTICLE{Straten+2011PASA...28....1V,
       author = {{van Straten}, W. and {Bailes}, M.},
        title = "{DSPSR: Digital Signal Processing Software for Pulsar Astronomy}",
      journal = {\pasa},
         year = 2011,
        month = jan,
       volume = {28},
       number = {1},
        pages = {1-14},
          doi = {10.1071/AS10021},
archivePrefix = {arXiv},
       eprint = {1008.3973},
 primaryClass = {astro-ph.IM},
       adsurl = {https://ui.adsabs.harvard.edu/abs/2011PASA...28....1V}
}

@ARTICLE{Hotan+2004PASA...21..302H,
       author = {{Hotan}, A.~W. and {van Straten}, W. and {Manchester}, R.~N.},
        title = "{PSRCHIVE and PSRFITS: An Open Approach to Radio Pulsar Data Storage and Analysis}",
      journal = {\pasa},
         year = 2004,
        month = jan,
       volume = {21},
       number = {3},
        pages = {302-309},
          doi = {10.1071/AS04022},
archivePrefix = {arXiv},
       eprint = {astro-ph/0404549},
 primaryClass = {astro-ph},
       adsurl = {https://ui.adsabs.harvard.edu/abs/2004PASA...21..302H}
}

@ARTICLE{Chen+2023RAA....23j4004C,
       author = {{Chen}, Xue and {Han}, J.~L. and {Su}, W.~Q. and {Yang}, Z.~L. and {Zhou}, D.~J.},
        title = "{Cleaning Radio Frequency Interference in Pulsar-Folded Data Based on the Conditional Random Fields with an Adaptive Prior}",
      journal = {Research in Astronomy and Astrophysics},
         year = 2023,
        month = oct,
       volume = {23},
       number = {10},
          eid = {104004},
        pages = {104004},
          doi = {10.1088/1674-4527/acd52b},
       adsurl = {https://ui.adsabs.harvard.edu/abs/2023RAA....23j4004C}
}

@ARTICLE{Chu2022MNRAS.509.1557C,
       author = {{Chu}, Qingbo and {Yu}, Shenghua and {Lu}, Youjun},
        title = "{Formation and evolution of binary neutron stars: mergers and their host galaxies}",
      journal = {\mnras},
         year = 2022,
        month = jan,
       volume = {509},
       number = {2},
        pages = {1557-1586},
          doi = {10.1093/mnras/stab2882},
archivePrefix = {arXiv},
       eprint = {2110.04687},
 primaryClass = {astro-ph.GA},
       adsurl = {https://ui.adsabs.harvard.edu/abs/2022MNRAS.509.1557C}
}

@ARTICLE{Moore+2015CQGra..32a5014M,
       author = {{Moore}, C.~J. and {Cole}, R.~H. and {Berry}, C.~P.~L.},
        title = "{Gravitational-wave sensitivity curves}",
      journal = {Classical and Quantum Gravity},
         year = 2015,
        month = jan,
       volume = {32},
       number = {1},
          eid = {015014},
        pages = {015014},
          doi = {10.1088/0264-9381/32/1/015014},
archivePrefix = {arXiv},
       eprint = {1408.0740},
 primaryClass = {gr-qc},
       adsurl = {https://ui.adsabs.harvard.edu/abs/2015CQGra..32a5014M}
}

@ARTICLE{Kramer+2021PhRvX..11d1050K,
       author = {{Kramer}, M. and {Stairs}, I.~H. and {Manchester}, R.~N. and {Wex}, N. and {Deller}, A.~T. and {Coles}, W.~A. and {Ali}, M. and {Burgay}, M. and {Camilo}, F. and {Cognard}, I. and {Damour}, T. and {Desvignes}, G. and {Ferdman}, R.~D. and {Freire}, P.~C.~C. and {Grondin}, S. and {Guillemot}, L. and {Hobbs}, G.~B. and {Janssen}, G. and {Karuppusamy}, R. and {Lorimer}, D.~R. and {Lyne}, A.~G. and {McKee}, J.~W. and {McLaughlin}, M. and {M{\"u}nch}, L.~E. and {Perera}, B.~B.~P. and {Pol}, N. and {Possenti}, A. and {Sarkissian}, J. and {Stappers}, B.~W. and {Theureau}, G.},
        title = "{Strong-Field Gravity Tests with the Double Pulsar}",
      journal = {PhRvX},
         year = 2021,
        month = oct,
       volume = {11},
       number = {4},
          eid = {041050},
        pages = {041050},
          doi = {10.1103/PhysRevX.11.041050},
archivePrefix = {arXiv},
       eprint = {2112.06795},
 primaryClass = {astro-ph.HE},
       adsurl = {https://ui.adsabs.harvard.edu/abs/2021PhRvX..11d1050K}
}

@BOOK{Poisson+2014grav.book.....P,
       author = {{Poisson}, Eric and {Will}, Clifford M.},
        title = "{Gravity}",
         year = 2014,
       adsurl = {https://ui.adsabs.harvard.edu/abs/2014grav.book.....P},
    publisher={Cambridge University Press},
      address={Cambridge}
}

@Inbook{Loutrel+2020arXiv200911332L,
author="Loutrel, Nicholas",
title="Repeated Bursts",
bookTitle="Handbook of Gravitational Wave Astronomy",
year="2020",
publisher="Springer Singapore",
address="Singapore",
pages="1--35",
isbn="978-981-15-4702-7",
doi="10.1007/978-981-15-4702-7_33-1",
url="https://doi.org/10.1007/978-981-15-4702-7_33-1"
}

@ARTICLE{Peters+1964PhRv..136.1224P,
       author = {{Peters}, P.~C.},
        title = "{Gravitational Radiation and the Motion of Two Point Masses}",
      journal = {Physical Review},
         year = 1964,
        month = nov,
       volume = {136},
       number = {4B},
        pages = {1224-1232},
          doi = {10.1103/PhysRev.136.B1224},
       adsurl = {https://ui.adsabs.harvard.edu/abs/1964PhRv..136.1224P}
}

@ARTICLE{Blandford+1976ApJ...205..580B,
       author = {{Blandford}, R. and {Teukolsky}, S.~A.},
        title = "{Arrival-time analysis for a pulsar in a binary system.}",
      journal = {\apj},
         year = 1976,
        month = apr,
       volume = {205},
        pages = {580-591},
          doi = {10.1086/154315},
       adsurl = {https://ui.adsabs.harvard.edu/abs/1976ApJ...205..580B}
}

@ARTICLE{Liu+2018ApJ...863...68L,
       author = {{Liu}, Bin and {Lai}, Dong},
        title = "{Black Hole and Neutron Star Binary Mergers in Triple Systems: Merger Fraction and Spin-Orbit Misalignment}",
      journal = {\apj},
         year = 2018,
        month = aug,
       volume = {863},
       number = {1},
          eid = {68},
        pages = {68},
          doi = {10.3847/1538-4357/aad09f},
archivePrefix = {arXiv},
       eprint = {1805.03202},
 primaryClass = {astro-ph.HE},
       adsurl = {https://ui.adsabs.harvard.edu/abs/2018ApJ...863...68L}
}

@ARTICLE{Liu+2015MNRAS.447..747L,
       author = {{Liu}, Bin and {Mu{\~n}oz}, Diego J. and {Lai}, Dong},
        title = "{Suppression of extreme orbital evolution in triple systems with short-range forces}",
      journal = {\mnras},
         year = 2015,
        month = feb,
       volume = {447},
       number = {1},
        pages = {747-764},
          doi = {10.1093/mnras/stu2396},
archivePrefix = {arXiv},
       eprint = {1409.6717},
 primaryClass = {astro-ph.EP},
       adsurl = {https://ui.adsabs.harvard.edu/abs/2015MNRAS.447..747L}
}

@ARTICLE{Damour+1992PhRvD..45.1840D,
       author = {{Damour}, Thibault and {Taylor}, J.~H.},
        title = "{Strong-field tests of relativistic gravity and binary pulsars}",
      journal = {\prd},
         year = 1992,
        month = mar,
       volume = {45},
       number = {6},
        pages = {1840-1868},
          doi = {10.1103/PhysRevD.45.1840},
       adsurl = {https://ui.adsabs.harvard.edu/abs/1992PhRvD..45.1840D}
}

@ARTICLE{Lower+2024MNRAS.534.3936L,
       author = {{Lower}, M.~E. and {Kramer}, M. and {Johnston}, S. and {Breton}, R.~P. and {Wex}, N. and {Bailes}, M. and {Buchner}, S. and {Camilo}, F. and {Oswald}, L.~S. and {Reardon}, D.~J. and {Shannon}, R.~M. and {Serylak}, M. and {Krishnan}, V. Venkatraman},
        title = "{MeerKAT observations of pair-plasma induced birefringence in the double pulsar eclipses}",
      journal = {\mnras},
         year = 2024,
        month = nov,
       volume = {534},
       number = {4},
        pages = {3936-3943},
          doi = {10.1093/mnras/stae2386},
archivePrefix = {arXiv},
       eprint = {2410.12510},
 primaryClass = {astro-ph.HE},
       adsurl = {https://ui.adsabs.harvard.edu/abs/2024MNRAS.534.3936L}
}

@ARTICLE{Hulse+1975ApJ...195L..51H,
       author = {{Hulse}, R.~A. and {Taylor}, J.~H.},
        title = "{Discovery of a pulsar in a binary system.}",
      journal = {\apjl},
         year = 1975,
        month = jan,
       volume = {195},
        pages = {L51-L53},
          doi = {10.1086/181708},
       adsurl = {https://ui.adsabs.harvard.edu/abs/1975ApJ...195L..51H}
}

@ARTICLE{meng2025doubleneutronstarpsr,
         author = {{Meng}, Lingqi and {Freire}, Paulo C.~C. and {Stovall}, Kevin and {Wex}, Norbert and {Miao}, Xueli and {Zhu}, Weiwei and {Kramer}, Michael and {Cordes}, James M. and {Hu}, Huanchen and {Jiang}, Jinchen and {Parent}, Emilie and {Shao}, Lijing and {Stairs}, Ingrid H. and {Xue}, Mengyao and {Brazier}, Adam and {Camilo}, Fernando and {Champion}, David J. and {Chatterjee}, Shami and {Crawford}, Fronefield and {Fang}, Ziyao and {Fu}, Qiuyang and {Guo}, Yanjun and {Hessels}, Jason W.~T. and {MacLaughlin}, Maura and {Miao}, Chenchen and {Niu}, Jiarui and {Wu}, Ziwei and {Yao}, Jumei and {Yuan}, Mao and {Yue}, Youlin and {Zhang}, Chengmin},
        title = "{The double neutron star PSR J1946+2052: I. Masses and tests of general relativity}",
      journal = {\aap},
         year = 2025,
        month = dec,
       volume = {704},
          eid = {A153},
        pages = {A153},
          doi = {10.1051/0004-6361/202555689},
archivePrefix = {arXiv},
       eprint = {2510.12506},
 primaryClass = {astro-ph.HE},
       adsurl = {https://ui.adsabs.harvard.edu/abs/2025A&A...704A.153M}
}

@ARTICLE{Zhao+2024ApJ...964L...7Z,
       author = {{Zhao}, D. and {Wang}, N. and {Yuan}, J.~P. and {Li}, D. and {Wang}, P. and {Xue}, M.~Y. and {Zhu}, W.~W. and {Miao}, C.~C. and {Yan}, W.~M. and {Wang}, J.~B. and {Yao}, J.~M. and {Wu}, Q.~D. and {Wang}, S.~Q. and {Sun}, S.~N. and {Kou}, F.~F. and {Chen}, Y.~T. and {Dang}, S.~J. and {Feng}, Y. and {Liu}, Z.~J. and {Miao}, X.~L. and {Meng}, L.~Q. and {Yuan}, M. and {Niu}, C.~H. and {Niu}, J.~R. and {Qian}, L. and {Wang}, S. and {Xie}, X.~Y. and {Xiao}, Y.~F. and {Yue}, Y.~L. and {You}, S.~P. and {Yu}, X.~H. and {Zhao}, R.~S. and {Yuen}, R. and {Zhou}, X. and {Zhang}, L. and {Wang}, Y.~B. and {Wu}, J.~F. and {Gan}, Z.~Y. and {Sun}, Z.~Y. and {Wang}, C.~J.},
        title = "{A Relativistic Double Neutron Star Binary PSR J1846-0513}",
      journal = {\apjl},
         year = 2024,
        month = mar,
       volume = {964},
       number = {1},
          eid = {L7},
        pages = {L7},
          doi = {10.3847/2041-8213/ad2fb3},
       adsurl = {https://ui.adsabs.harvard.edu/abs/2024ApJ...964L...7Z}
}

@ARTICLE{Bernadich+2023AandA...678A.187C,
       author = {{Colom i Bernadich}, M. and {Balakrishnan}, V. and {Barr}, E. and {Berezina}, M. and {Burgay}, M. and {Buchner}, S. and {Champion}, D.~J. and {Chen}, W. and {Desvignes}, G. and {Freire}, P.~C.~C. and {Grunthal}, K. and {Kramer}, M. and {Men}, Y. and {Padmanabh}, P.~V. and {Parthasarathy}, A. and {Pillay}, D. and {Rammala}, I. and {Sengupta}, S. and {Venkatraman Krishnan}, V.},
        title = "{The MPIfR-MeerKAT Galactic Plane Survey. II. The eccentric double neutron star system PSR J1208{\ensuremath{-}}5936 and a neutron star merger rate update}",
      journal = {\aap},
         year = 2023,
        month = oct,
       volume = {678},
          eid = {A187},
        pages = {A187},
          doi = {10.1051/0004-6361/202346953},
archivePrefix = {arXiv},
       eprint = {2308.16802},
 primaryClass = {astro-ph.HE},
       adsurl = {https://ui.adsabs.harvard.edu/abs/2023A&A...678A.187C}
}

@ARTICLE{Balakrishnan+2023ApJ...942L..35B,
       author = {{Balakrishnan}, Vishnu and {Freire}, Paulo C.~C. and {Ransom}, S.~M. and {Ridolfi}, Alessandro and {Barr}, E.~D. and {Chen}, W. and {Krishnan}, Vivek Venkatraman and {Champion}, D. and {Kramer}, M. and {Gautam}, T. and {Padmanabh}, Prajwal V. and {Men}, Yunpeng and {Abbate}, F. and {Stappers}, B.~W. and {Stairs}, I. and {Keane}, E. and {Possenti}, A.},
        title = "{Missing for 20 yr: MeerKAT Redetects the Elusive Binary Pulsar M30B}",
      journal = {\apjl},
         year = 2023,
        month = jan,
       volume = {942},
       number = {2},
          eid = {L35},
        pages = {L35},
          doi = {10.3847/2041-8213/acae99},
archivePrefix = {arXiv},
       eprint = {2301.04983},
 primaryClass = {astro-ph.HE},
       adsurl = {https://ui.adsabs.harvard.edu/abs/2023ApJ...942L..35B}
}

@ARTICLE{Padmanabh+2024AandA...686A.166P,
       author = {{Padmanabh}, P.~V. and {Ransom}, S.~M. and {Freire}, P.~C.~C. and {Ridolfi}, A. and {Taylor}, J.~D. and {Choza}, C. and {Clark}, C.~J. and {Abbate}, F. and {Bailes}, M. and {Barr}, E.~D. and {Buchner}, S. and {Burgay}, M. and {DeCesar}, M.~E. and {Chen}, W. and {Corongiu}, A. and {Champion}, D.~J. and {Dutta}, A. and {Geyer}, M. and {Hessels}, J.~W.~T. and {Kramer}, M. and {Possenti}, A. and {Stairs}, I.~H. and {Stappers}, B.~W. and {Venkatraman Krishnan}, V. and {Vleeschower}, L. and {Zhang}, L.},
        title = "{Discovery and timing of ten new millisecond pulsars in the globular cluster Terzan 5}",
      journal = {\aap},
         year = 2024,
        month = jun,
       volume = {686},
          eid = {A166},
        pages = {A166},
          doi = {10.1051/0004-6361/202449303},
archivePrefix = {arXiv},
       eprint = {2403.17799},
 primaryClass = {astro-ph.HE},
       adsurl = {https://ui.adsabs.harvard.edu/abs/2024A&A...686A.166P}
}

@ARTICLE{Gao+2022SCPMA..6529705G,
       author = {{Gao}, XuYang and {Reich}, Wolfgang and {Sun}, XiaoHui and {Zhao}, He and {Hong}, Tao and {Yuan}, ZhongSheng and {Reich}, Patricia and {Han}, JinLin},
        title = "{Peering into the Milky Way by FAST: IV. Identification of two new Galactic supernova remnants G203.1+6.6 and G206.7+5.9}",
      journal = {Science China Physics, Mechanics, and Astronomy},
         year = 2022,
        month = dec,
       volume = {65},
       number = {12},
          eid = {129705},
        pages = {129705},
          doi = {10.1007/s11433-022-2031-7},
archivePrefix = {arXiv},
       eprint = {2211.11408},
 primaryClass = {astro-ph.GA},
       adsurl = {https://ui.adsabs.harvard.edu/abs/2022SCPMA..6529705G}
}

@ARTICLE{Jiang+2020RAA....20...64J,
       author = {{Jiang}, Peng and {Tang}, Ning-Yu and {Hou}, Li-Gang and {Liu}, Meng-Ting and {Kr{\v{c}}o}, Marko and {Qian}, Lei and {Sun}, Jing-Hai and {Ching}, Tao-Chung and {Liu}, Bin and {Duan}, Yan and {Yue}, You-Ling and {Gan}, Heng-Qian and {Yao}, Rui and {Li}, Hui and {Pan}, Gao-Feng and {Yu}, Dong-Jun and {Liu}, Hong-Fei and {Li}, Di and {Peng}, Bo and {Yan}, Jun and {FAST Collaboration}},
        title = "{The fundamental performance of FAST with 19-beam receiver at L band}",
      journal = {Research in Astronomy and Astrophysics},
         year = 2020,
        month = may,
       volume = {20},
       number = {5},
          eid = {064},
        pages = {064},
          doi = {10.1088/1674-4527/20/5/64},
archivePrefix = {arXiv},
       eprint = {2002.01786},
 primaryClass = {astro-ph.IM},
       adsurl = {https://ui.adsabs.harvard.edu/abs/2020RAA....20...64J}
}

@ARTICLE{Xuan+2025PhRvD.111d3018X,
       author = {{Xuan}, Zeyuan and {Naoz}, Smadar and {Li}, Alvin K.~Y. and {Kocsis}, Bence and {Petigura}, Erik and {Knee}, Alan M. and {McIver}, Jess and {Kremer}, Kyle and {Farr}, Will M.},
        title = "{Extracting astrophysical information of highly eccentric binaries in the millihertz gravitational wave band}",
      journal = {\prd},
         year = 2025,
        month = feb,
       volume = {111},
       number = {4},
          eid = {043018},
        pages = {043018},
          doi = {10.1103/PhysRevD.111.043018},
archivePrefix = {arXiv},
       eprint = {2409.15413},
 primaryClass = {astro-ph.HE},
       adsurl = {https://ui.adsabs.harvard.edu/abs/2025PhRvD.111d3018X}
}

@ARTICLE{zhang2000deathline,
       author = {{Zhang}, Bing and {Harding}, Alice K. and {Muslimov}, Alexander G.},
        title = "{Radio Pulsar Death Line Revisited: Is PSR J2144-3933 Anomalous?}",
      journal = {\apjl},
         year = 2000,
        month = mar,
       volume = {531},
       number = {2},
        pages = {L135-L138},
          doi = {10.1086/312542},
archivePrefix = {arXiv},
       eprint = {astro-ph/0001341},
 primaryClass = {astro-ph},
       adsurl = {https://ui.adsabs.harvard.edu/abs/2000ApJ...531L.135Z}
}

@ARTICLE{Taylor+1982ApJ...253..908T,
       author = {{Taylor}, J.~H. and {Weisberg}, J.~M.},
        title = "{A new test of general relativity - Gravitational radiation and the binary pulsar PSR 1913+16}",
      journal = {\apj},
         year = 1982,
        month = feb,
       volume = {253},
        pages = {908-920},
          doi = {10.1086/159690},
       adsurl = {https://ui.adsabs.harvard.edu/abs/1982ApJ...253..908T}
}

@ARTICLE{Guo+2024MNRAS.530.4461G,
       author = {{Guo}, Yun-Lang and {Wang}, Bo and {Chen}, Wen-Cong and {Li}, Xiang-Dong and {Ge}, Hong-Wei and {Jiang}, Long and {Han}, Zhan-Wen},
        title = "{Electron-capture supernovae in NS + He star systems and the double neutron star systems}",
      journal = {\mnras},
         year = 2024,
        month = jun,
       volume = {530},
       number = {4},
        pages = {4461-4473},
          doi = {10.1093/mnras/stae1112},
archivePrefix = {arXiv},
       eprint = {2401.05103},
 primaryClass = {astro-ph.HE},
       adsurl = {https://ui.adsabs.harvard.edu/abs/2024MNRAS.530.4461G}
}

@ARTICLE{Sotomayor+2013A&A...552A..58S,
       author = {{Sotomayor-Beltran}, C. and {Sobey}, C. and {Hessels}, J.~W.~T. and {de Bruyn}, G. and {Noutsos}, A. and {Alexov}, A. and {Anderson}, J. and {Asgekar}, A. and {Avruch}, I.~M. and {Beck}, R. and {Bell}, M.~E. and {Bell}, M.~R. and {Bentum}, M.~J. and {Bernardi}, G. and {Best}, P. and {Birzan}, L. and {Bonafede}, A. and {Breitling}, F. and {Broderick}, J. and {Brouw}, W.~N. and {Br{\"u}ggen}, M. and {Ciardi}, B. and {de Gasperin}, F. and {Dettmar}, R. -J. and {van Duin}, A. and {Duscha}, S. and {Eisl{\"o}ffel}, J. and {Falcke}, H. and {Fallows}, R.~A. and {Fender}, R. and {Ferrari}, C. and {Frieswijk}, W. and {Garrett}, M.~A. and {Grie{\ss}meier}, J. and {Grit}, T. and {Gunst}, A.~W. and {Hassall}, T.~E. and {Heald}, G. and {Hoeft}, M. and {Horneffer}, A. and {Iacobelli}, M. and {Juette}, E. and {Karastergiou}, A. and {Keane}, E. and {Kohler}, J. and {Kramer}, M. and {Kondratiev}, V.~I. and {Koopmans}, L.~V.~E. and {Kuniyoshi}, M. and {Kuper}, G. and {van Leeuwen}, J. and {Maat}, P. and {Macario}, G. and {Markoff}, S. and {McKean}, J.~P. and {Mulcahy}, D.~D. and {Munk}, H. and {Orru}, E. and {Paas}, H. and {Pandey-Pommier}, M. and {Pilia}, M. and {Pizzo}, R. and {Polatidis}, A.~G. and {Reich}, W. and {R{\"o}ttgering}, H. and {Serylak}, M. and {Sluman}, J. and {Stappers}, B.~W. and {Tagger}, M. and {Tang}, Y. and {Tasse}, C. and {ter Veen}, S. and {Vermeulen}, R. and {van Weeren}, R.~J. and {Wijers}, R.~A.~M.~J. and {Wijnholds}, S.~J. and {Wise}, M.~W. and {Wucknitz}, O. and {Yatawatta}, S. and {Zarka}, P.},
        title = "{Calibrating high-precision Faraday rotation measurements for LOFAR and the next generation of low-frequency radio telescopes}",
      journal = {\aap},
         year = 2013,
        month = apr,
       volume = {552},
          eid = {A58},
        pages = {A58},
          doi = {10.1051/0004-6361/201220728},
archivePrefix = {arXiv},
       eprint = {1303.6230},
 primaryClass = {astro-ph.IM},
       adsurl = {https://ui.adsabs.harvard.edu/abs/2013A&A...552A..58S}
}

@INPROCEEDINGS{Paczynski+1976IAUS...73...75P,
       author = {{Paczynski}, B.},
        title = "{Common Envelope Binaries}",
    booktitle = {Structure and Evolution of Close Binary Systems},
         year = 1976,
       editor = {{Eggleton}, Peter and {Mitton}, Simon and {Whelan}, John},
       volume = {73},
        month = jan,
        pages = {75},
       series = {IAU Symposium},
       adsurl = {https://ui.adsabs.harvard.edu/abs/1976IAUS...73...75P}
}

@ARTICLE{Dewi+2002MNRAS.331.1027D,
       author = {{Dewi}, J.~D.~M. and {Pols}, O.~R. and {Savonije}, G.~J. and {van den Heuvel}, E.~P.~J.},
        title = "{The evolution of naked helium stars with a neutron star companion in close binary systems}",
      journal = {\mnras},
         year = 2002,
        month = apr,
       volume = {331},
       number = {4},
        pages = {1027-1040},
          doi = {10.1046/j.1365-8711.2002.05257.x},
archivePrefix = {arXiv},
       eprint = {astro-ph/0201239},
 primaryClass = {astro-ph},
       adsurl = {https://ui.adsabs.harvard.edu/abs/2002MNRAS.331.1027D}
}

@ARTICLE{Iben+1993PASP..105.1373I,
       author = {{Iben}, Icko, Jr. and {Livio}, Mario},
        title = "{Common Envelopes in Binary Star Evolution}",
      journal = {\pasp},
         year = 1993,
        month = dec,
       volume = {105},
        pages = {1373},
          doi = {10.1086/133321},
       adsurl = {https://ui.adsabs.harvard.edu/abs/1993PASP..105.1373I}
}

@ARTICLE{Ivanova+2013A&ARv..21...59I,
       author = {{Ivanova}, N. and {Justham}, S. and {Chen}, X. and {De Marco}, O. and {Fryer}, C.~L. and {Gaburov}, E. and {Ge}, H. and {Glebbeek}, E. and {Han}, Z. and {Li}, X. -D. and {Lu}, G. and {Marsh}, T. and {Podsiadlowski}, P. and {Potter}, A. and {Soker}, N. and {Taam}, R. and {Tauris}, T.~M. and {van den Heuvel}, E.~P.~J. and {Webbink}, R.~F.},
        title = "{Common envelope evolution: where we stand and how we can move forward}",
      journal = {\aapr},
         year = 2013,
        month = feb,
       volume = {21},
          eid = {59},
        pages = {59},
          doi = {10.1007/s00159-013-0059-2},
archivePrefix = {arXiv},
       eprint = {1209.4302},
 primaryClass = {astro-ph.HE},
       adsurl = {https://ui.adsabs.harvard.edu/abs/2013A&ARv..21...59I}
}

@BOOK{Tauris+2023pbse.book.....T,
     ISBN = {9780691179087},
     URL = {http://www.jstor.org/stable/jj.8595663},
     author = {Thomas M. Tauris and Edward P. J. van den Heuvel},
     series = {Princeton Series in Astrophysics},
     publisher = {Princeton University Press},
     title = {Physics of Binary Star Evolution: From Stars to X-ray Binaries and Gravitational Wave Sources},
     urldate = {2024-05-02},
     volume = {68},
     year = {2023},
     address = {Princeton}
}

@ARTICLE{Tauris+2017ApJ...846..170T,
       author = {{Tauris}, T.~M. and {Kramer}, M. and {Freire}, P.~C.~C. and {Wex}, N. and {Janka}, H. -T. and {Langer}, N. and {Podsiadlowski}, Ph. and {Bozzo}, E. and {Chaty}, S. and {Kruckow}, M.~U. and {van den Heuvel}, E.~P.~J. and {Antoniadis}, J. and {Breton}, R.~P. and {Champion}, D.~J.},
        title = "{Formation of Double Neutron Star Systems}",
      journal = {\apj},
         year = 2017,
        month = sep,
       volume = {846},
       number = {2},
          eid = {170},
        pages = {170},
          doi = {10.3847/1538-4357/aa7e89},
archivePrefix = {arXiv},
       eprint = {1706.09438},
 primaryClass = {astro-ph.HE},
       adsurl = {https://ui.adsabs.harvard.edu/abs/2017ApJ...846..170T}
}

@ARTICLE{Han+2025RAA....25a4001H,
       author = {{Han}, J.~L. and {Zhou}, D.~J. and {Wang}, C. and {Su}, W.~Q. and {Yan}, Yi and {Jing}, W.~C. and {Yang}, Z.~L. and {Wang}, P.~F. and {Wang}, T. and {Xu}, J. and {Cai}, N.~N. and {Sun}, J.~H. and {Yang}, Q.~L. and {Xu}, R.~X. and {Wang}, H.~G. and {You}, X.~P.},
        title = "{The FAST Galactic Plane Pulsar Snapshot Survey. VI. The Discovery of 473 New Pulsars}",
      journal = {Research in Astronomy and Astrophysics},
         year = 2025,
        month = jan,
       volume = {25},
       number = {1},
          eid = {014001},
        pages = {014001},
          doi = {10.1088/1674-4527/ada3b7},
archivePrefix = {arXiv},
       eprint = {2411.15961},
 primaryClass = {astro-ph.HE},
       adsurl = {https://ui.adsabs.harvard.edu/abs/2025RAA....25a4001H}
}

@ARTICLE{Wang+2024arXiv241203062W,
       author = {{Wang}, P.~F. and {Han}, J.~L. and {Yang}, Z.~L. and {Wang}, T. and {Wang}, C. and {Su}, W.~Q. and {Xu}, J. and {Zhou}, D.~J. and {Yan}, Yi and {Jing}, W.~C. and {Cai}, N.~N. and {Yuan}, J.~P. and {Xu}, R.~X. and {Wang}, H.~G. and {You}, X.~P.},
        title = "{The FAST Galactic Plane Pulsar Snapshot Survey. VIII. 116 Binary Pulsars}",
      journal = {Research in Astronomy and Astrophysics},
         year = 2025,
        month = jan,
       volume = {25},
       number = {1},
          eid = {014003},
        pages = {014003},
          doi = {10.1088/1674-4527/ada3b8},
archivePrefix = {arXiv},
       eprint = {2412.03062},
 primaryClass = {astro-ph.HE},
       adsurl = {https://ui.adsabs.harvard.edu/abs/2025RAA....25a4003W}
}

@ARTICLE{Damour+1986AIHPA..44..263D,
       author = {{Damour}, T. and {Deruelle}, N.},
        title = "{General relativistic celestial mechanics of binary systems. II. The post-Newtonian timing formula.}",
      journal = {Annales de L'Institut Henri Poincare Section (A) Physique Theorique},
         year = 1986,
        month = jan,
       volume = {44},
       number = {3},
        pages = {263-292},
       adsurl = {https://ui.adsabs.harvard.edu/abs/1986AIHPA..44..263D}
}

@ARTICLE{Bhattacharyya+2008MNRAS.387..273B,
       author = {{Bhattacharyya}, Bhaswati and {Nityananda}, Rajaram},
        title = "{Determination of the orbital parameters of binary pulsars}",
      journal = {\mnras},
         year = 2008,
        month = jun,
       volume = {387},
       number = {1},
        pages = {273-278},
          doi = {10.1111/j.1365-2966.2008.13213.x},
archivePrefix = {arXiv},
       eprint = {0803.1907},
 primaryClass = {astro-ph},
       adsurl = {https://ui.adsabs.harvard.edu/abs/2008MNRAS.387..273B}
}

@ARTICLE{Manchester+2005AJ....129.1993M,
       author = {{Manchester}, R.~N. and {Hobbs}, G.~B. and {Teoh}, A. and {Hobbs}, M.},
        title = "{The Australia Telescope National Facility Pulsar Catalogue}",
      journal = {\aj},
         year = 2005,
        month = apr,
       volume = {129},
       number = {4},
        pages = {1993-2006},
          doi = {10.1086/428488},
archivePrefix = {arXiv},
       eprint = {astro-ph/0412641},
 primaryClass = {astro-ph},
       adsurl = {https://ui.adsabs.harvard.edu/abs/2005AJ....129.1993M}
}

@ARTICLE{Martinez+2015ApJ...812..143M,
       author = {{Martinez}, J.~G. and {Stovall}, K. and {Freire}, P.~C.~C. and {Deneva}, J.~S. and {Jenet}, F.~A. and {McLaughlin}, M.~A. and {Bagchi}, M. and {Bates}, S.~D. and {Ridolfi}, A.},
        title = "{Pulsar J0453+1559: A Double Neutron Star System with a Large Mass Asymmetry}",
      journal = {\apj},
         year = 2015,
        month = oct,
       volume = {812},
       number = {2},
          eid = {143},
        pages = {143},
          doi = {10.1088/0004-637X/812/2/143},
archivePrefix = {arXiv},
       eprint = {1509.08805},
 primaryClass = {astro-ph.HE},
       adsurl = {https://ui.adsabs.harvard.edu/abs/2015ApJ...812..143M}
}

@ARTICLE{Barr+2024Sci...383..275B,
       author = {{Barr}, Ewan D. and {Dutta}, Arunima and {Freire}, Paulo C.~C. and {Cadelano}, Mario and {Gautam}, Tasha and {Kramer}, Michael and {Pallanca}, Cristina and {Ransom}, Scott M. and {Ridolfi}, Alessandro and {Stappers}, Benjamin W. and {Tauris}, Thomas M. and {Venkatraman Krishnan}, Vivek and {Wex}, Norbert and {Bailes}, Matthew and {Behrend}, Jan and {Buchner}, Sarah and {Burgay}, Marta and {Chen}, Weiwei and {Champion}, David J. and {Chen}, C. -H. Rosie and {Corongiu}, Alessandro and {Geyer}, Marisa and {Men}, Y.~P. and {Padmanabh}, Prajwal Voraganti and {Possenti}, Andrea},
        title = "{A pulsar in a binary with a compact object in the mass gap between neutron stars and black holes}",
      journal = {Science},
         year = 2024,
        month = jan,
       volume = {383},
       number = {6680},
        pages = {275-279},
          doi = {10.1126/science.adg3005},
archivePrefix = {arXiv},
       eprint = {2401.09872},
 primaryClass = {astro-ph.HE},
       adsurl = {https://ui.adsabs.harvard.edu/abs/2024Sci...383..275B}
}

@ARTICLE{Leeuwen+2015ApJ...798..118V,
       author = {{van Leeuwen}, J. and {Kasian}, L. and {Stairs}, I.~H. and {Lorimer}, D.~R. and {Camilo}, F. and {Chatterjee}, S. and {Cognard}, I. and {Desvignes}, G. and {Freire}, P.~C.~C. and {Janssen}, G.~H. and {Kramer}, M. and {Lyne}, A.~G. and {Nice}, D.~J. and {Ransom}, S.~M. and {Stappers}, B.~W. and {Weisberg}, J.~M.},
        title = "{The Binary Companion of Young, Relativistic Pulsar J1906+0746}",
      journal = {\apj},
         year = 2015,
        month = jan,
       volume = {798},
       number = {2},
          eid = {118},
        pages = {118},
          doi = {10.1088/0004-637X/798/2/118},
archivePrefix = {arXiv},
       eprint = {1411.1518},
 primaryClass = {astro-ph.SR},
       adsurl = {https://ui.adsabs.harvard.edu/abs/2015ApJ...798..118V}
}

@ARTICLE{Lynch+2012ApJ...745..109L,
       author = {{Lynch}, Ryan S. and {Freire}, Paulo C.~C. and {Ransom}, Scott M. and {Jacoby}, Bryan A.},
        title = "{The Timing of Nine Globular Cluster Pulsars}",
      journal = {\apj},
         year = 2012,
        month = feb,
       volume = {745},
       number = {2},
          eid = {109},
        pages = {109},
          doi = {10.1088/0004-637X/745/2/109},
archivePrefix = {arXiv},
       eprint = {1112.2612},
 primaryClass = {astro-ph.HE},
       adsurl = {https://ui.adsabs.harvard.edu/abs/2012ApJ...745..109L}
}

@ARTICLE{Breton+2012ApJ...747...89B,
       author = {{Breton}, R.~P. and {Kaspi}, V.~M. and {McLaughlin}, M.~A. and {Lyutikov}, M. and {Kramer}, M. and {Stairs}, I.~H. and {Ransom}, S.~M. and {Ferdman}, R.~D. and {Camilo}, F. and {Possenti}, A.},
        title = "{The Double Pulsar Eclipses. I. Phenomenology and Multi-frequency Analysis}",
      journal = {\apj},
         year = 2012,
        month = mar,
       volume = {747},
       number = {2},
          eid = {89},
        pages = {89},
          doi = {10.1088/0004-637X/747/2/89},
archivePrefix = {arXiv},
       eprint = {1201.2637},
 primaryClass = {astro-ph.GA},
       adsurl = {https://ui.adsabs.harvard.edu/abs/2012ApJ...747...89B}
}

@ARTICLE{McEwen+2024ApJ...962..167M,
       author = {{McEwen}, A.~E. and {Swiggum}, J.~K. and {Kaplan}, D.~L. and {Tan}, C.~M. and {Meyers}, B.~W. and {Fonseca}, E. and {Agazie}, G.~Y. and {Chawla}, P. and {Crowter}, K. and {DeCesar}, M.~E. and {Dolch}, T. and {Dong}, F.~A. and {Fiore}, W. and {Fonseca}, E. and {Good}, D.~C. and {Istrate}, A.~G. and {Kaspi}, V.~M. and {Kondratiev}, V.~I. and {van Leeuwen}, J. and {Levin}, L. and {Lewis}, E.~F. and {Lynch}, R.~S. and {Masui}, K.~W. and {McKee}, J.~W. and {McLaughlin}, M.~A. and {Al Noori}, H. and {Parent}, E. and {Ransom}, S.~M. and {Siemens}, X. and {Spiewak}, R. and {Stairs}, I.~H.},
        title = "{The Green Bank North Celestial Cap Survey. IX. Timing Follow-up for 128 Pulsars}",
      journal = {\apj},
         year = 2024,
        month = feb,
       volume = {962},
       number = {2},
          eid = {167},
        pages = {167},
          doi = {10.3847/1538-4357/ad11f0},
archivePrefix = {arXiv},
       eprint = {2312.07471},
 primaryClass = {astro-ph.HE},
       adsurl = {https://ui.adsabs.harvard.edu/abs/2024ApJ...962..167M}
}

@ARTICLE{Sengar+2022MNRAS.512.5782S,
       author = {{Sengar}, R. and {Balakrishnan}, V. and {Stevenson}, S. and {Bailes}, M. and {Barr}, E.~D. and {Bhat}, N.~D.~R. and {Burgay}, M. and {Bernadich}, M.~C. i. and {Cameron}, A.~D. and {Champion}, D.~J. and {Chen}, W. and {Flynn}, C.~M.~L. and {Jameson}, A. and {Johnston}, S. and {Keith}, M.~J. and {Kramer}, M. and {Morello}, V. and {Ng}, C. and {Possenti}, A. and {Stappers}, B. and {Shannon}, R.~M. and {van Straten}, W. and {Wongphechauxsorn}, J.},
        title = "{The High Time Resolution Universe Pulsar Survey - XVII. PSR J1325-6253, a low eccentricity double neutron star system from an ultra-stripped supernova}",
      journal = {\mnras},
         year = 2022,
        month = jun,
       volume = {512},
       number = {4},
        pages = {5782-5792},
          doi = {10.1093/mnras/stac821},
archivePrefix = {arXiv},
       eprint = {2204.06869},
 primaryClass = {astro-ph.HE},
       adsurl = {https://ui.adsabs.harvard.edu/abs/2022MNRAS.512.5782S}
}

@ARTICLE{Janssen+2008AandA...490..753J,
       author = {{Janssen}, G.~H. and {Stappers}, B.~W. and {Kramer}, M. and {Nice}, D.~J. and {Jessner}, A. and {Cognard}, I. and {Purver}, M.~B.},
        title = "{Multi-telescope timing of PSR J1518+4904}",
      journal = {\aap},
         year = 2008,
        month = nov,
       volume = {490},
       number = {2},
        pages = {753-761},
          doi = {10.1051/0004-6361:200810076},
archivePrefix = {arXiv},
       eprint = {0808.2292},
 primaryClass = {astro-ph},
       adsurl = {https://ui.adsabs.harvard.edu/abs/2008A&A...490..753J}
}

@ARTICLE{Corongiu+2007AandA...462..703C,
       author = {{Corongiu}, A. and {Kramer}, M. and {Stappers}, B.~W. and {Lyne}, A.~G. and {Jessner}, A. and {Possenti}, A. and {D'Amico}, N. and {L{\"o}hmer}, O.},
        title = "{The binary pulsar PSR J1811-1736: evidence of a low amplitude supernova kick}",
      journal = {\aap},
         year = 2007,
        month = feb,
       volume = {462},
       number = {2},
        pages = {703-709},
          doi = {10.1051/0004-6361:20054385},
archivePrefix = {arXiv},
       eprint = {astro-ph/0611436},
 primaryClass = {astro-ph},
       adsurl = {https://ui.adsabs.harvard.edu/abs/2007A&A...462..703C}
}

@ARTICLE{Weisberg+2016ApJ...829...55W,
       author = {{Weisberg}, J.~M. and {Huang}, Y.},
        title = "{Relativistic Measurements from Timing the Binary Pulsar PSR B1913+16}",
      journal = {\apj},
         year = 2016,
        month = sep,
       volume = {829},
       number = {1},
          eid = {55},
        pages = {55},
          doi = {10.3847/0004-637X/829/1/55},
archivePrefix = {arXiv},
       eprint = {1606.02744},
 primaryClass = {astro-ph.HE},
       adsurl = {https://ui.adsabs.harvard.edu/abs/2016ApJ...829...55W}
}

@ARTICLE{Swiggum+2015ApJ...805..156S,
       author = {{Swiggum}, J.~K. and {Rosen}, R. and {McLaughlin}, M.~A. and {Lorimer}, D.~R. and {Heatherly}, S. and {Lynch}, R. and {Scoles}, S. and {Hockett}, T. and {Filik}, E. and {Marlowe}, J.~A. and {Barlow}, B.~N. and {Weaver}, M. and {Hilzendeger}, M. and {Ernst}, S. and {Crowley}, R. and {Stone}, E. and {Miller}, B. and {Nunez}, R. and {Trevino}, G. and {Doehler}, M. and {Cramer}, A. and {Yencsik}, D. and {Thorley}, J. and {Andrews}, R. and {Laws}, A. and {Wenger}, K. and {Teter}, L. and {Snyder}, T. and {Dittmann}, A. and {Gray}, S. and {Carter}, M. and {McGough}, C. and {Dydiw}, S. and {Pruett}, C. and {Fink}, J. and {Vanderhout}, A.},
        title = "{PSR J1930-1852: a Pulsar in the Widest Known Orbit around Another Neutron Star}",
      journal = {\apj},
         year = 2015,
        month = jun,
       volume = {805},
       number = {2},
          eid = {156},
        pages = {156},
          doi = {10.1088/0004-637X/805/2/156},
archivePrefix = {arXiv},
       eprint = {1503.06276},
 primaryClass = {astro-ph.HE},
       adsurl = {https://ui.adsabs.harvard.edu/abs/2015ApJ...805..156S}
}

@ARTICLE{Lazarus+2016ApJ...831..150L,
       author = {{Lazarus}, P. and {Freire}, P.~C.~C. and {Allen}, B. and {Aulbert}, C. and {Bock}, O. and {Bogdanov}, S. and {Brazier}, A. and {Camilo}, F. and {Cardoso}, F. and {Chatterjee}, S. and {Cordes}, J.~M. and {Crawford}, F. and {Deneva}, J.~S. and {Eggenstein}, H. -B. and {Fehrmann}, H. and {Ferdman}, R. and {Hessels}, J.~W.~T. and {Jenet}, F.~A. and {Karako-Argaman}, C. and {Kaspi}, V.~M. and {Knispel}, B. and {Lynch}, R. and {van Leeuwen}, J. and {Machenschalk}, B. and {Madsen}, E. and {McLaughlin}, M.~A. and {Patel}, C. and {Ransom}, S.~M. and {Scholz}, P. and {Seymour}, A. and {Siemens}, X. and {Spitler}, L.~G. and {Stairs}, I.~H. and {Stovall}, K. and {Swiggum}, J. and {Venkataraman}, A. and {Zhu}, W.~W.},
        title = "{Einstein@Home Discovery of a Double Neutron Star Binary in the PALFA Survey}",
      journal = {\apj},
         year = 2016,
        month = nov,
       volume = {831},
       number = {2},
          eid = {150},
        pages = {150},
          doi = {10.3847/0004-637X/831/2/150},
archivePrefix = {arXiv},
       eprint = {1608.08211},
 primaryClass = {astro-ph.HE},
       adsurl = {https://ui.adsabs.harvard.edu/abs/2016ApJ...831..150L}
}

@ARTICLE{Haniewicz+2021MNRAS.500.4620H,
       author = {{Haniewicz}, H.~T. and {Ferdman}, R.~D. and {Freire}, P.~C.~C. and {Champion}, D.~J. and {Bunting}, K.~A. and {Lorimer}, D.~R. and {McLaughlin}, M.~A.},
        title = "{Precise mass measurements for the double neutron star system J1829+2456}",
      journal = {\mnras},
         year = 2021,
        month = feb,
       volume = {500},
       number = {4},
        pages = {4620-4627},
          doi = {10.1093/mnras/staa3466},
archivePrefix = {arXiv},
       eprint = {2007.07565},
 primaryClass = {astro-ph.SR},
       adsurl = {https://ui.adsabs.harvard.edu/abs/2021MNRAS.500.4620H}
}

@ARTICLE{Ferdman+2014MNRAS.443.2183F,
       author = {{Ferdman}, R.~D. and {Stairs}, I.~H. and {Kramer}, M. and {Janssen}, G.~H. and {Bassa}, C.~G. and {Stappers}, B.~W. and {Demorest}, P.~B. and {Cognard}, I. and {Desvignes}, G. and {Theureau}, G. and {Burgay}, M. and {Lyne}, A.~G. and {Manchester}, R.~N. and {Possenti}, A.},
        title = "{PSR J1756-2251: a pulsar with a low-mass neutron star companion}",
      journal = {\mnras},
         year = 2014,
        month = sep,
       volume = {443},
       number = {3},
        pages = {2183-2196},
          doi = {10.1093/mnras/stu1223},
archivePrefix = {arXiv},
       eprint = {1406.5507},
 primaryClass = {astro-ph.SR},
       adsurl = {https://ui.adsabs.harvard.edu/abs/2014MNRAS.443.2183F}
}

@ARTICLE{Fonseca+2014ApJ...787...82F,
       author = {{Fonseca}, Emmanuel and {Stairs}, Ingrid H. and {Thorsett}, Stephen E.},
        title = "{A Comprehensive Study of Relativistic Gravity Using PSR B1534+12}",
      journal = {\apj},
         year = 2014,
        month = may,
       volume = {787},
       number = {1},
          eid = {82},
        pages = {82},
          doi = {10.1088/0004-637X/787/1/82},
archivePrefix = {arXiv},
       eprint = {1402.4836},
 primaryClass = {astro-ph.HE},
       adsurl = {https://ui.adsabs.harvard.edu/abs/2014ApJ...787...82F}
}

@ARTICLE{Martinez+2017ApJ...851L..29M,
       author = {{Martinez}, J.~G. and {Stovall}, K. and {Freire}, P.~C.~C. and {Deneva}, J.~S. and {Tauris}, T.~M. and {Ridolfi}, A. and {Wex}, N. and {Jenet}, F.~A. and {McLaughlin}, M.~A. and {Bagchi}, M.},
        title = "{Pulsar J1411+2551: A Low-mass Double Neutron Star System}",
      journal = {\apjl},
         year = 2017,
        month = dec,
       volume = {851},
       number = {2},
          eid = {L29},
        pages = {L29},
          doi = {10.3847/2041-8213/aa9d87},
archivePrefix = {arXiv},
       eprint = {1711.09804},
 primaryClass = {astro-ph.HE},
       adsurl = {https://ui.adsabs.harvard.edu/abs/2017ApJ...851L..29M}
}

@ARTICLE{DeCesar+2015ApJ...807L..23D,
       author = {{DeCesar}, Megan E. and {Ransom}, Scott M. and {Kaplan}, David L. and {Ray}, Paul S. and {Geller}, Aaron M.},
        title = "{A Highly Eccentric 3.9 Millisecond Binary Pulsar in the Globular Cluster NGC 6652}",
      journal = {\apjl},
         year = 2015,
        month = jul,
       volume = {807},
       number = {2},
          eid = {L23},
        pages = {L23},
          doi = {10.1088/2041-8205/807/2/L23},
archivePrefix = {arXiv},
       eprint = {1506.03367},
 primaryClass = {astro-ph.HE},
       adsurl = {https://ui.adsabs.harvard.edu/abs/2015ApJ...807L..23D}
}

@ARTICLE{Chu:2024rwt,
       author = {{Chu}, Qingbo and {Lu}, Youjun and {Yu}, Shenghua},
        title = "{Spin Evolution and Mass Distribution of Galactic Binary Neutron Stars}",
      journal = {\apj},
         year = 2025,
        month = feb,
       volume = {980},
       number = {2},
          eid = {181},
        pages = {181},
          doi = {10.3847/1538-4357/ad90e2},
archivePrefix = {arXiv},
       eprint = {2412.15464},
 primaryClass = {astro-ph.HE},
       adsurl = {https://ui.adsabs.harvard.edu/abs/2025ApJ...980..181C}
}

@ARTICLE{2021A&A...650A.107M,
       author = {{Marchant}, Pablo and {Pappas}, Kaliro{\"e} M.~W. and {Gallegos-Garcia}, Monica and {Berry}, Christopher P.~L. and {Taam}, Ronald E. and {Kalogera}, Vicky and {Podsiadlowski}, Philipp},
        title = "{The role of mass transfer and common envelope evolution in the formation of merging binary black holes}",
      journal = {\aap},
         year = 2021,
        month = jun,
       volume = {650},
          eid = {A107},
        pages = {A107},
          doi = {10.1051/0004-6361/202039992},
archivePrefix = {arXiv},
       eprint = {2103.09243},
 primaryClass = {astro-ph.SR},
       adsurl = {https://ui.adsabs.harvard.edu/abs/2021A&A...650A.107M}
}

@ARTICLE{Sgalletta+2023MNRAS.526.2210S,
       author = {{Sgalletta}, Cecilia and {Iorio}, Giuliano and {Mapelli}, Michela and {Artale}, M. Celeste and {Boco}, Lumen and {Chattopadhyay}, Debatri and {Lapi}, Andrea and {Possenti}, Andrea and {Rinaldi}, Stefano and {Spera}, Mario},
        title = "{Binary neutron star populations in the Milky Way}",
      journal = {\mnras},
         year = 2023,
        month = dec,
       volume = {526},
       number = {2},
        pages = {2210-2229},
          doi = {10.1093/mnras/stad2768},
archivePrefix = {arXiv},
       eprint = {2305.04955},
 primaryClass = {astro-ph.HE},
       adsurl = {https://ui.adsabs.harvard.edu/abs/2023MNRAS.526.2210S}
}

@article{10.1093/mnras/sty2463,
       author = {{Vigna-G{\'o}mez}, Alejandro and {Neijssel}, Coenraad J. and {Stevenson}, Simon and {Barrett}, Jim W. and {Belczynski}, Krzysztof and {Justham}, Stephen and {de Mink}, Selma E. and {M{\"u}ller}, Bernhard and {Podsiadlowski}, Philipp and {Renzo}, Mathieu and {Sz{\'e}csi}, Dorottya and {Mandel}, Ilya},
        title = "{On the formation history of Galactic double neutron stars}",
      journal = {\mnras},
         year = 2018,
        month = dec,
       volume = {481},
       number = {3},
        pages = {4009-4029},
          doi = {10.1093/mnras/sty2463},
archivePrefix = {arXiv},
       eprint = {1805.07974},
 primaryClass = {astro-ph.SR},
       adsurl = {https://ui.adsabs.harvard.edu/abs/2018MNRAS.481.4009V}
}

@article{10.1111/j.1365-2966.2005.09087.x,
       author = {{Hobbs}, G. and {Lorimer}, D.~R. and {Lyne}, A.~G. and {Kramer}, M.},
        title = "{A statistical study of 233 pulsar proper motions}",
      journal = {\mnras},
         year = 2005,
        month = jul,
       volume = {360},
       number = {3},
        pages = {974-992},
          doi = {10.1111/j.1365-2966.2005.09087.x},
archivePrefix = {arXiv},
       eprint = {astro-ph/0504584},
 primaryClass = {astro-ph},
       adsurl = {https://ui.adsabs.harvard.edu/abs/2005MNRAS.360..974H}
}

@article{Wen:2002km,
    author = {{Wen}, Linqing},
        title = "{On the Eccentricity Distribution of Coalescing Black Hole Binaries Driven by the Kozai Mechanism in Globular Clusters}",
      journal = {\apj},
         year = 2003,
        month = nov,
       volume = {598},
       number = {1},
        pages = {419-430},
          doi = {10.1086/378794},
archivePrefix = {arXiv},
       eprint = {astro-ph/0211492},
 primaryClass = {astro-ph},
       adsurl = {https://ui.adsabs.harvard.edu/abs/2003ApJ...598..419W}
}

@article{LISA:2017pwj,
    author = {{Amaro-Seoane}, Pau and {Audley}, Heather and {Babak}, Stanislav and {Baker}, John and {Barausse}, Enrico and {Bender}, Peter and {Berti}, Emanuele and {Binetruy}, Pierre and {Born}, Michael and {Bortoluzzi}, Daniele and {Camp}, Jordan and {Caprini}, Chiara and {Cardoso}, Vitor and {Colpi}, Monica and {Conklin}, John and {Cornish}, Neil and {Cutler}, Curt and {Danzmann}, Karsten and {Dolesi}, Rita and {Ferraioli}, Luigi and {Ferroni}, Valerio and {Fitzsimons}, Ewan and {Gair}, Jonathan and {Gesa Bote}, Lluis and {Giardini}, Domenico and {Gibert}, Ferran and {Grimani}, Catia and {Halloin}, Hubert and {Heinzel}, Gerhard and {Hertog}, Thomas and {Hewitson}, Martin and {Holley-Bockelmann}, Kelly and {Hollington}, Daniel and {Hueller}, Mauro and {Inchauspe}, Henri and {Jetzer}, Philippe and {Karnesis}, Nikos and {Killow}, Christian and {Klein}, Antoine and {Klipstein}, Bill and {Korsakova}, Natalia and {Larson}, Shane L and {Livas}, Jeffrey and {Lloro}, Ivan and {Man}, Nary and {Mance}, Davor and {Martino}, Joseph and {Mateos}, Ignacio and {McKenzie}, Kirk and {McWilliams}, Sean T and {Miller}, Cole and {Mueller}, Guido and {Nardini}, Germano and {Nelemans}, Gijs and {Nofrarias}, Miquel and {Petiteau}, Antoine and {Pivato}, Paolo and {Plagnol}, Eric and {Porter}, Ed and {Reiche}, Jens and {Robertson}, David and {Robertson}, Norna and {Rossi}, Elena and {Russano}, Giuliana and {Schutz}, Bernard and {Sesana}, Alberto and {Shoemaker}, David and {Slutsky}, Jacob and {Sopuerta}, Carlos F. and {Sumner}, Tim and {Tamanini}, Nicola and {Thorpe}, Ira and {Troebs}, Michael and {Vallisneri}, Michele and {Vecchio}, Alberto and {Vetrugno}, Daniele and {Vitale}, Stefano and {Volonteri}, Marta and {Wanner}, Gudrun and {Ward}, Harry and {Wass}, Peter and {Weber}, William and {Ziemer}, John and {Zweifel}, Peter},
        title = "{Laser Interferometer Space Antenna}",
      journal = {arXiv e-prints},
         year = 2017,
        month = feb,
          eid = {arXiv:1702.00786},
        pages = {arXiv:1702.00786},
          doi = {10.48550/arXiv.1702.00786},
archivePrefix = {arXiv},
       eprint = {1702.00786},
 primaryClass = {astro-ph.IM},
       adsurl = {https://ui.adsabs.harvard.edu/abs/2017arXiv170200786A}
}

@article{Hu:2017mde,
     author = {{Hu}, Wen-Rui and {Wu}, Yue-Liang},
        title = "{The Taiji Program in Space for gravitational wave physics and the nature of gravity}",
      journal = {National Science Review},
         year = 2017,
        month = sep,
       volume = {4},
       number = {5},
        pages = {685-686},
          doi = {10.1093/nsr/nwx116},
       adsurl = {https://ui.adsabs.harvard.edu/abs/2017NSRev...4..685H}
}

@article{TianQin:2015yph,
       author = {{Luo}, Jun and {Chen}, Li-Sheng and {Duan}, Hui-Zong and {Gong}, Yun-Gui and {Hu}, Shoucun and {Ji}, Jianghui and {Liu}, Qi and {Mei}, Jianwei and {Milyukov}, Vadim and {Sazhin}, Mikhail and {Shao}, Cheng-Gang and {Toth}, Viktor T. and {Tu}, Hai-Bo and {Wang}, Yamin and {Wang}, Yan and {Yeh}, Hsien-Chi and {Zhan}, Ming-Sheng and {Zhang}, Yonghe and {Zharov}, Vladimir and {Zhou}, Ze-Bing},
        title = "{TianQin: a space-borne gravitational wave detector}",
      journal = {Classical and Quantum Gravity},
         year = 2016,
        month = feb,
       volume = {33},
       number = {3},
          eid = {035010},
        pages = {035010},
          doi = {10.1088/0264-9381/33/3/035010},
archivePrefix = {arXiv},
       eprint = {1512.02076},
 primaryClass = {astro-ph.IM},
       adsurl = {https://ui.adsabs.harvard.edu/abs/2016CQGra..33c5010L}
}

@article{Gong:2021gvw,
       author = {{Gong}, Yungui and {Luo}, Jun and {Wang}, Bin},
        title = "{Concepts and status of Chinese space gravitational wave detection projects}",
      journal = {Nature Astronomy},
         year = 2021,
        month = sep,
       volume = {5},
        pages = {881-889},
          doi = {10.1038/s41550-021-01480-3},
archivePrefix = {arXiv},
       eprint = {2109.07442},
 primaryClass = {astro-ph.IM},
       adsurl = {https://ui.adsabs.harvard.edu/abs/2021NatAs...5..881G}
}

@ARTICLE{Chen+2025MNRAS.544L..89C,
       author = {{Chen}, Zhiwei and {Zhang}, Jihui and {Lu}, Youjun and {Liu}, Jifeng and {Zeng}, Changwen},
        title = "{Constraining common envelope evolution in binary neutron star formation with combined galactic and gravitational-wave observations}",
      journal = {\mnras},
         year = 2025,
        month = nov,
       volume = {544},
       number = {1},
        pages = {L89-L95},
          doi = {10.1093/mnrasl/slaf105},
archivePrefix = {arXiv},
       eprint = {2508.14397},
 primaryClass = {astro-ph.HE},
       adsurl = {https://ui.adsabs.harvard.edu/abs/2025MNRAS.544L..89C}
}


\newpage 

\section*{Acknowledgments}
%
%
This work made use of the data from FAST (https://cstr.cn/31116.02.FAST). FAST is a Chinese national mega-science facility, built and operated by the National Astronomical Observatories, Chinese Academy of Sciences. We all appreciate the excellent performance of the FAST and the operation team. 
%

The authors were supported by the National Natural Science Foundation of China (NSFC), grant numbers 12588202, 12041303 and 12225304, the Chinese Academy of Sciences via project JZHKYPT-2021-06. 

JLH organized the FAST GPPS survey, processed the survey data and discovered the pulsar, organized the team to work on all aspects of this project, and finalized the paper; 
Under the supervision by JLH,  ZLY processed the data for timing and polarization profile, identified the DNS nature of PSR J0653+0443, made the follow-up observations and drafted the paper; 
LGH got the FAST survey data of this part of the sky region via a normal FAST project, requested and consulted by JLH; 
WQS made some follow-up FAST observations;
CW arranged some FAST observation schedules; 
TW developed the procedures and relative codes for FAST data preparation for fits files; 
YY developed the procedures to verify pulsar detection; 
PFW developed the procedures for FAST pulsar polarization and timing processing, and made the ionosphere RM correction; 
ZWC, YLG, YJL and BW carried out simulations for the evolution history of such binaries;
YGG, CZ, QG, KYS and YJL 
verified the relativistic effects and estimated the strain amplitudes of gravitational wave radiations;  
BL checked all possible terms of the relativistic effect for this specific DNS and found other higher order terms are not yet measurable;
NNC and JX maintained the computer platform and data processing systems;
Other authors joined discussions of the results and contributed to the finalization of the paper.

\newpage

\section*{Methods}

\setcounter{table}{0}
\setcounter{figure}{0}
\setcounter{section}{0}
\renewcommand\thefigure{S\arabic{figure}}
\renewcommand\thetable{S\arabic{table}}
\renewcommand\thesection{Sect. S\arabic{section}}

\section{PSR~J0653+0443: FAST observations and basic properties}
\label{sec:pulsartiming}

The FAST, the most sensitive L-band radio telescope \cite{Nan+2006ScChG..49..129N, Jiang+2020RAA....20...64J}, has conducted the Galactic Plane Pulsar Snapshot (GPPS) survey since 2020 as one of its five major key science projects \cite{Han+2021RAA....21..107H, Han+2025RAA....25a4001H}. Additional FAST observation projects by co-authors were also used to collect pulsar data (Table~\ref{tab:obslist}).

PSR J0653+0443 was first detected on 2022 November 20 (20221120) via the coordinated FAST project PT2022\_0060, and confirmed on 20230107 by the GPPS survey observations. Subsequent FAST timing observations employed the L-band 19-beam receiver \cite{Jiang+2020RAA....20...64J}, covering 1.0 to 1.5 GHz with 1024 or 2048 frequency channels. Most observations recorded four polarization channels ($XX$, $YY$, Re[$X^{*}Y$], Im[$X^{*}Y$]) at a 49.152~{\textmu}s sampling rate. Pulsar data calibration used 1.1 K modulated noise signals (2.01326 s period) injected into receiver feeds for 2 minutes before or after observations.

\begin{table*}[b]
    \centering
    \caption{{\bf FAST observations of PSR~J0653+0443.} This table lists the observations dates, FAST project numbers, project principal investigator (PI) names, FAST observation beam names, observation durations (in minutes), and beam offsets from the finally determined position of the pulsar, and range of mean anomaly $M$, and range or value (if the range is too small) of true anomaly $\theta$.}
    \label{tab:obslist}
    \footnotesize
    \setlength{\tabcolsep}{1.5mm}
    \begin{tabular}{clllcccc}
    \hline \hline
      Obs. Date & FAST Project & PI & FAST observation Beam & Duration & Offset & $M$ & $\theta$ \\
    (yyyymmdd) &  &  &        &  (min)   &   (arcmin)  & (deg)  &  (deg)\\
	\hline
	20221120 & PT2022\_0060	& L.~G. Hou & G209.04+2.63P1M13 & 5  & 2.0 & 214.52 to 214.54 & 181.09  \\
	20221120 & PT2022\_0060	& L.~G. Hou & G209.04+2.63P3M04 & 5  & 1.0 & 214.56 to 214.57 & 181.09  \\
	20230107 & ZD2022\_2 & J.~L. Han & J065303+044312M01 & 15    & 0.2 & 72.69 to 72.73 & 176.02  \\
    20230731 & ZD2022\_2 & J.~L. Han & J0653+0443M01 & 5         & 0.2 & 287.81 to 287.86 & 184.01  \\
    20230822 & PT2023\_0162 & W.~Q. Su & J0653+0443M01 & 5       & 0.2 & 27.19 to 27.21 & 172.11  \\
    20230923 & PT2023\_0162 & W.~Q. Su & J0653+0443M01 & 5       & 0.2 & 172.95 to 172.96 & 179.78  \\
    20231221 & PT2023\_0162 & W.~Q. Su & J0653+0443M01 & 5       & 0.2 & 216.94 to 216.96 & 181.17  \\
    20240331 & ZD2023\_2 & J.~L. Han & J0653+0443gM01 & 5        & 0.2 & 320.05 to 320.07 & 186.32  \\
    20240513 & ZD2023\_2 & J.~L. Han & J0653+0443gM01 & 5        & 0.2 & 155.21 to 155.22 & 179.22 to 179.23  \\
    20240518 & ZD2023\_2 & J.~L. Han & J0653+0443gM01 & 5        & 0.2 & 177.98 to 177.99 & 179.94  \\
    20240622 & ZD2023\_2 & J.~L. Han & J0653+0443gM01 & 5        & 0.2 & 336.97 to 336.99 & 188.61  \\
    20240626 & ZD2023\_2 & J.~L. Han & J0653+0443gM01 & 5        & 0.2 & 354.97 to 354.98 & 196.65 to 196.67  \\
    20240627 & ZD2023\_2 & J.~L. Han & J0653+0443gM01 & 5        & 0.2 & 359.59 to 359.61 & 223.43 to 224.09  \\
    20240628 & ZD2023\_2 & J.~L. Han & J0653+0443gM01 & 5        & 0.2 & 4.17 to 4.19 & 162.08 to 162.11  \\
    20240811 & ZD2024\_2 & J.~L. Han & J0653+0443gM01 & 5        & 0.2 & 203.89 to 203.91 & 180.75  \\
    20240821 & PT2024\_0020 & Z.~L. Yang & J0653+0443gM01 & 15   & 0.2 & 248.81 to 248.86 & 182.28  \\
    20240911 & PT2024\_0020 & Z.~L. Yang & J0653+0443gM01 & 15   & 0.2 & 344.18 to 344.23 & 190.31 to 190.32  \\
    20240912 & PT2024\_0020 & Z.~L. Yang & J0653+0443gM01 & 15   & 0.2 & 348.75 to 348.80 & 191.99 to 192.01  \\
    20240913 & PT2024\_0020 & Z.~L. Yang & J0653+0443gM01 & 15   & 0.2 & 353.95 to 354.00 & 195.49 to 195.54  \\
    20240914 & PT2024\_0020 & Z.~L. Yang & J0653+0443gM01 & 30   & 0.2 & 358.18 to 358.28 & 204.60 to 205.10  \\
    20241002 & PT2024\_0020 & Z.~L. Yang & J0653+0443gM01 & 40   & 0.0 & 79.85 to 79.98 & 176.38 to 176.39  \\
    20241103 & PT2024\_0020 & Z.~L. Yang & J0653+0443gM01 & 40   & 0.0 & 224.97 to 225.09 & 181.43 to 181.44  \\
    20241202 & PT2024\_0020 & Z.~L. Yang & J0653+0443gM01 & 40   & 0.0 & 356.66 to 356.79 & 199.53 to 199.82  \\
    20241203 & PT2024\_0020 & Z.~L. Yang & J0653+0443gM01 & 60   & 0.0 & 1.28 to 1.47 & 151.92 to 153.34  \\
    20250219 & PT2024\_0020 & Z.~L. Yang & J0653+0443gM01 & 120  & 0.0 & --0.23 to +0.15 & --125.33 to +114.02  \\
    20250302 & PT2024\_0239 & Z.~L. Yang & J0653+0443M01 & 41    & 0.2 & 49.83 to 49.96 & 174.55 to 174.56  \\
    20250509 & PT2024\_0020 & Z.~L. Yang & J0653+0443M01 & 60    & 0.0 & 358.75 to 358.94 & 208.29 to 210.10  \\
    \hline
    \end{tabular}
\end{table*}

We extracted barycentric periods for all observation epochs using {\sc Presto} \cite{Ransom+2001PhDT.......123R}; their small variations are indications of the nature of a binary system. We then performed a four-dimensional search over orbital period $P_{\rm orb}$, time of periastron passage $T_0$, orbital eccentricity $e$, and longitude of periastron $\omega$ to fit the observed barycentric periods across epochs, adopting the linear least-squares method from \cite{Bhattacharyya+2008MNRAS.387..273B}. The pulsar’s intrinsic spin period $P$ and approximate projected semi-major axis $x$ were derived from fitting the first seven observations. We found that PSR~J0653+0443 moves in a highly eccentric orbit ($e>0.9$) with a long period $P_{\rm orb}\simeq79$~days.

\begin{table*}[t]
    \centering
    \caption{{\bf Observation opportunities for periastron passages of PSR~J0653+0443.} 
    This table lists the dates and Beijing Time (UTC + 08:00) of periastron passages of PSR~J0653+0443, along with the corresponding zenith angles for FAST. From 2025 to 2028, periastron passages are only observable by FAST on 20250219 and 20270421, where the zenith angle is \(<28.^{\circ}5\).
    }
    \label{tab:periastron_passage} \footnotesize
    \begin{tabular}{lcr}
    \hline \hline
     Date   &  Time   & Zenith angle \\
     yyyymmdd  &   (UTC + 08:00) &  ($^{\circ}$)  \\
	\hline
	{\bf 20250219} & 22:12  & {\bf 22} \\
    20250510 & 00:13  & 110 \\
    20250728 & 02:14  & 129 \\
    20251015 & 04:15  & 35 \\
    20260102 & 06:16  & 78 \\
    20260322 & 08:17  & 149\\
    20260609 & 10:18  & 65 \\
    20260827 & 12:19  & 47 \\
    20261114 & 14:20  & 139 \\
    20270201 & 16:21  & 97 \\
    {\bf 20270421} & 18:23  & {\bf 23} \\
    20270709 & 20:24  & 112 \\
    20270926 & 22:25  & 127 \\
    20271215 & 00:26  & 33 \\
    20280303 & 02:27  & 80 \\
    20280521 & 04:28  & 148 \\
    20280808 & 06:29  & 63 \\
    20281026 & 08:30  & 49 \\
    \hline
    \end{tabular}
\end{table*}

\begin{figure}[bt]
    \centering
    \includegraphics[width=0.35\columnwidth]{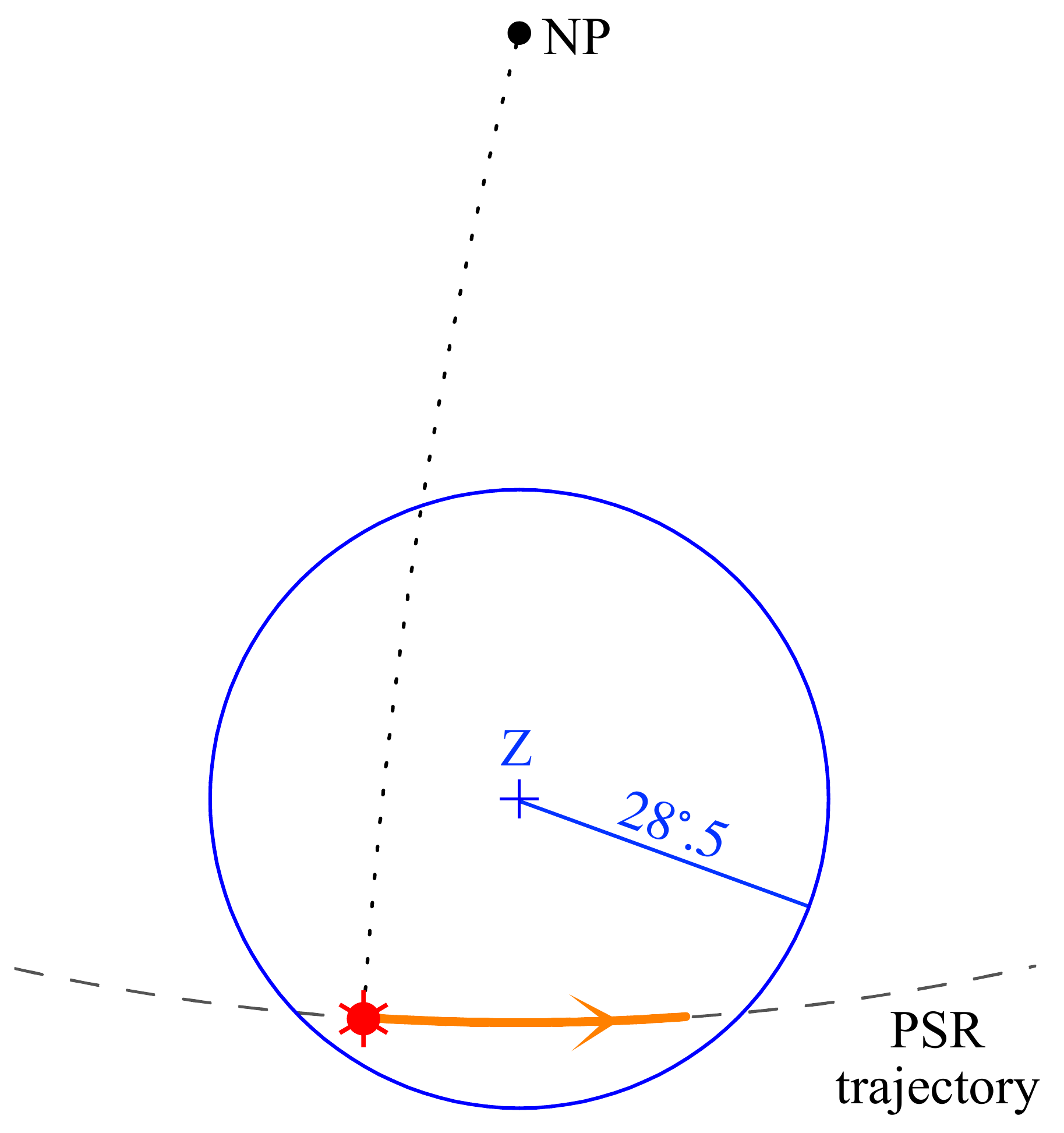}
    \caption{{\bf PSR J0653+0443 trajectory on 20250219 in the FAST view on the sky.} The north pole (NP) is indicated by the black dot, and the zenith of FAST (Z) is marked by the cross. The solid circle indicates the zenith angles of 28$^\circ$.5. PSR J0653+0443 moves on its trajectory from left to right along the dashed line. The observation duration is indicated by thick orange line. Dotted line connects the commencing position of the long tracking observation and the NP.
    }
    \label{fig:motion_trajectory}
\end{figure}

From 2022 to 2025, we conducted a series of FAST timing observations for the pulsar (see Table~\ref{tab:obslist}). Using all available FAST data and initial orbital parameters, we folded pulsar profiles into 256 bins (one profile every 1 minutes sub-integration) via the {\sc Dspsr} package \cite{Straten+2011PASA...28....1V}, then extracted pulse times of arrival (TOAs) from these folded profiles using {\sc Psrchive} \cite{Hotan+2004PASA...21..302H}.
Data processing included several steps: (1) polarization calibration under the ideal feed assumption using injected calibration signals via the \texttt{pac} command in {\sc Psrchive}; (2) automatic radio frequency interference removal following \cite{Chen+2023RAA....23j4004C}; (3) time-integration of all sub-integration profiles and dividing into 4 sub-bands across the 500 MHz bandwidth to determine the dispersion measure (DM); (4) comparing profiles to a frequency-averaged template (generated via \texttt{paas} command) and extracting TOAs from all profiles using the \texttt{pat} command.

Using {\sc TEMPO2} \cite{Hobbs+2006MNRAS.369..655H}, we obtained phase-connected timing solutions after four successive observations near the periastron passage in June 2024 (Table~\ref{tab:obslist}), yielding the exact position of PSR J0653+0443. On 20250219, we conducted a 2-hour on-source observation within the FAST zenith field (Fig.~\ref{fig:motion_trajectory}), covering periastron passage (Fig.~\ref{fig:geo}). For this long session, we summed all frequency channels via the \texttt{pam} command while retaining a 30-s integration time to minimize true anomaly ($\theta$) separation between adjacent TOAs.
All data were refolded using updated timing solutions to improve precision, and this process was iterated to derive final phase-connected timing solutions from all available FAST observations. The final timing solution is summarized in Table~\ref{tab:timingPara}, with corresponding timing residuals shown in Fig.~\ref{timing_res}; the binary geometry is illustrated in Fig.~\ref{fig:geo}. With an extremely large orbital eccentricity of $e=0.99238692$, PSR~J0653+0443 is the most eccentric stellar binary ever known (the second highest one is WDS~05354$-$3316 with $e= 0.985\pm0.002$ \cite{Tokovinin+2020AJ....160....7T}).

\begin{figure}[t]
    \centering
    \includegraphics[width=0.55\columnwidth]{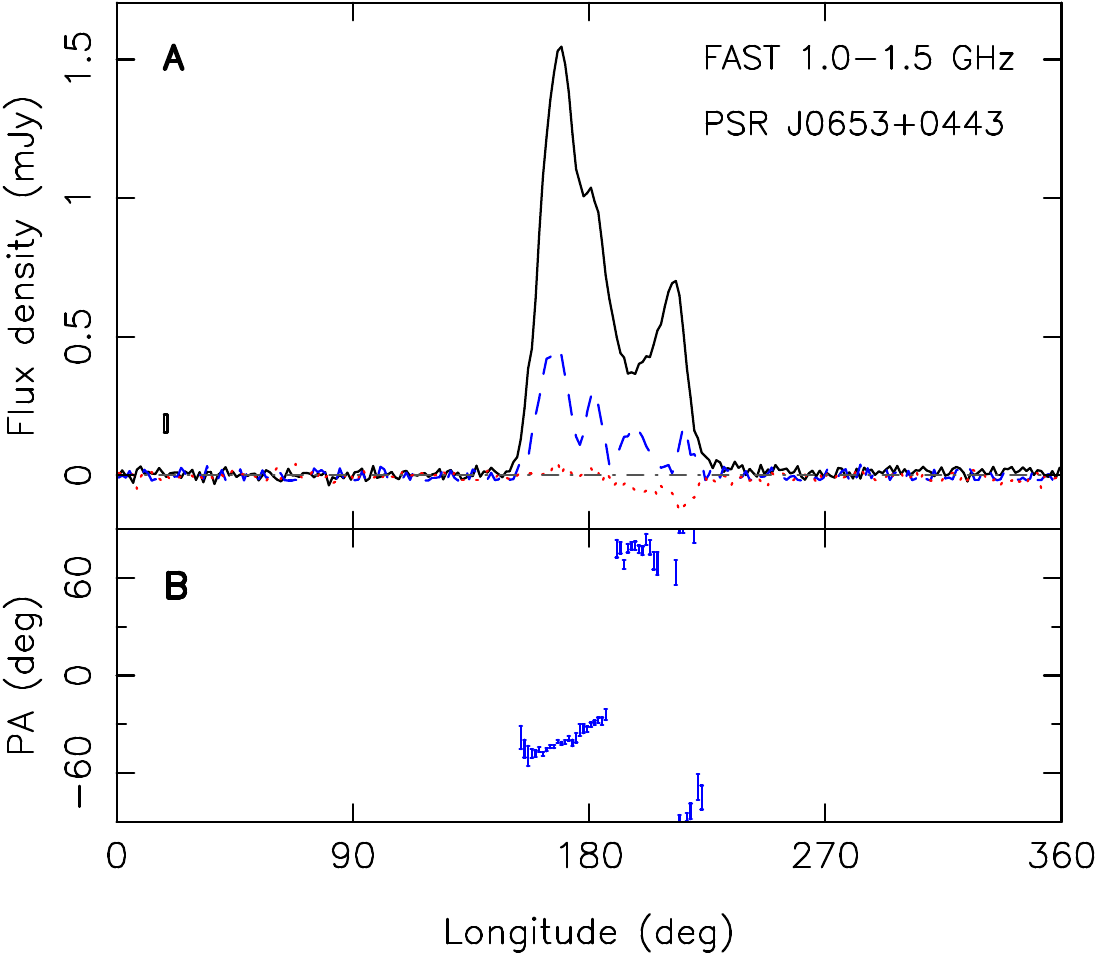}
    \caption{\textbf{Polarization profiles of PSR~J0653+0443.} The 256-bin profiles were obtained from the 2-hr observation on 2025 February 19. Panel ({\bf A}) shows the total intensity $I$ (black line), linear $L$ (blue dash line) and circular $V$ (red dotted line, positive for the left-hand sense) polarization profiles, with a scale-mark for $\pm2\sigma_{\rm I,off}$ height and 1-bin width. Panel ({\bf B}) is the position angles (PAs) of the linear polarization at reference frequency 1250 MHz with uncertainties of $\pm1\sigma$.}
    \label{pol}
\end{figure}

Based on timing solutions, a precise binary mass function for PSR J0653+0443 was found to be $f(m_{\rm p},m_{\rm c},i) = (4\pi^2/G)x^3/P_{\rm orb}^2=(m_{\rm c}\sin i)^3/(m_{\rm p}+m_{\rm c})^2=0.33966(21)~M_\odot$. 
Since $\sin i \leq 1$, the companion mass is constrained to $>1.38\ M_\odot$, indicating the DNS nature for the binary. The total system mass derived from the periastron advance is 2.81(2)$~M_\odot$.
Using the relation between the relative orbital semi-major axis $a_{\rm pc}$ and total mass $m_{\rm t}$:
\begin{equation}
(4\pi^2/G)a_{\rm pc}^3/P_{\rm orb}^2=m_{\rm t},
\end{equation}
we find $a_{\rm pc} = 254\ \text{lt-s}$ --- about twice the pulsar's projected semi-major axis $x = a_{\rm pc} m_{\rm c} \sin i/ (m_{\rm p}+m_{\rm c}) = 125.5\ \text{lt-s}$. From empirical DNS mass distributions --- where recycled neutron star masses are typically comparable to or exceed their companions \cite{Farrow+2019ApJ...876...18F} --- we constrain the orbital inclination to $\sin i > 0.98$, indicating a nearly edge-on orbit. 

The polarization profiles of PSR~J0653+0443 were obtained from the 2-hr observation on 2025 February 19, as showed in Fig.~\ref{pol}. The Faraday rotation measure (RM) was determined using \texttt{rmfit} and then polarization data were RM-corrected using \texttt{pam}. The fitting result of RM is 26.5$\pm$0.5 rad\,m$^{-2}$. The RM of the pulsar should be 21.4$\pm$0.6 rad\,m$^{-2}$ after subtracting the contribution from the Earth’s ionosphere 5.1$\pm$0.3 rad\,m$^{-2}$ which is obtained using the vertical total electron content maps of the ionosphere, CODE\footnote{ftp.aiub.unibe.ch/CODE/} and the International Geomagnetic Reference Field (IGRF-13)\footnote{https://www.ngdc.noaa.gov/IAGA/vmod/igrf.html} with an updated code IONFR\footnote{https://sourceforge.net/projects/ionfarrot/} \cite{Sotomayor+2013A&A...552A..58S}. Assuming a telescope gain $G$ of 16.1 K\,Jy$^{-1}$ \cite{Jiang+2020RAA....20...64J,Gao+2022SCPMA..6529705G}, the flux density of PSR~J0653+0443 is $S_{\rm 1250MHz} = 102.2\pm 0.4$~{\textmu}Jy, which was extracted using \texttt{psrflux}. These parameters are also listed in Table~\ref{tab:timingPara}.

\begin{table}[tb]
    \centering
    \caption{{\bf Parameters of DNS systems sorted by orbital eccentricity ($e$).} Data are taken from the references listed in the last column. Candidate DNS systems are marked with * after their names. Based on the periastron advance rate, we derived the relative separation $S_{\rm pc,min}$, velocity $v_{\rm pc,max}$ and acceleration $Acc_{\rm pc,max}$ between the pulsar and its companion at the time of periastron passage for each system, and calculated the range of value for $\Delta_{\rm RB,parv}$. 
    }
    \label{tab:DNSlist}
    \footnotesize 
    \centering
    \setlength\tabcolsep{3pt}
    \begin{tabular}{lrrrrrrrrcc}
    \hline\\[-1mm]
    PSR      & \multicolumn{1}{c}{$P_{\rm orb}$} &  \multicolumn{1}{c}{$x$}    & \multicolumn{1}{c}{$e$} & \multicolumn{1}{c}{$\omega$}    & \multicolumn{1}{c}{$\dot{\omega}$} & \multicolumn{1}{c}{$S_{\rm pc,min}$} & \multicolumn{1}{c}{$v_{\rm pc,max}$} & \multicolumn{1}{c}{$Acc_{\rm pc,max}$} & $\Delta_{\rm RB,parv}$ & Ref.\\
         & \multicolumn{1}{c}{(day)}  &  \multicolumn{1}{c}{(lt-s)} &  & \multicolumn{1}{c}{(deg)} & \multicolumn{1}{c}{(deg\,yr$^{-1}$)} & \multicolumn{1}{c}{($R_\odot$)} & \multicolumn{1}{c}{(km\,s$^{-1}$)} & \multicolumn{1}{c}{(m\,s$^{-2}$)} &(\textmu s) & \\
    \hline\\ [-1mm]
{\bf J0653+0443}   & 79.084062  & 125.540  & 0.9923869 & 44.02 & 0.0179         & 0.83  &  1.14$\times10^3$ &  1.11$\times10^3$  & (-2.91 to +2.49)$\times10^3$  &  -  \\[1mm] 
\hline  
\multicolumn{11}{c}{DNS systems in the Galactic field}\\ \hline
J1811$-$1736 & 18.7792 & 34.7827 & 0.8280  & 127.66 & 0.0090          &  7.02 & 359  &   14   & -80 to +31 & \cite{Corongiu+2007AandA...462..703C}\\
B1913+16     &     0.3230   &   2.3418 & 0.6171  & 292.54 & 4.22659   &  1.07 & 902  &  674   & -3 to +30 & \cite{Weisberg+2016ApJ...829...55W} \\ 
J1757$-$1854 &     0.1835   &   2.2378 & 0.6058  & 301.99 & 10.36497  &  0.75 & 1057 & 1336   & -8 to +42 & \cite{Cameron+2023MNRAS.523.5064C} \\                   
J2150+3427*   &    10.5921   &  25.4880 & 0.6015  &  262.79 &  0.0115 & 11.10 & 267  &    6   & -0 to +26 & \cite{Wu+2023ApJ...958L..17W} \\
J0509+3801   & 0.3796   &   2.0505 & 0.5864  & 144.20 & 3.03476       &  1.29 & 813  &  465   & -23 to +10  & \cite{McEwen+2024ApJ...962..167M} \\ 
J1930$-$1852 &    45.0600   &  86.8903 & 0.3989  & 292.08 & 0.00078   & 43.91 & 125  &  0.3   & -1 to +19 & \cite{Swiggum+2015ApJ...805..156S} \\ 
J1901+0658   &    14.4548   &  32.4020 & 0.3662  &  43.63 &   0.00531 & 22.27 & 181  &    2   & -14 to +4 & \cite{Su+2024MNRAS.530.1506S} \\
J1208$-$5936 &    0.6316    &  4.2570  & 0.3480  &  247.98 &  0.918   &  2.77 & 490  &   92   & -1 to +13 &  \cite{Bernadich+2023AandA...678A.187C}\\
J1759+5036*  &   2.0430    &  6.8246 & 0.3083     &  92.13 & 0.12890  &  6.51 & 320  &   17   & -8.2 to +0.0 & \cite{McEwen+2024ApJ...962..167M}\\
J1753$-$2240*  &  13.6376    & 18.1154 & 0.3036   & -49.31            & -     & -    &   -    & - & - &  \cite{Keith+2009MNRAS.393..623K} \\
J0528+3529   & 11.7262  &  31.4347 & 0.2901 & 184.94 &  0.00720       & 21.83 & 179  &    2   & -7.5 to +8.6 &  \cite{Wang+2024arXiv241203062W}\\
B1534+12     &     0.4207   &   3.7295 & 0.2737  & 283.31 & 1.75579   &  2.38 & 523  &  129   & -0 to +12 & \cite{Fonseca+2014ApJ...787...82F} \\
J1155$-$6529 &  3.67        & 15.34    & 0.26     & -     & -         &  -    & -    &  -     & -         & \cite{Padmanabh+2023MNRAS.524.1291P}\\
J1518+4904   &     8.6340   &  20.0440 & 0.2495  & 342.55 & 0.01137   & 18.54 & 187  &    2   & -3.6 to +6.1 & \cite{Janssen+2008AandA...490..753J} \\  
J1844$-$0128 &  10.6003  & 20.9678  & 0.2349 & 255.70 & 0.0059        & 18.59 & 147  &    1   & -0.1 to +4.7 & \cite{Wang+2024arXiv241203062W} \\
J1018$-$1523*  &   8.9840    & 26.1566 & 0.2277   & 60.01 & 0.010     & 19.10 & 176  &    2   & -8.2 to +0.8 & \cite{Swiggum+2023ApJ...944..154S} \\
J1846$-$0513 &     0.6130   &   4.7565 & 0.2086  & 328.84 & 0.8956    &  3.32 & 427  &   66   & -2.7 to +7.3 & \cite{Zhao+2024ApJ...964L...7Z} \\
J1756$-$2251 &     0.3196   &   2.7565 & 0.1806  & 327.82 & 2.58240   &  2.21 & 512  &  145   & -1.9 to +5.5 & \cite{Ferdman+2014MNRAS.443.2183F} \\
J1411+2551   &     2.6159   &   9.2051 & 0.1699 & 81.54 & 0.07680     &  9.05 & 250  &    9   & -4.8 to +0.0  & \cite{Martinez+2017ApJ...851L..29M} \\
J1829+2456   &     1.1760   &   7.2380 & 0.1391  & 229.92 & 0.29319   &  5.54 & 319  &   23   & -0.8 to +5.0 & \cite{Haniewicz+2021MNRAS.500.4620H} \\ 
J0453+1559   & 4.0725   &  14.4668 & 0.1125  & 223.07 & 0.03793       & 13.31 & 209  &    4   & -0.7 to +3.5  & \cite{Martinez+2015ApJ...812..143M} \\  
J1913+1102   &     0.2063   &   1.7546 & 0.0895 & 264.28 & 5.632      &  1.90 & 561  &  218   & 0.0 to +2.8 & \cite{Lazarus+2016ApJ...831..150L}  \\
J1755$-$2550*  &   9.6963    & 12.2844 & 0.0894  & 129.68 & -         & -     & -    &   -    & -          & \cite{2018MNRAS.476.4315N}\\
J0737$-$3039B &     0.1023   &   1.5161 & 0.0878 & 267.03 & 16.89950  &  1.15 & 683  &  534   & -0.0 to +3.5 & \cite{Kramer+2021PhRvX..11d1050K} \\
J0737$-$3039A & 0.1023   &   1.4150 & 0.0878 & 87.03 & 16.89950       &  1.15 & 683  &  534   & -3.3 to +0.0  & \cite{Kramer+2021PhRvX..11d1050K} \\
J1906+0746*  &   0.1660    &  1.4200 & 0.0853    &  76.33 & 7.5841    &  1.60 & 581  &  279   & -2.3 to +0.0 & \cite{Leeuwen+2015ApJ...798..118V} \\
J1325$-$6253 &     1.8156   &   7.5739 & 0.0640 & 190.23 & 0.138      &  8.01 & 254  &   11   & -0.8 to +1.2  & \cite{Sengar+2022MNRAS.512.5782S} \\ 
J1946+2052   &     0.0785   &   1.1545 & 0.0638 & 130.36 & 25.7921    &  0.98 & 722  &  717   & -2.1 to +0.3 & \cite{meng2025doubleneutronstarpsr} \\
\hline
\multicolumn{11}{c}{DNS systems in the Globular clusters}\\  \hline 
J1835$-$3259A* & 9.25      & 19.6    & 0.968      & 291   & -         &  -    & -    & -      & -        & \cite{DeCesar+2015ApJ...807L..23D}\\
J0514$-$4002A* &  18.7852  & 36.2903 & 0.8880     & 82.34 &  0.01296  &  4.50 & 445  &   33   & -76 to +4  & \cite{Ridolfi+2019MNRAS.490.3860R} \\
J2140$-$2311B* &  6.2157   & 19.5222 & 0.8794     & 160.80 & 0.078    &  2.34 & 625  &  127   & -159 to +128 & \cite{Balakrishnan+2023ApJ...942L..35B} \\
J1807$-$2459B* &   9.9567    & 28.9204 & 0.7470   &  11.33 & 0.01832  &  6.74 & 356  &   16   & -68 to +56 & \cite{Lynch+2012ApJ...745..109L} \\
J0514$-$4002E* &     7.4479   &  27.8190 & 0.7079  & 65.32 & 0.03468  &  7.36 & 414  &   20   & -77 to +12  & \cite{Barr+2024Sci...383..275B} \\  
B2127+11C    &     0.3353   &   2.5185 & 0.6814  & 345.31 & 4.4644    &  0.90 & 982  &  913   & -32 to +43 & \cite{Jacoby+2006ApJ...644L.113J} \\ 
J1748$-$2446ao* & 57.5557  & 62.3139 & 0.3249     & 245.29 & 0.000560 & 62.22 & 114  &  0.2   & -1 to +10 & \cite{Padmanabh+2024AandA...686A.166P}   \\
\hline
    \end{tabular}
\end{table}


\section{Striking features of PSR~J0653+0443 DNS system compared with others}
\label{compareDNS}

We listed the parameters of known DNS systems in Table~\ref{tab:DNSlist}. Among them, the DNS system of PSR~J0653+0443 has {\it striking features}: the longest orbital period $P_{\rm orb} = 79.084062$~days, the longest projected semi-major axis $x = 125.540$~lt-s, the largest orbital eccentricity $e=0.9923869$, the second shortest relative separation $S_{\rm pc,min}=0.83R_\odot$ near the periastron, the largest relative velocity $v_{\rm pc,max} = 1.14\times10^3~ {\rm km~s}^{-3}$ near the periastron, the second largest relative acceleration velocity $Acc_{\rm pc,max} = 1.11\times10^3 ~ {\rm m~s}^{-2}$ near the periastron, and the largest TOA modulation $\Delta_{\rm RB,parv} = (-2.91,\ 2.49)$~ms caused by the orbital-phase-dependent periastron advance. 
The DNS system of PSR J1757$-$1854 has the largest relative acceleration $Acc_{\rm pc,max}$ and the shortest relative separation $S_{\rm pc,min}$ near the periastron \cite{Cameron+2018MNRAS.475L..57C, Cameron+2023MNRAS.523.5064C}. In contrast, PSR J1946+2052 has the shortest orbital period among all DNS systems \cite{Stovall+2018ApJ...854L..22S,meng2025doubleneutronstarpsr}. 
No doubt that the DNS system of PSR J0653+0443 is highly relativistic around periastron passage.

\begin{figure}
    \centering
    \includegraphics[width=0.65\columnwidth]{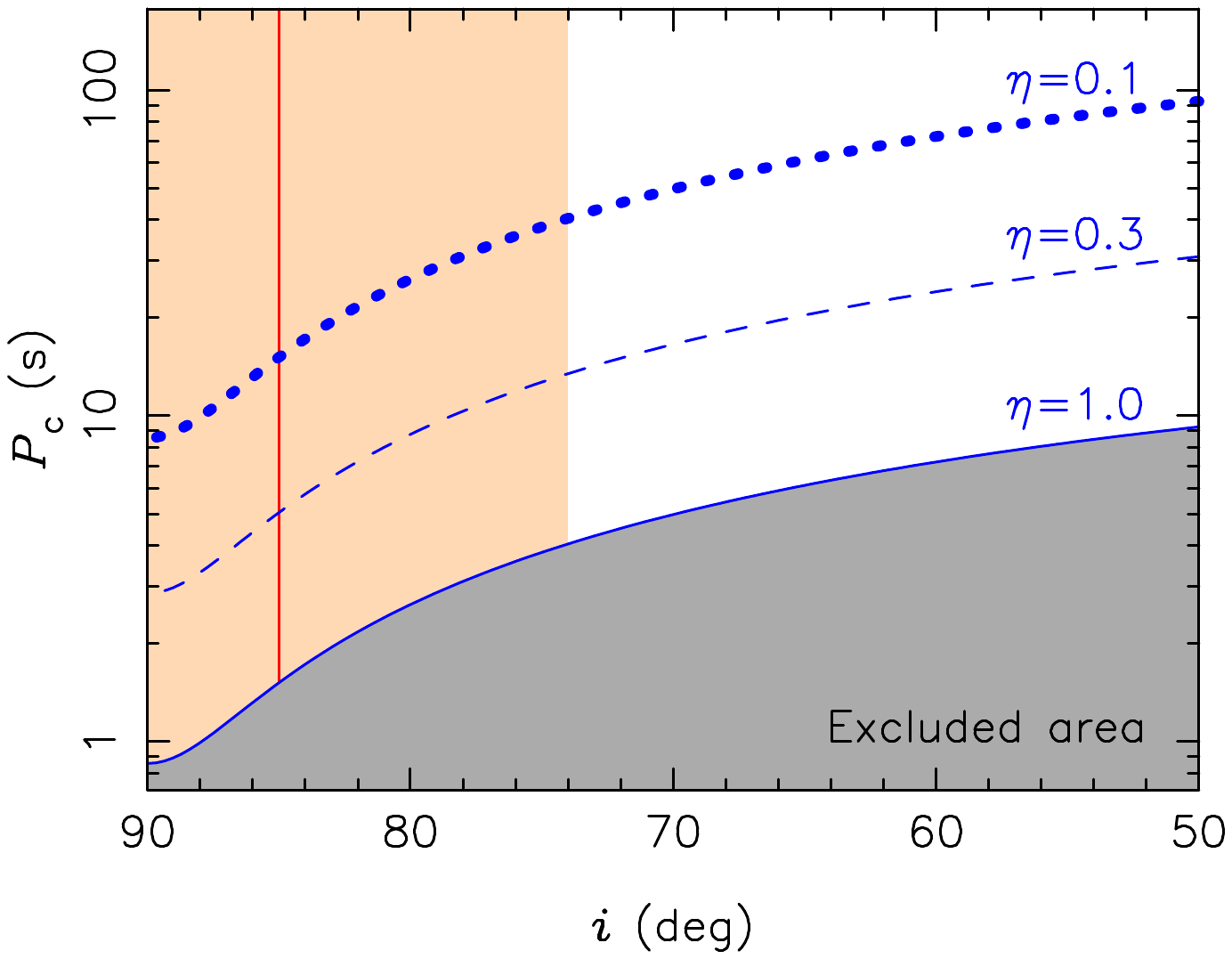}
    \caption{\textbf{Parameter space of orbital inclination angle $i$ and companion spin period $P_{\rm c}$ constrained by the pulsar eclipse.} The three blue solid lines show the fraction $\eta=d/r_{\rm LC} = 0.1$, 0.2 and 0.3 of the companion magnetosphere responsible for eclipse, that is 0.1 for the double pulsar PSR~J0737$-$3039A/B \cite{Breton+2012ApJ...747...89B,Lower+2024A&A...682A..26L}. Some parameter space can be excluded because $\eta$ cannot be larger that 1, and the companion's spin period must be longer than 0.86 s. The red line marks the best fitting $i$ from pulsar timing, and the orange area is for the uncertainty of $i$ at a 68.3\% confidence level. }
    \label{fig:ipcns}
\end{figure}

\section{Constraints on the companion properties by the observed eclipse of PSR J0653+0443}
\label{pulsar_eclipse}

During the 2-hour tracking observation of PSR~J0653+0443 on 20250219 (see Fig.~\ref{fig:geo}) --- conducted to cover the periastron passage --- pulsar signals were eclipsed for approximately 85~s (see Fig.~\ref{fig:eclipse}). This eclipse corresponds to a true anomaly span of $\Delta\theta = 6.84^\circ $ and occurs around the conjunction phase $\theta = 90^\circ - \omega = 46^\circ $, a phenomenon similar to that of the double pulsar PSR~J0737$-$3039A \cite{Lyne+2004Sci...303.1153L} that has an orbital inclination angle of $\sim 89.4^\circ $ \cite{Kramer+2021PhRvX..11d1050K}. The eclipse is caused by the pulsar passing behind the magnetosphere of its companion neutron star, confirming that PSR~J0653+0443 resides in a nearly edge-on orbit. At the conjunction phase of the PSR~J0653+0443 DNS system, the separation between the two neutron stars is $ {S}_{\rm pc} = a_{\rm pc}(1-e^2)(1+e\sin\omega)^{-1} = 2.28\ \text{lt-s} $  and can be approximated as a constant in the short eclipse duration.

For a given companion spin period $ P_{\rm c} $ (in seconds), the light-cylinder diameter is $ D_{\rm LC} = cP_{\rm c}/\pi \sim 0.32 P_{\rm c}~\text{lt-s} $. The effective magnetosphere responsible for eclipse is only a fraction of the light-cylinder, written as $ \eta D_{\rm LC}$, and $ \eta \sim 0.1 $ for the double pulsar PSR~J0737$-$3039A/B \cite{Breton+2012ApJ...747...89B,Lower+2024A&A...682A..26L}. According to binary evolution theories, the companion of PSR J0653+0443 was not a recycled neutron star. Given that PSR J0653+0443 has a large characteristic age of 6.5~Gyr, its companion neutron star probably has an unknown long spin period $P_{\rm c}$.

For such a companion to produce the eclipse, the effective magnetosphere must intersect with the LoS to the pulsar. 
The LoS has an inclination angle of $i$ from the normal direction of the orbital plane, and can be expressed as the unit vector$\overrightarrow{\rm LoS}$, approximated as $(\sin i \cos(\Delta\theta/2),\  \pm\sin(\Delta\theta/2)\sin i ,\ \cos i) \simeq (\sin i,\ \pm\Delta\theta/2\sin i,\ \cos i)$ at the start or the end of the eclipse (here assuming that the $x$-axis is along the direction from the pulsar to its companion, the $z$-axis is along the normal direction of the orbital plane). Deviating from the vector from the pulsar to the companion, $\vec{S}_{\rm pc} = (S_{\rm pc}, 0, 0)$, the maximum offset of the companion $d_{\rm max}$ from the LoS to the pulsar is
\begin{equation}
d_{\rm max} = |\vec{S}_{\rm pc} \times \overrightarrow{\rm LoS}| = {S}_{\rm pc} \left( \cos^2 i +  (\Delta\theta/2)^2 \sin^2i \right)^{1/2}  = \eta (D_{\rm LC}/2).
\label{er2}
\end{equation}
One then gets $ \cos^2 i [1-(\Delta\theta/2)^2] + (\Delta\theta/2)^2 = (\eta D_{\rm LC}/2S_{\rm pc})^2$, and finally finds an interplay between $i$ and $P_{\rm c}$ as $\cos i = 0.07\sqrt{\eta^2P_{\rm c}^2 - 0.73}$ for different values of $\eta$, as shown in Fig.~\ref{fig:ipcns}. The companion has a spin period of $P_{\rm c}>0.86$~s (found with $\eta=1$ and $\cos i =0$). For the best-fit orbital inclination angle from timing, $i = 76^\circ$  (Table~\ref{tab:timingPara}), the companion should have a spin period of $P_{\rm c} \simeq 3.5$~s for $\eta=1$. A slowly rotating companion has a larger magnetosphere, and more easily causes the eclipse even in a less edge-on orbit. With $\eta=0.1$ (the value of the double pulsar PSR J0737$-$3039B), PSR J0653+0443’s companion should have a period $P_{\rm c} \simeq 8.6$~s for $i = 90^\circ$, and $P_{\rm c} \simeq 35$~s for the best-fit inclination angle of $i=76^\circ$. 

Although the companion could in principle be an observable radio pulsar, searches for its pulsed emission from our FAST data yielded no detection. With the longest observation session lasting 2~hr and a signal-to-noise ratio threshold of 10 for pulsar detection \cite{Han+2021RAA....21..107H}, the maximum flux density of the companion’s radio emission is $6.1 \sqrt{W / (P_{\rm c} - W)}$ {\textmu}Jy, where $W$ is the effective pulse width. For $W/P_{\rm c} = 0.1$ and 0.01, the corresponding upper limits on the companion’s flux density are 2.0 and 0.6~{\textmu}Jy, respectively.

\section{Keplerian parameters and modified timing models for pulsars in a high-eccentricity orbit}
\label{sec:KandPK}

In general, the mean anomaly $M$, the eccentric anomaly $u$ and the true anomaly $\theta$ are employed to describe the motion of an object in an elliptic orbit. For a constant orbital period $P_{\rm orb}$,  the mean anomaly $M$ at time $t$ (after the periastron passage $T_0$) is given by $ M=2\pi (t-T_0)/P_{\rm orb}. $  The eccentric anomaly $u$ is the angle between the auxiliary circle and the semi-major axis of the ellipse, related to $M$ by the Kepler equation $ u-e\sin u=M, $ here $e$ is the orbital eccentricity. The true anomaly $\theta$ is the angle between the current position of an object and its closest approach (periastron) relative to the focus of the orbit, which can be calculated from $u$ using $ \tan(\theta/2)=[(1+e)/(1-e)]^{1/2}\tan(u/2), $, with with its time derivative given by $    \Dot{\theta}=(2\pi/P_{\rm orb})(1-e^2)^{-\frac{3}{2}}(1+e\cos\theta)^2. $
PSR~J0653+0443 moves in an extremely high-eccentricity orbit: it takes only 46 minutes to travel from $\theta = -90^\circ$ to $90^\circ$, while for most of its 79-day orbital period, the pulsar resides in the orbital phase with true anomaly $\theta \sim 180^\circ$.

The R\"{o}mer delay accounting for pulsar orbital motion (Eq.~\ref{Romer_delay}) is incorporated into standard pulsar timing models, such as the DD model \cite{Damour+1985AIHPA..43..107D, Damour+1986AIHPA..44..263D} and the DDGR model \cite{Taylor+1987grg..conf..209T, Taylor+1989ApJ...345..434T}.  For high-eccentricity orbits, the Keplerian parameters ---  projected semi-major axis $x$, longitude of periastron $\omega$, and orbital eccentricity $e$  --- exhibit strong covariance when the DD or DDGR models are applied to pulsar timing analyses. Analogous to the ELL1 model designed for low-eccentricity cases \cite{Lange+2001MNRAS.326..274L}, we take  
\begin{equation}
    \left\{
    \begin{aligned}
    &\zeta_1 \equiv x \sin \omega / e, \\ 
    &\zeta_2 \equiv x \cos \omega \sqrt{1-e^2},\\
    \end{aligned}
    \right.
\end{equation}
to replace Keplerian parameters $x$ and $\omega$ in the timing model for pulsars in high-eccentricity orbits. Here, 
$\zeta_1$ denotes the projected length of the distance from the orbital center to the directrix onto the LoS; 
$\zeta_2$ denotes the projected length of the semi-minor axis onto the LoS.
We then revised the DD model and implemented it in the \textsc{TEMPO2} timing software as the binary model "DDe". Similarly, we revised the DDGR model to get the "DDeGR" model.
%
%
With this replacement, the R\"{o}mer delay given in Eq.~\eqref{Romer_delay} can be expressed as
\begin{equation}
\Delta_{\rm RB}=[\zeta_1e(\cos u-e)+\zeta_2\sin u]/c,
\label{Romer_delay_new}
\end{equation}
and $\zeta_1$, $\zeta_2$, and $e$ are independent when working for timing solution. Such independence can be verified by examining the partial derivatives of the so-expressed R\"{o}mer delay. Taken the relation $\partial u/\partial e=\sin u/(1-e\cos u)$ derived from $u-e\sin u = M$ into account, the partial derivatives of the R\"{o}mer delay $\Delta_{\rm RB}$ for $\zeta_1$, $\zeta_2$, and $e$ are given by
\begin{equation}
\left\{
\begin{aligned}
    {c}\frac{\partial \Delta_{\rm RB}}{\partial \zeta_1} &=e(\cos u-e),\\
    {c}\frac{\partial\Delta_{\rm RB}}{\partial \zeta_2} &=\sin u,\\
    {c}\frac{\partial \Delta_{\rm RB}}{\partial e} &=\zeta_1(1-e^2)\frac{\cos u}{1-e\cos u} + \zeta_2\frac{\sin u\cos u}{1 - e\cos u} - 3\zeta_1e.
    \end{aligned}
\right.
    \label{romer_delay_u}
\end{equation}
One can substitute $u$ by the true anomaly $\theta$ with the expressions of $\cos u = (e + \cos \theta)/(1 + e \cos \theta)$ and $\sin u = \sqrt{1 - e^2} \sin \theta / (1 + e \cos \theta)$, the partial derivatives in Eq.~\ref{romer_delay_u} can be rewritten as:
\begin{equation}
\left\{
\begin{aligned}
    {c}\frac{\partial \Delta_{\rm RB}}{\partial \zeta_1} &=\frac{e(1-e^2)\cos\theta}{1+e \cos\theta},\\
    {c}\frac{\partial \Delta_{\rm RB}}{\partial \zeta_2} &=\frac{\sqrt{1-e^2}\sin\theta}{1+e\cos \theta},\\
    {c}\frac{\partial \Delta_{\rm RB}}{\partial e} &=\zeta_1(e+\cos\theta)+\frac{\zeta_2}{\sqrt{1-e^2}}\frac{\sin\theta(e+\cos\theta)}{1+e\cos \theta}-3\zeta_1e.
    \end{aligned}
\right.
    \label{romer_delay_f}
\end{equation}
These partial derivatives of the R\"{o}mer delay vary significantly over the orbital period. The first two partial derivatives in Eq.(12) describe the variations in the two orthogonal projections of $u$, indicating the orthogonality of $\zeta_1$ and $\zeta_2$ across most orbital phases --- though not near periastron. For high-eccentricity orbits, near periastron these two partial derivatives approximate to zero within a small range around $u=0^\circ$. In contrast, the last partial derivative ($\frac{\partial \Delta_{\rm RB}}{\partial e}$) exhibits substantial variation, so that observation data taken around periastron can constrain the eccentricity $e$ with high accuracy. In practice, we also replace $e$ with the parameter $\delta_{\rm e} \equiv \sqrt{1 - e^2}$ in the timing model-fitting.

The original orbital parameters $x$, $\omega$, and $e$ can then be expressed in terms of the transformed set $\zeta_1$, $\zeta_2$, and $\delta_{\rm e}$ through the following relations
\begin{equation}
\left\{
\begin{aligned}
    & x^2 = \zeta^2_1 (1 - \delta_{\rm e}^2) + \zeta^2_2 / \delta_{\rm e}^2, \\
    & \sin \omega = \frac{\zeta_1 \sqrt{1 - \delta_{\rm e}^2}}{\sqrt{\zeta^2_1 (1 - \delta_{\rm e}^2) + \zeta^2_2 / \delta_{\rm e}^2}}, \\
    & \cos \omega = \frac{\zeta_2 / \delta_{\rm e}}{\sqrt{\zeta^2_1 (1 - \delta_{\rm e}^2) + \zeta^2_2 / \delta_{\rm e}^2}}, \\
    & e^2 = 1 - \delta_{\rm e}^2.
    \label{transform}
\end{aligned}
\right.
\end{equation}
In \textsc{Tempo2} \cite{Hobbs+2006MNRAS.369..655H}, parameter uncertainties are estimated via their partial derivatives. 
We thus obtain uncertainties for $\zeta_1$, $\zeta_2$, and $\delta_{\rm e}$ from model-fitting --- these uncertainties are nearly independent. Finally, we use Eq.~\ref{transform} to convert the fitted parameters $\zeta_1$, $\zeta_2$, and  $\delta_{\rm e}$ to the conventional orbital parameters $x$, $\omega$ and $e$, and  derive their 1$\sigma$ uncertainties via the law of propagation of uncertainties.

A large modulation $\Delta_{\rm RB,parv}$ can be used to constrain the non-relativistic periastron advance.
As proposed in \cite{Damour+1986AIHPA..44..263D}, we take the relativistic periastron advance as the predication given by the General Relativity $\dot{\omega}_{\rm GR}$ and any non-relativistic contribution written as $\dot{\omega}_{\rm EX}$, then the observed periastron advance $\dot{\omega} = \dot{\omega}_{\rm GR}+\dot{\omega}_{\rm EX}$. We then have
\begin{equation}
\begin{aligned}
    \omega-\omega_0 = &~\dot{\omega}_{\rm GR}(P_{\rm orb}/2\pi)\theta + \dot{\omega}_{\rm EX}(P_{\rm orb}/2\pi)M\\
                    = &~\dot{\omega}(P_{\rm orb}/2\pi) M + \dot{\omega}_{\rm GR} (P_{\rm orb}/2\pi) (\theta - M),
\end{aligned}
\label{dom_3}
\end{equation}
therefore both $\dot{\omega}=\dot{\omega}_{\rm GR}+\dot{\omega}_{\rm EX}$ and $\dot{\omega}_{\rm GR}$ can be measured because of their different dependence on orbital phases. By combining Eqs.~\ref{Romer_delay}, \ref{omdot}, and \ref{dom_3}, we derived the corresponding periodical modification of the R\"{o}mer delay ($\Delta_{\rm RB,parv}$, Eq.~\ref{delta_T}) caused by periastron advance rate variations $\dot{\omega}_{\rm GR} (P_{\rm orb}/2\pi) (\theta - M)$. 
Including Eq.~\ref{dom_3} for periastron advance descriptions into the DDe, DDGR, and DDeGR models, we get the DDe+, DDGR+, and DDeGR+ timing models used for pulsar timing. 

\begin{table*}[tb]
    \centering
        \caption{{\bf The correlation coefficients $r$ between orbital parameters for pulsars in high-eccentricity orbits.} Comparing to conventional orbital parameters $x$, $\omega$ and $e$, the new orbital parameters $\zeta_1$, $\zeta_2$ and $\delta_{\rm e}$ have much lower correlation coefficients between each other when fitting for the R\"{o}mer delay in \textsc{Tempo2} \cite{Hobbs+2006MNRAS.369..655H}.}
    \begin{tabular}{l|c|c|c|c|c|c}
     \hline
     \hline
     Model    & \multicolumn{3}{c|}{DDGR+} & \multicolumn{3}{c}{DDeGR+} \\
     \hline
    Parameters & $x$ \& $\omega$ & $x$ \& $e$ & $e$ \& $\omega$ & $\zeta_1$ \& $\zeta_2$ & $\zeta_1$ \& $\delta_{\rm e}$ & $\delta_{\rm e}$ \& $\zeta_2$ \\
     \hline 
     $r$ & $-0.99960$ & 0.98665 & $-0.98665$ & $0.12007$ & $0.28475$ & $-0.00134$ \\
     \hline
    \end{tabular}
    \label{tab:cor-param}
\end{table*}

\begin{table*}[tb]
	\centering
	\caption{\textbf{Parameters of PSR~J0653+0443 obtained from best-fitting of DDGR, DDe+ and DDGR+ models.} The uncertainties given in brackets are at 68.3$\%$ C.L. (1$\sigma$ uncertainty) multiplied by $\sqrt{\chi^2_{\rm min}/n_{\rm free}}$. 
	}
    \scriptsize
	\setlength{\tabcolsep}{1.5mm}
	\begin{tabular}{lccc} 
	\hline
		\hline \\[-2.5mm]
        \multicolumn{4}{c}{General information} \\
        \hline\\[-2.5mm]
		Number of TOAs      &     \multicolumn{3}{c}{342} \\
		MJD range           &    \multicolumn{3}{c}{59902 to 60804} \\
        MJD of period determination & \multicolumn{3}{c}{60300} \\
        UNITS       &   \multicolumn{3}{c}{TCB} \\
        Solar System ephemeris   &  \multicolumn{3}{c}{DE440}\\[1mm]
        \hline \\[-2.5mm]
        Binary model & DDGR & DDe+ & DDGR+  \\
        \hline \\[-2.5mm]
        Right ascension, $\alpha$ (J2000) &  06:53:04.5746(2) &  06:53:04.5746(2) & 06:53:04.5746(2) \\
		Declination, $\delta$ (J2000) & +04:43:16.357(14) & +04:43:16.350(13) & +04:43:16.349(13) \\
		Proper motion in RA (mas yr$^{-1}$) & -10(5)  & -11(5) & -10(5) \\
		Proper motion in DEC (mas yr$^{-1}$)& -12(12) & -5(12) & -4(12) \\
		Spin frequency, $\nu$ (Hz) & 21.036552768032(14) & 21.03655276804(2) & 21.036552768034(14)\\ 
		Spin frequency derivative, $\Dot{\nu}$ (Hz $\rm s^{-1}$) & -5.08(6)$\times$10$^{-17}$ & -5.09(6)$\times$10$^{-17}$ & -5.09(6)$\times$10$^{-17}$\\ 
		Dispersion measure, DM (pc $\rm cm^{-3}$)  & 27.420(4) & 27.420(4) &   27.420(4)\\
		Orbital period, $P_{\rm orb}$ (day) & 79.08406177(10) & 79.08406180(14) & 79.08406184(12) \\
		Epoch of periastron, $T_0$ (MJD) & 60725.59202197(8) & 60725.59202207(14) & 60725.59202195(7) \\
		Projected semi-major axis, $x$ (lt-s) & 125.52(6) & - & 125.539(26) \\
        Longitude of periastron, $\omega$ (deg) & 44.03(2) & - & 44.019(11) \\
        Orbital eccentricity, $e$  &  0.99238696(4) & - & 0.99238692(5) \\
	    $\zeta_1\equiv x\sin\omega/e$ (lt-s) & - & 87.905439(17) & - \\
	    $\zeta_2\equiv x\cos\omega\sqrt{1-e^2}$ (lt-s) & - & 11.118736(16) & -\\
	    $\delta_{\rm e}\equiv\sqrt{1-e^2}$ & - & 0.1231601(10)  & - \\
	    Sin($i$)         & - & 0.93(10) & - \\
	    Companion mass, $m_{\rm c}$ ($M_\odot$) & 1.5(2) & 2.2(8) & 1.44(10) \\
	    Total mass, $m_{\rm c}$ ($M_\odot$) & 2.780(4) & - & 2.814(20) \\
	    $\Dot{\omega}$ (deg\,yr$^{-1}$) & - & 0.017785(11) & 0.017791(10) \\ 
	    $\Dot{\omega}_{\rm GR}$ (deg\,yr$^{-1}$) & - & 0.01790(10) & - \\[1mm]
		\hline \hline
    \end{tabular}
    \label{timing model2}
\end{table*}

\section{Timing PSR~J0653+0443 and relativistic effects}

We applied the new binary models (DDe and DDeGR) with replaced parameters to timing data of PSR~J0653+0443.
In \textsc{Tempo2} \cite{Hobbs+2006MNRAS.369..655H},  the covariance matrix $C$ is calculated for the best-fitted model, from which the correlation coefficient $r$ between model-fitted parameter No.1 and parameter No.2 can be calculated as $r=C_{12}/\sqrt{C_{11}C_{22}}$. The correlation coefficients $r$ between $\zeta_1$, $\zeta_1$ and $\delta_{\rm e}$ for fitting for the R\"{o}mer delay are listed in Table.~\ref{tab:cor-param}, which are much smaller than those for conventional orbital parameters ($x$, $e$ and $\omega$). Therefore, $\zeta_1$, $\zeta_2$ and $e$ offer a sophisticated parameterization which can be used for fitting the high-eccentricity orbits of $\sqrt{1-e^2} \ll 1$. The derived values of PSR~J0653+0443 orbital parameters are presented in Table~\ref{tab:timingPara}. 

\subsection{Post-Keplerian parameters}
\label{sec:PK}

In the DD model for pulsar timing, the relativistic effects are described using post-Keplerian parameters such as periastron advance rate $\dot{\omega}$, Shapiro delay amplitude $r$ and shape $s$, Einstein delay parameter $\gamma$, orbital period change $\dot{P}_{\rm orb}$, periodic relativistic corrections to the orbital motion $\delta_{\rm r}$ and $\delta_\theta$, and the two aberration parameters $A$ and $B$  
\cite{Damour+1985AIHPA..43..107D, Damour+1986AIHPA..44..263D}. 
%
In the BT model \cite{Blandford+1976ApJ...205..580B},  periastron advance is assumed to be at a constant rate $\dot{\omega}$. 
Naturally, the periastron advance occurs more rapidly at smaller orbital separations when the general relativistic effects are stronger. 
Damour and Taylor (1992) \cite{Damour+1992PhRvD..45.1840D} replaced $\omega-\omega_0 = \dot{\omega}t$ in the BT model \cite{Blandford+1976ApJ...205..580B} with $\omega-\omega_0 = \dot{\omega}(P_{\rm orb}/2\pi)\theta$, and got a significantly improved fit to the PSR B1913+16 data, assuming that the periastron advance are fully relativistic. 
Assuming the validity of general relativity, these post-Keplerian parameters are the functions of total mass $m_{\rm t}$ and companion mass $m_{\rm c}$, and 
combining the measurements of these relativistic effects together can constrain the masses of binaries to a high accuracy \cite{Kramer+2021PhRvX..11d1050K, meng2025doubleneutronstarpsr}, as realized in the DDGR model \cite{Taylor+1987grg..conf..209T, Taylor+1989ApJ...345..434T}. 

From timing data of PSR~J0653+0443, we measured both the total periastron advance rate that is $\dot{\omega}=0.017785(11)$ deg\,yr$^{-1}$ and the part contributed by the relativistic effect that is $\dot{\omega}_{\rm GR}=0.1790(10)$ deg\,yr$^{-1}$ (see Table~\ref{timing model2}).    
%
%
In general relativity, $r=Gm_{\rm c}/c^3$ and $s=\sin i$, so the Shapiro delay can be calculated using Eq.~\ref{Delta_S}. The two Shapiro delay parameters, $r$ and $s$, were also marginally detected in PSR~J0653+0443.  Applying the DDe+ model to PSR~J0653+0443 timing data, we obtained $\sin i=0.93(10)$ and $m_{\rm c}=2.2(8)~M_\odot$ (see Table~\ref{timing model2}).

The Einstein delay is described by the post-Keplerian parameter $\gamma$ as $\Delta_{\rm EB}=\gamma\sin u$. Using the mass measurements from the DDeGR+ model, $\gamma$ is estimated to be 45 ms, which would induce a significant periodic modulation in the arrival times. 
However, comparing Einstein delay with Eq.~\ref{Romer_delay_new}, it's clear that without large changes of $\omega$, the post-Keplerian parameter $\gamma$ and the orbital parameter $\zeta_2/c$ are highly correlated and their uncertainties are the same. 
Currently, $\gamma$ can not be directly measured, we treated it as a fixed parameter when $\gamma$ is a parameter of the timing model (see Table~\ref{timing model2}). 

Besides the post-Keplerian parameter $\dot{\omega}$, $r$, $s$ and $\gamma$, other post-Keplerian parameters, 
including the orbital period derivative $\dot{P}_{\rm orb}$, the periodic relativistic corrections to the orbital motion $\delta_{\rm r}$ and $\delta_{\theta}$, and the two aberration parameters $A$ and $B$ \cite{Damour+1986AIHPA..44..263D}, 
are currently too small to be detected and are therefore not yet required in the timing model.

Currently, the contributions of higher post-Newtonian (PN) order general relativistic effects and neutron star deformation to periastron advance are negligible. The rotation of the neutron star, which is coupled to the orbital angular momentum vector, can influence the evolution of the eccentricity vector's orientation. This spin-orbit coupling arises from general relativistic effects at the 1.5 PN order, as well as from NS deformation caused by tidal and rotational bulges \cite{Liu+2015MNRAS.447..747L,Liu+2018ApJ...863...68L}. However, the contributions of these effects to the periastron advance are significantly weaker than the 1 PN effect by at least six orders of magnitude, therefore are currently negligible.

\begin{figure}
    \centering
    \includegraphics[width=0.45\columnwidth]{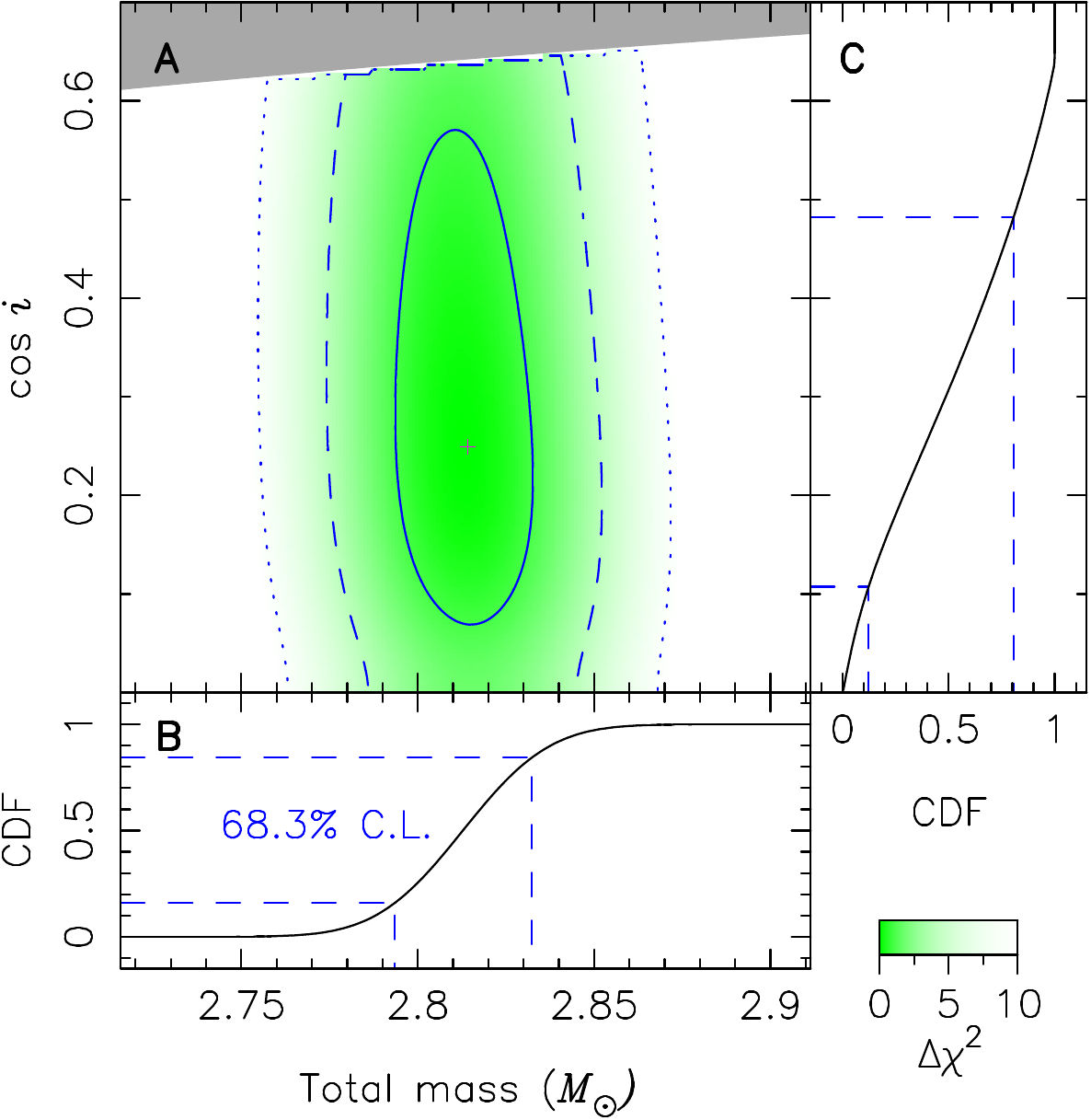}
    \hspace{2mm}
    \includegraphics[width=0.45\columnwidth]{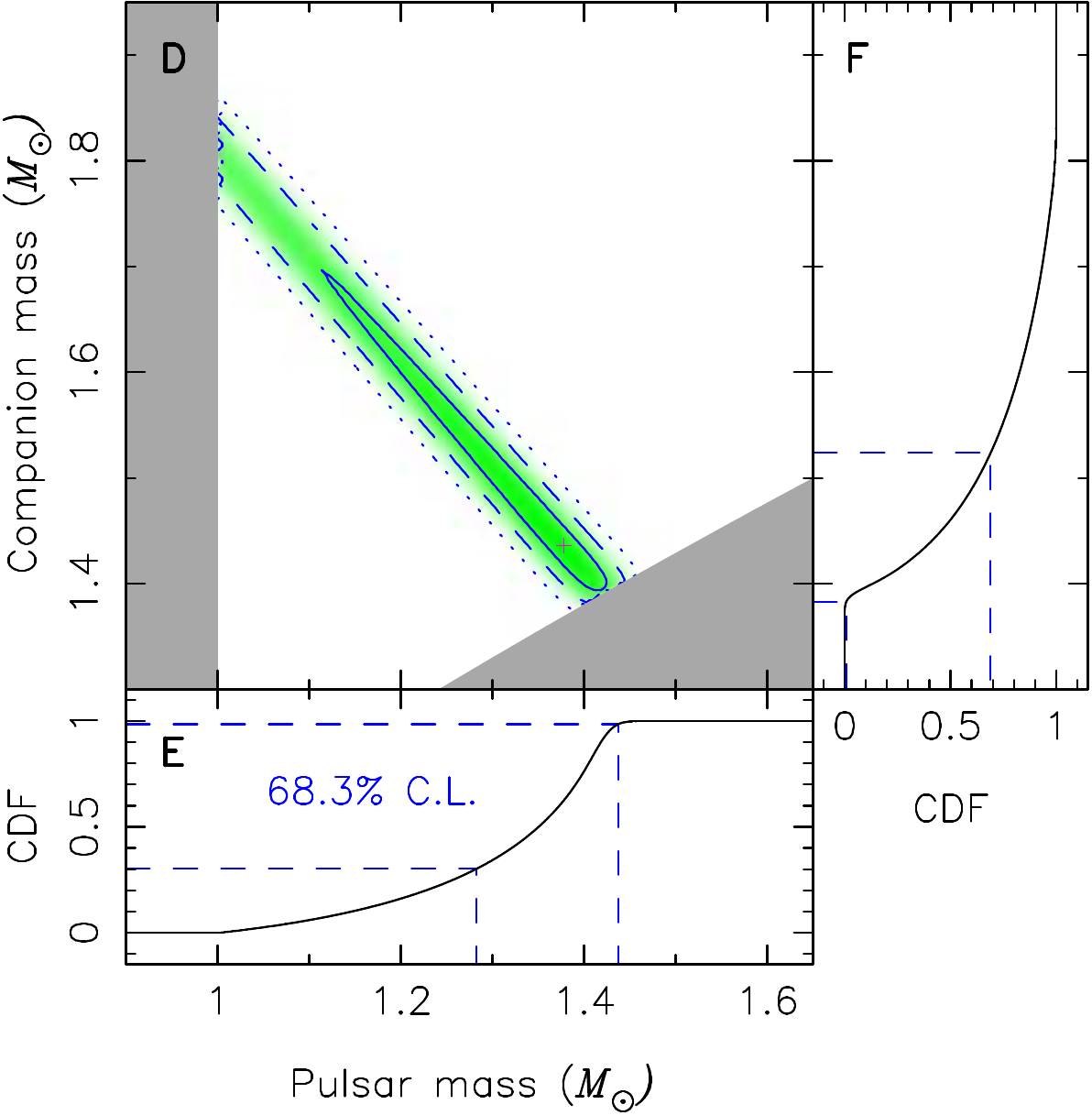}
    \caption{ 
    {\bf $\Delta\chi^2$ maps for masses and orbital inclination angle after the best-fitting the timing model to timing data of PSR~J0653+0443.}
    Panels ({\bf A}) and ({\bf D}) show the probability distributions with contours overplotted at $\Delta\chi^2 =1$, 4, and 9. The gray-shaded region is excluded by the constraints of $m_{\rm p}<1.0~M_\odot$ and $\sin i>1.0$. 
    The best-fitted parameters are marked by a red cross, corresponding to $m_{\rm t} = 2.814~M_\odot$, $i=76^\circ$, $m_{\rm p} = 1.38~M_\odot$ and $m_{\rm c}=1.44~M_\odot$. Panels {\bf B, C, E, and F} show the cumulative possibility density functions (CDF).  These CDFs yield the following parameter ranges (indicated by blue dashed lines): total mass $M_{\rm{tot}} = 2.793~\text{to}~2.832~M_\odot$, orbital inclination $i = 61~\text{to}~84^{\circ}$, pulsar mass $M_{\rm{p}} = 1.28~\text{to}~1.44~M_\odot$, and and companion mass $M_{\rm{c}} = 1.38 ~\text{to}~1.52~M_\odot$.}
    \label{mtot-cosi}
\end{figure}

\begin{figure}
    \centering
    \includegraphics[width=0.45\columnwidth]{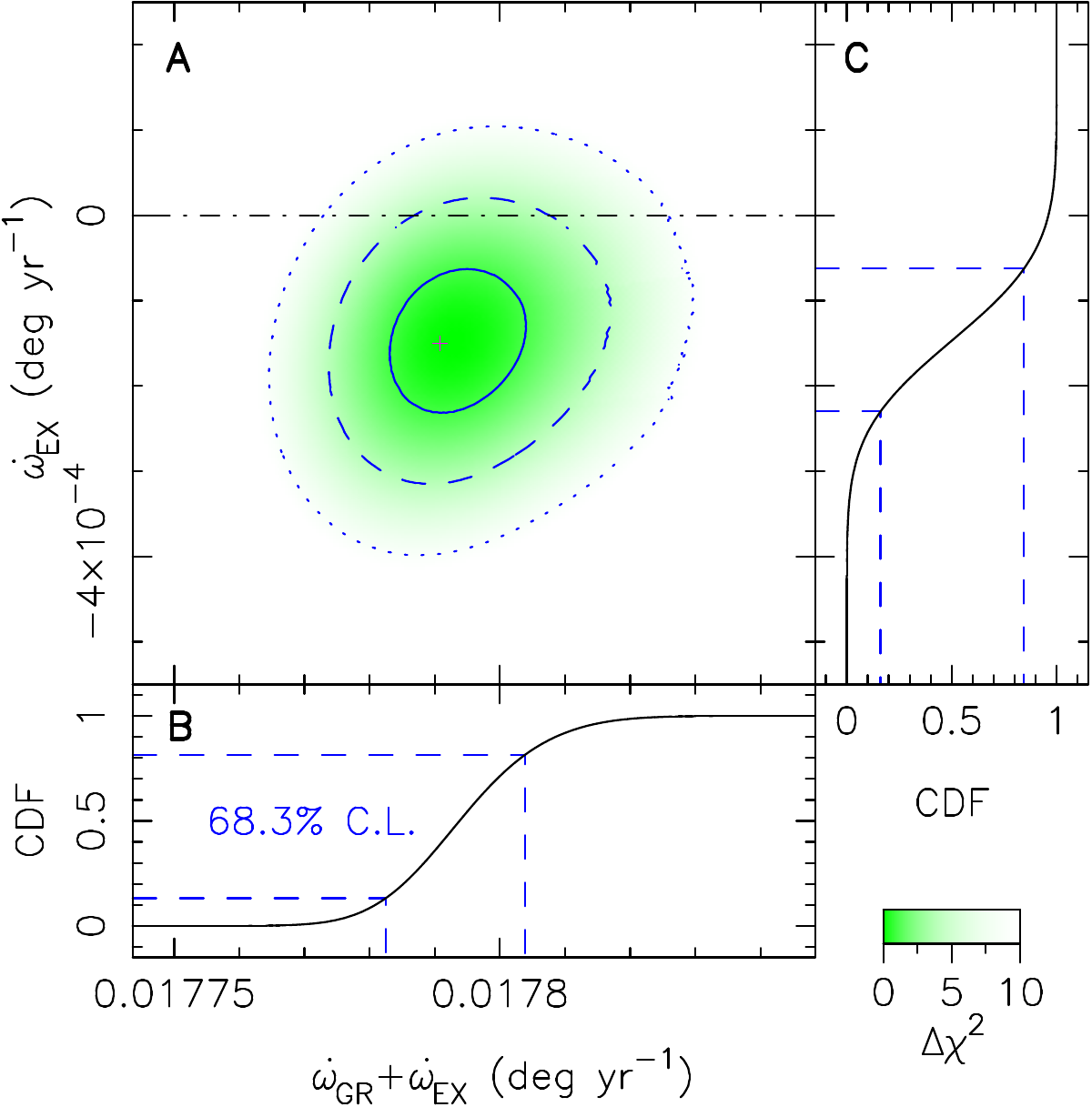}
    \caption{\textbf{$\Delta\chi^2$ map for $\Dot{\omega}_{\rm GR}$ and $\Dot{\omega}_{\rm EX}$ after the best-fitting the timing model to timing data of PSR~J0653+0443.} 
    Panel ({\bf A}) shows the probability distribution with the contours at the $\Delta\chi^2=1$, 4 and 9, and 
    the red cross marks the best-fit value. 
    Panels ({\bf B}) and ({\bf C}) show the 1D cumulative possibility density distribution functions (CDF), and the best derived $\Dot{\omega}_{\rm GR}+\Dot{\omega}_{\rm EX} =  (1.7791^{+0.0013}_{-0.0009})\times10^{-2}$ deg\,yr$^{-1}$ and 
    $\Dot{\omega}_{\rm EX} =  (-1.50^{+0.88}_{-0.80})\times10^{-4}$~deg\,yr$^{-1}$. }
    \label{omdot1-omdot2}
\end{figure}

\subsection[Bayesian analysis]{Measurement of $m_{\mathrm{p}}$, $m_{\mathrm{c}}$ and $\dot{\omega}_{\mathrm{EX}}$ via Bayesian analysis}

After we found the timing solution of PSR~J0653+0443, we performed Bayesian analyses to constrain the system masses using these relativistic effects, adopting the approach described in \cite{Splaver+2002ApJ...581..509S}. 

The corresponding \( \Delta\chi^2 \) map is shown in Fig.~\ref{mtot-cosi}; all \( \chi^2 \) values were scaled such that \( \chi^2_{\rm min} \) equals the number of free degrees of freedom \( n_{\rm free} \).
We adopted a uniform distribution as the prior and additionally imposed a constraint \( m_{\rm p}>1.0~M_\odot \) based on current knowledge of neutron star masses \cite{PulsarAstronomy2022}. The posterior probability density \( p(m_{\rm t},\cos i) \) is thus given by:
\begin{equation}
p(m_{\rm t},\cos i)=
\begin{cases}
C e^{-\Delta\chi^2/2} & \text{if $m_{\rm p} > 1~M_\odot$}, \\
0 & \text{otherwise},
\end{cases}
\end{equation}
where $C$ is a normalization constant ensuring a total probability of unity. From the \( \Delta\chi^2 \) map, we derived the following constraints on the system parameters: \( m_{\rm t} = 2.814^{+0.018}_{-0.021}~M_\odot \), \( i = {76^{\circ}}^{+8^{\circ}}_{-15^{\circ}} \), \( m_{\rm p} = 1.38^{+0.06}_{-0.10}~M_\odot \), and \( m_{\rm c} = 1.44^{+0.08}_{-0.06}~M_\odot \) (see Fig.~\ref{mtot-cosi}).

In addition, we performed a second Bayesian analysis targeting $\dot{\omega}$ and $\dot{\omega}_{\rm EX}$ (Fig.~\ref{omdot1-omdot2}), following a similar procedure using the DDeGR+ model. This analysis yields the probability density distribution of $\dot{\omega}_{\rm EX}$, with a constraint of $\dot{\omega}_{\rm EX} = (-1.5^{+0.9}_{-0.8}) \times 10^{-4} $ deg\,yr$^{-1}$.
One potential contribution to $\dot{\omega}_{\rm EX}$ arises from the proper motion of PSR J0653+0443. Proper motion alters our line of sight, inducing changes in both the longitude of periastron $\omega$ and the projected semi-major axis $x$ \cite{lori2012}. However, the proper motion of PSR J0653+0443 is currently unmeasurable, with an upper limit on the order of tens of mas\,yr$^{-1}$. The resulting contribution to $\dot{\omega}_{\rm EX}$ is expected to be less than $1 \times 10^{-5} \, \text{deg}\,\text{yr}^{-1}$, which is smaller than the derived constraint on $\dot{\omega}_{\rm EX}$ above.

\section{Evolution of PSR~J0653+0443-like binaries} 

In principle, binaries in an extremely eccentric orbit could be formed near collisions with passing stars \cite{PulsarAstronomy2022}, if they are in very dynamical stellar environments such as globular clusters \cite{DeCesar+2015ApJ...807L..23D}. The PSR~J0653+0443 DNS system is located about 1~kpc from us in the Perseus arm, with almost the same mass for the pulsar and the companion neutron star. It is generally believed that the progenitors of such DNS systems in the Galactic field are NS-helium (He) star binaries. The He star companion undergoes an ultra-stripped supernova (USSNe) to form a NS, either via an electron-capture supernova (ECSNe) or an iron-core collapse supernova (Fe CCSNe) \cite{2003MNRAS.344..629D, Tauris+2017ApJ...846..170T}. If the binary system remains bound following the He star’s SN explosion, a DNS system is formed.

\begin{figure}
    \centering
    \includegraphics[width=0.49\columnwidth]{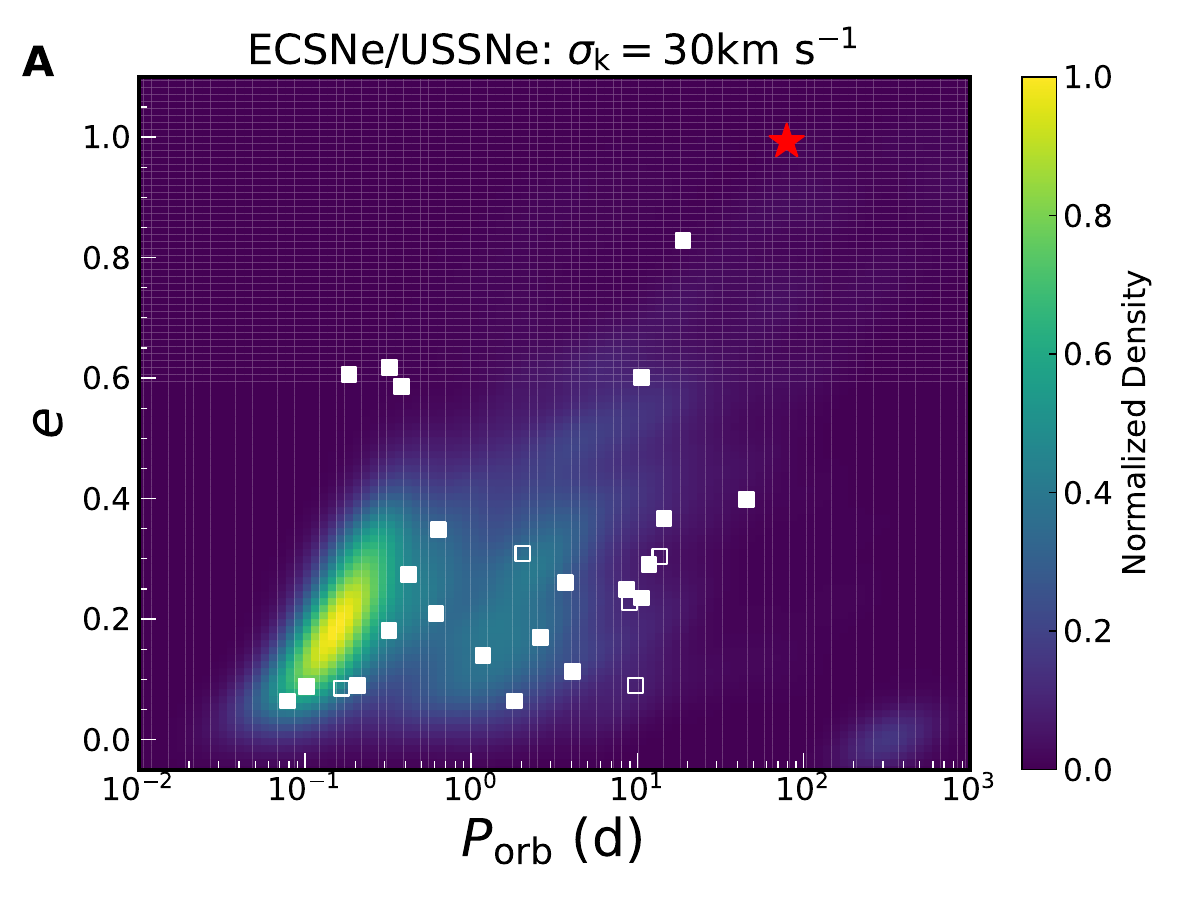}
    \includegraphics[width=0.49\columnwidth]{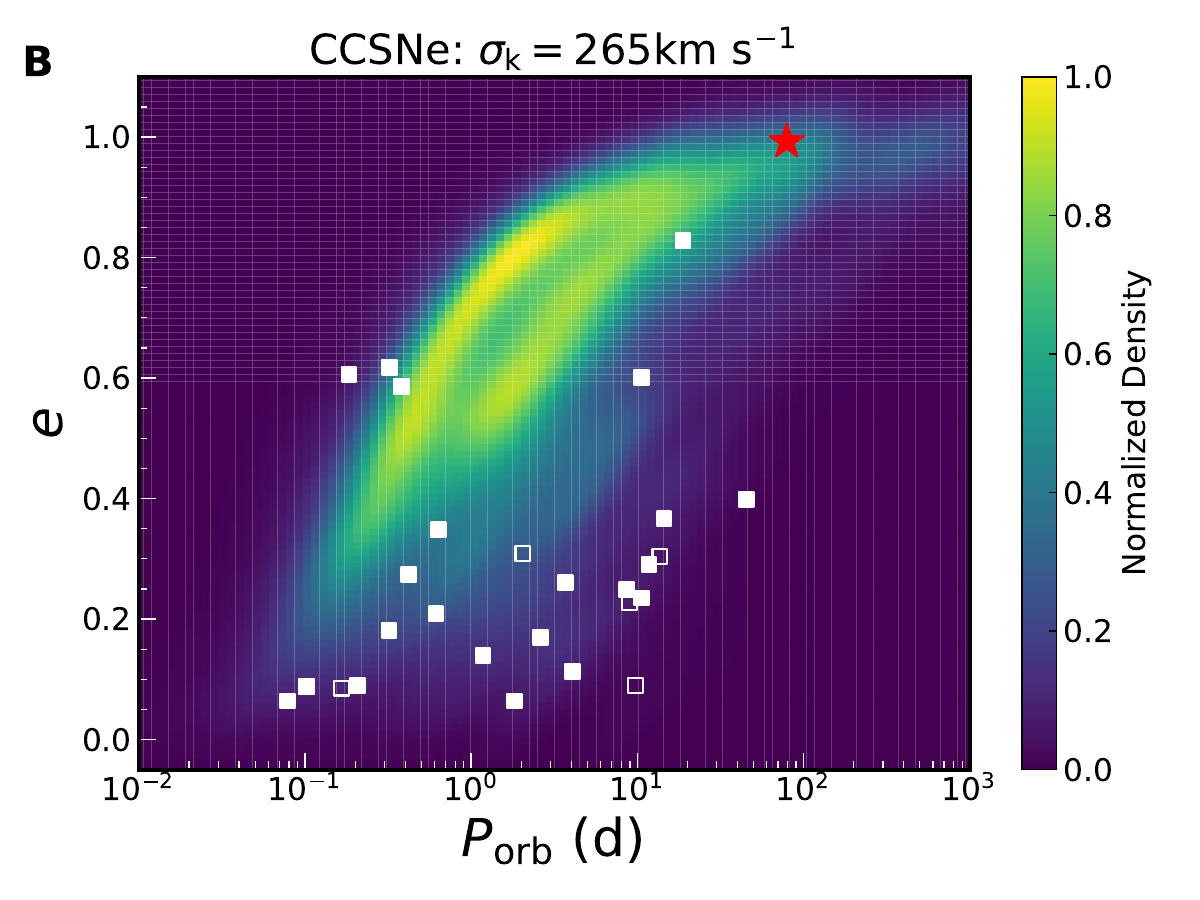}
    \caption{{\bf Normalized probability density distribution of DNS systems in the parameter space of orbital period and eccentricity}. Panel ({\bf A}) show the probability density distribution of DNS systems produced in electron-capture supernova (ECSNe) or ultra-stripped supernovae (USSNe) in the population synthesis, with a kick velocity dispersion of $\sigma_{\rm k}=30~{\rm km~s^{-1}}$. Squares represent DNS systems in the ATNF Pulsar Catalogue version 2.7.0 (not including those in global clusters), open squares indicate DNS candidates \cite{Manchester+2005AJ....129.1993M}, and the red star marks PSR~J0653+0443. Panel ({\bf B}) is similar to Panel ({\bf A}) but for DNS systems produced in an iron-core collapse supernova (CCSNe) models with $\sigma_{\rm k}=265~{\rm km~s^{-1}}$. 
    }
    \label{fig:pe2}
\end{figure}

\begin{figure}
    \centering
    \includegraphics[width=0.55\columnwidth]{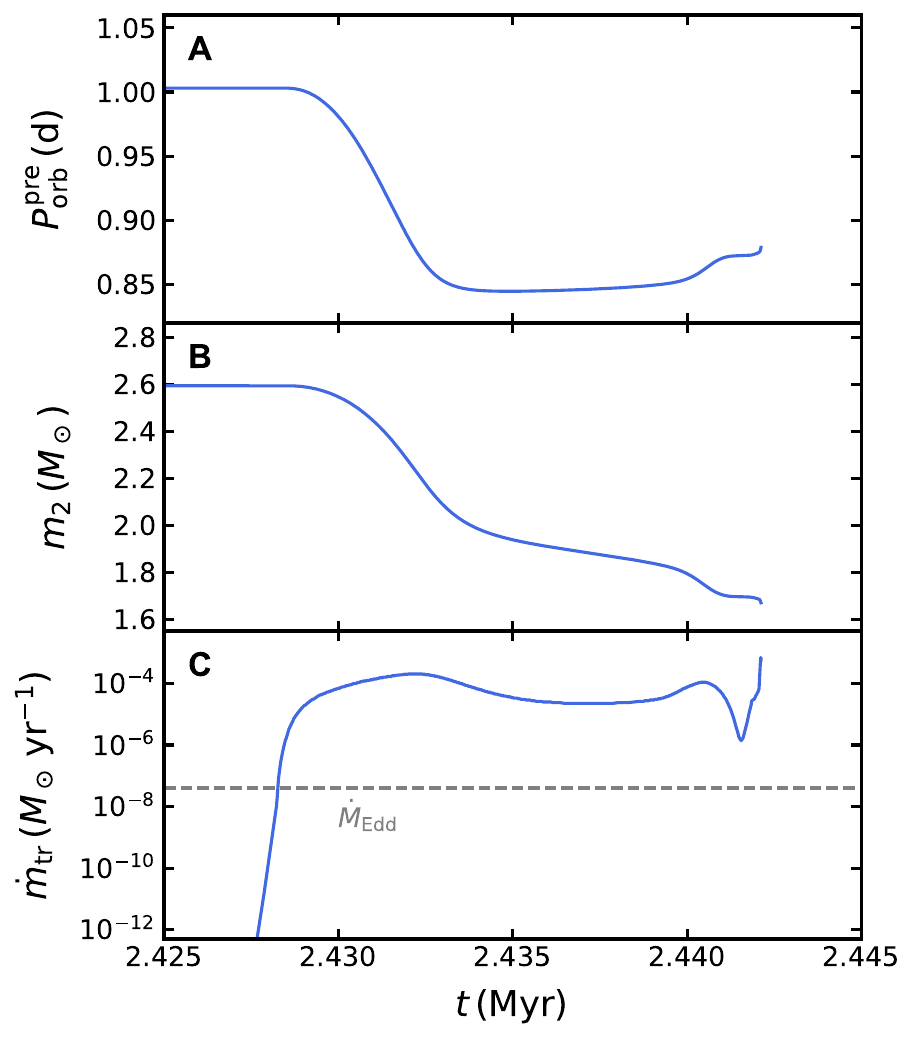}
    \caption{
    {\bf Evolution of the J0653+0443 progenitor binary}. The progenitor binary should include the NS and He star. Assuming the initial orbital period is $P_{\rm orb}^{\rm pre} \simeq 1\,$d and the He star companion mass of $2.6\,M_\odot$, we demonstrate the parameter changes over the stellar evolution in the code \textsc{MESA} \cite{Paxton+2011ApJS..192....3P, Paxton+2019ApJS..243...10P}. 
    The orbital period $P_{\rm orb}$, companion mass $m_2$, and mass-transfer rate $\dot m_{\rm tr}$
    are evolved with time, as shown in panels ({\bf A}), ({\bf B}) and ({\bf C}), respectively. 
    The Eddington accretion rate ($\dot{M}_{\rm Edd}=4\times10^{-8}~M_\odot$\,yr$^{-1}$) is indicated by the dashed line in panel ({\bf C}). During the mass transfer phase, the transfer rate exceeds $\dot{M}_{\rm Edd}$, and is highly super-Eddington.
    }
    \label{casebb}
\end{figure}

\subsection{Evolution history of the PSR J0653+0443 progenitor binary}

The observational properties of PSR~J0653+0443 indicates that it originated from an Fe CCSNe rather than an ECSNe or USSNe (Fig.~\ref{fig:pe2}), 
since low kick velocities and ejecta masses of ECSNe or USSNe cannot produce such a high eccentricity
\cite{Guo+2024MNRAS.530.4461G}.
%
%
We employed the stellar evolution code MESA to model the evolutionary example of the progenitor system of PSR~J0653+0443, a binary consisting of a NS and a He star companion (see also \cite{wang2021MNRAS.506.4654W}).
The NS is treated as a point mass with an initial mass of $1.4\,M_\odot$,
while the companion is a He star with a metallicity of $0.02$ and an initial mass of $2.6\,M_\odot$. The initial orbital period is set to 1 day.
We set the Eddington accretion rate ($\dot{M}_{\rm Edd}$) to be $4.0\times10^{-8}\,M_\odot\rm\,yr^{-1}$,
and the fraction of transferred material accreted by NS is set to be $0.5$.
In addition, we employed the Ledoux criterion and set the mixing-length parameter to be $2.0$, and set the overshooting parameter to be $0.014$.
Fig.~\ref{casebb} shows the evolution of key properties of the progenitor system. The central helium-burning phase of the He star companion lasts for $\sim2.36\,$Myr. Subsequently, expansion caused by helium shell burning leads the companion to fill its Roche lobe at $\sim2.427\,$Myr, triggering Case BB mass-transfer process. The code stops when the companion develops an ONe core with a mass of $\sim1.47\,M_\odot$, and an Fe CCSN will be triggered within \textless$100\,$yr. The Case BB phase lasts for $\sim0.015\,$Myr, and the mass-transfer rate is highly super-Eddington. NSs can spin up by gaining angular momentum through accreted material. It has been suggested that Case BB allows first-born NS to accrete material at mass-accretion rates of $\sim2~\text{to}~3$ $\dot{M}_{\rm Edd}$  \cite{Tauris+2017ApJ...846..170T}. Accordingly, the NS can accrete approximately $\sim0.0015\,M_\odot$ of material, which is sufficient to spin it up to a spin period of $47\,$ms. The orbital period of the progenitor system does not change significantly before the supernova explosion, with $P_{\rm orb}^{\rm pre}$ approximately $0.9\,$d. According to Fig.~\ref{fig:space}, if the kick velocity added to the second-born NS is $\gtrsim400\,$km\,s$^{-1}$, a DNS with a long orbital period and high eccentricity like PSR~J0653+0443 can form.

\subsection{Formation of PSR J0653+0443-like system via SN explosion of the companion star}
\label{sec:sn}

The eccentricity ($e$) and orbital period ($P_{\rm orb}$) of the resulting NS binary are altered by the SN explosion of the companion star, primarily due to instantaneous mass loss and the velocity kick ($V_{\rm k}$). The properties of the formed DNS are also influenced by the orientation of the kick velocity, which is determined by two angles: the angle between $V_{\rm k}$ and the pre-SN orbital velocity ($0\leq\psi\leq180^{\circ}$), and the positional angle of $V_{\rm k}$ out of the orbital plane ($-180^{\circ}<\phi\leq180^{\circ}$). 
The ratio of the pre-SN orbital semi-major axis ($a_0$) to the post-SN orbital semi-major axis ($a$) can be expressed as \cite{1983ApJ...267..322H}:
    \begin{equation}
        \frac{a_0}{a} = 2 - \frac {m_0}{m_0-\Delta m}(1+\nu^2+2\nu\,\cos\psi),
    \end{equation}
where $m_0$ and $\Delta m$ are the pre-SN total mass and the ejecta mass during the SN explosion, respectively. We adopt an ejecta mass appropriate for ultra-stripped SNe: $\Delta m=0.3\,M_\odot$ \cite{Tauris+2017ApJ...846..170T}. We define $\nu = V_{\rm k}/V_0$, where $V_0$ is the relative velocity between the two stars. Accordingly, the eccentricity after the SN explosion is given by:
    \begin{equation}
        1-e^2 = \frac{a_0}{a}\frac {m_0}{m_0-\Delta m}[1+2\nu\,\cos\psi
        +\nu^2(\cos^2\psi+\sin^2\psi\sin^2\phi)].
    \end{equation}

\begin{figure}
    \centering
    \includegraphics[width=0.65\columnwidth]{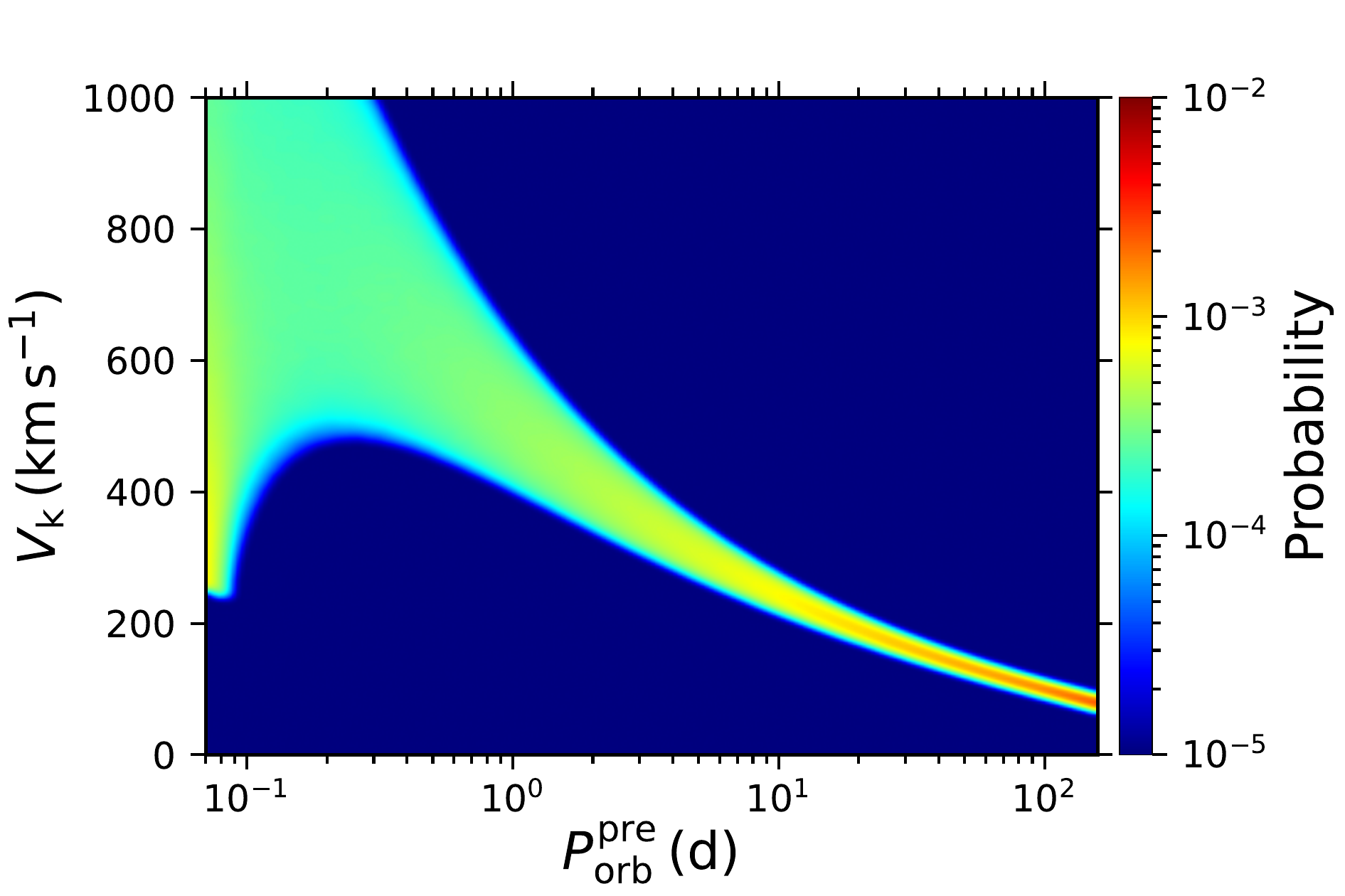}
    \caption{
    {\bf Parameter space for producing J0653+0443 in the log$P_{\rm orb}^{\rm pre}-V_{\rm k}$ plane.}
This figure presents the parameter space for forming PSR J0653+0443, visualized in the log\(P_{\rm orb}^{\rm pre}\)-\(V_{\rm k}\) plane (where \(P_{\rm orb}^{\rm pre}\) is the pre-SN orbital period and \(V_{\rm k}\) is the neutron star kick velocity).
We set 200 values of \(P_{\rm orb}^{\rm pre}\) to be logarithmically distributed over the range of 0.07 -- 120~d, and 200 NS kick velocities to be uniformly distributed between 0 -- 1000 km\,s\(^{-1}\). For each pre-SN system (defined by a given \(P_{\rm orb}^{\rm pre}\) and \(V_{\rm k}\)), we performed 5,000,000 calculation cycles with randomly oriented NS kicks.
The colorbar in the figure represents the probability of forming double neutron star (DNS) systems that match the key orbital properties of PSR J0653+0443, specifically with orbital eccentricity in the range \(0.990<e<1.0\) and orbital period in \(70~{\rm d} <P_{\rm orb} < 90~{\rm d}\).
        }
    \label{fig:space}
\end{figure}


Fig.~\ref{fig:space} shows the parameter space for producing J0653+0443 in the log$P_{\rm orb}^{\rm pre}-V_{\rm k}$ plane,
where the $P_{\rm orb}^{\rm pre}$ is the pre-SN orbital period of the NS binary.
We set the $P_{\rm orb}^{\rm pre}$ ranges from $0.07$ to $120\,$d
because the shortest $P_{\rm orb}^{\rm pre}$ for forming a DNS is $\sim0.07\,$d.
Moreover,
if $P_{\rm orb}^{\rm pre}$ exceeds $\sim120\,$d,
the He star companion can not fill its Roche lobe,
and the first-born NS will avoid recycling.
From Fig.~\ref{fig:space},
we can see that progenitor systems with longer $P_{\rm orb}^{\rm pre}$ can more easily reproduce J0653+0443 with relatively lower kick velocities.
Meanwhile, it has been suggested that the kick velocities for producing DNSs are typically $\lesssim 800\,$km\,s$^{-1}$
\cite{Tauris+2017ApJ...846..170T}.
On the other hand,
DNSs with longer $P_{\rm orb}$ are preferentially accompanied by the higher spin periods ($P_{\rm spin}$) of the first-born NSs
\cite{Tauris+2017ApJ...846..170T, Guo+2024MNRAS.530.4461G}.
To form a pulsar with $P_{\rm spin}$ $\sim 47\,$ms, the pre-SN orbital period likely needs to be $\lesssim10\,$d.
Additionally,
the pre-SN orbital period of J0653+0443 is unlikely to be $\lesssim0.5\,$d,
as shorter periods would result in a more rapidly spinning pulsar ($P_{\rm spin}\lesssim30\,$ms)
\cite{Tauris+2017ApJ...846..170T, Guo+2024MNRAS.530.4461G}.
Therefore,
we speculate that the $P_{\rm orb}^{\rm pre}$ of J0653+0443 is likely in the range of $\sim 0.5~\text{to}~10$\,d,
and the kick velocity of the second-born SN is $\sim 200~\text{to}~800$\,km\,s$^{-1}$. 

\begin{figure}
    \centering
    \includegraphics[width=0.6\columnwidth]{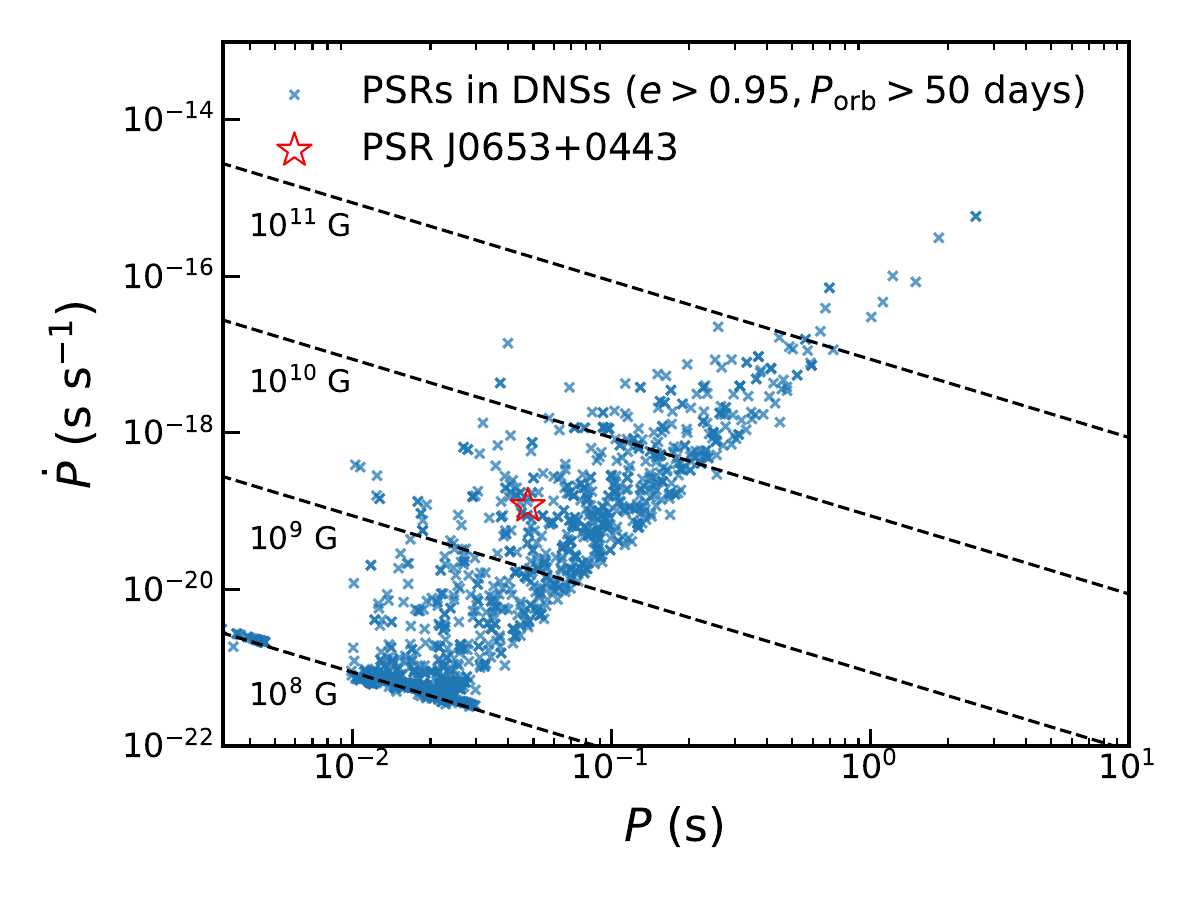}
    \caption{{\bf Spin period ($P$) and its derivative ($\dot{P}$) of pulsars in  DNS systems survived in the  population synthesis}. We simulated the evolution of $\sim 10^9$ binaries of massive main sequence stars, and identified 10$^6$ survived DNS systems. Among them there are $\sim 660$ systems having $e>0.95$ and $P_{\rm orb}>50$ days. The red star represents PSR~J0653+0443 with $P=0.0475 \rm ~s$ and $\dot{P}=1.17\times 10^{-19}\rm~ s\,s^{-1}$. The black dashed lines indicate the surface magnetic field of neutron stars for $B=10^{8}$, $10^9$, $10^{10}$, and $10^{11}$~G, respectively.}
    \label{fig:ppdot}
\end{figure}

\subsection{Population synthesis for binaries and probability to form PSR~J0653+0443-like DNS systems}
\label{sec:pop} 

We conducted the population synthesis to analysis the probability of formation of DNS systems as  PSR~J0653+0443, using the model in \cite{Chu2022MNRAS.509.1557C, Chu:2024rwt, Chen+2025MNRAS.544L..89C}. In the model, we considered the common-envelope (CE) evolution phase with the CE ejection efficiency $\alpha$, in the form of 
\begin{equation}
    \frac{G(m_{1}-m_{\rm 1c})m_1}{\lambda r_{\rm L}}=\alpha\bigg(\frac{Gm_{\rm 1c}m_2}{2a_{\rm f}}-\frac{Gm_1m_2}{2a_{i}}\bigg),
\end{equation}
where $r_{\rm L}$ is the Roche lobe radius, $a_{i}$ and $a_{f}$ are the semi-major axis of the binary before and after the CE stage, $m_1$ and $m_{\rm 1c}$ are the masses of the star and its core, $m_{2}$ is the companion mass, $\alpha$ is assumed to be $3$ as in \cite{Sgalletta+2023MNRAS.526.2210S}, $\lambda$ is the structure parameter depending on the evolutionary state of the donor and taken the same value as \cite{2021A&A...650A.107M}. We considered various types of supernova explosions, including USSNe, ECSNe and CCSNe, depending on the helium core mass. For these supernovae explosion mechanisms, we adopt the following Maxwellian distribution for the dispersion $\sigma_{\rm k}$ of kick velocities,
\begin{equation}
    \frac{dN}{Ndv_{\rm k}}=\bigg(\frac{2}{\pi}\bigg)^{1/2}\frac{v_{\rm k}^2}{\sigma_{\rm k}^3}\exp{\bigg(-\frac{v_{\rm k}^2}{2\sigma_{\rm k}^2}\bigg)},
\end{equation}
where $dN/N$ is the normalization factor in the kick velocity bin $dv_{\rm k}$. For the ECSNe and USSNe models, we assumed $\sigma_{\rm K}=30$~km~s$^{-1}$ \cite{10.1093/mnras/sty2463} and for the CCSNe model, we assumed $\sigma_{\rm K}=265$~km~s$^{-1}$ \cite{10.1111/j.1365-2966.2005.09087.x}. %
The spin evolution of the first-born NS caused by the accretion from the donor star has been considered by treating them as high-mass X-ray binaries (HMXB). For the OB-type and Be-type HMXB systems, the accretion could act as either the matter overflow from the RL or the stellar wind. 
As done in \cite{Chu:2024rwt}, we assume that the star formation rate density inside the Milky Way is a constant, and the metallicity is set to be $Z=0.71 Z_{\odot}$. We also assume that NS above the ``death-line" is a radio-loud pulsar, so that 
$ 
 P < P_{\rm death}=\sqrt{B / (1.7\times 10^{11}\rm G )}~\rm s
$ 
\cite{zhang2000deathline}.

In the population synthesis, we have both the spin period and the derivative of spin period recorded, as plotted in Figure~\ref{fig:ppdot}.  Here we assumed that  
 \begin{equation}
     \dot{P}=9.73\times 10^{-16}\mu_{30}^2I_{45}^{-1}P^{-1} \rm s ~s^{-1},
 \end{equation}
where $\mu_{30}$ and $I_{45}$ are the magnetic dipole moment in units of $10^{30} ~\rm G\,cm^3$ and the moment of inertial in units of $10^{45} ~\rm g cm^2$, respectively. We indeed find one pulsar with a similar $P$ and $\dot{P}$ among the $\sim 660$ survived DNS systems with $e>0.95$ and $P_{\rm orb}>50$ days (Fig.\ref{fig:pe2}). 


Through the population synthesis, we obtained the parameter distribution of survived DNS systems at the present time, as shown in Figure~\ref{fig:pe2}. 
The PSR~J0653+0443-like DNS with an extremely high eccentricity and a large orbital period is extremely rare. The fraction of DNS systems with an eccentricity of $e>0.99$ and an orbital period in the range of $P_{\rm orb}\in [70,90]$~days is only about $\lesssim 2.5\times 10^{-5}$ among survived DNS systems with an observable pulsar. For DNS systems with a larger parameter space, such as $e>0.95$ and $P_{\rm orb}\in [50,110]$ days, the fraction is only about $\lesssim 5 \times 10^{-3}$. Notably this fraction may change with the dispersion of kick velocities $\sigma_{\rm K}$. If $\sigma_{\rm K}=500~\rm km~ s^{-1}$, much higher than the observational constraint, is assumed for the CCSNe case, the fraction within $e>0.95$ and $P_{\rm orb}\in [50,110]$ days will increase to $\sim 2\%$. 

The fact that PSR~J0653+0443 with an extremely high eccentricity and a large orbital period has been detected among 28 known Galactic field DNS system systems imposes a challenge to the current understanding of not only binary evolution for the DNS production but also the core collapse SN processes.

\begin{figure}
    \centering
    \includegraphics[width=0.55\columnwidth]{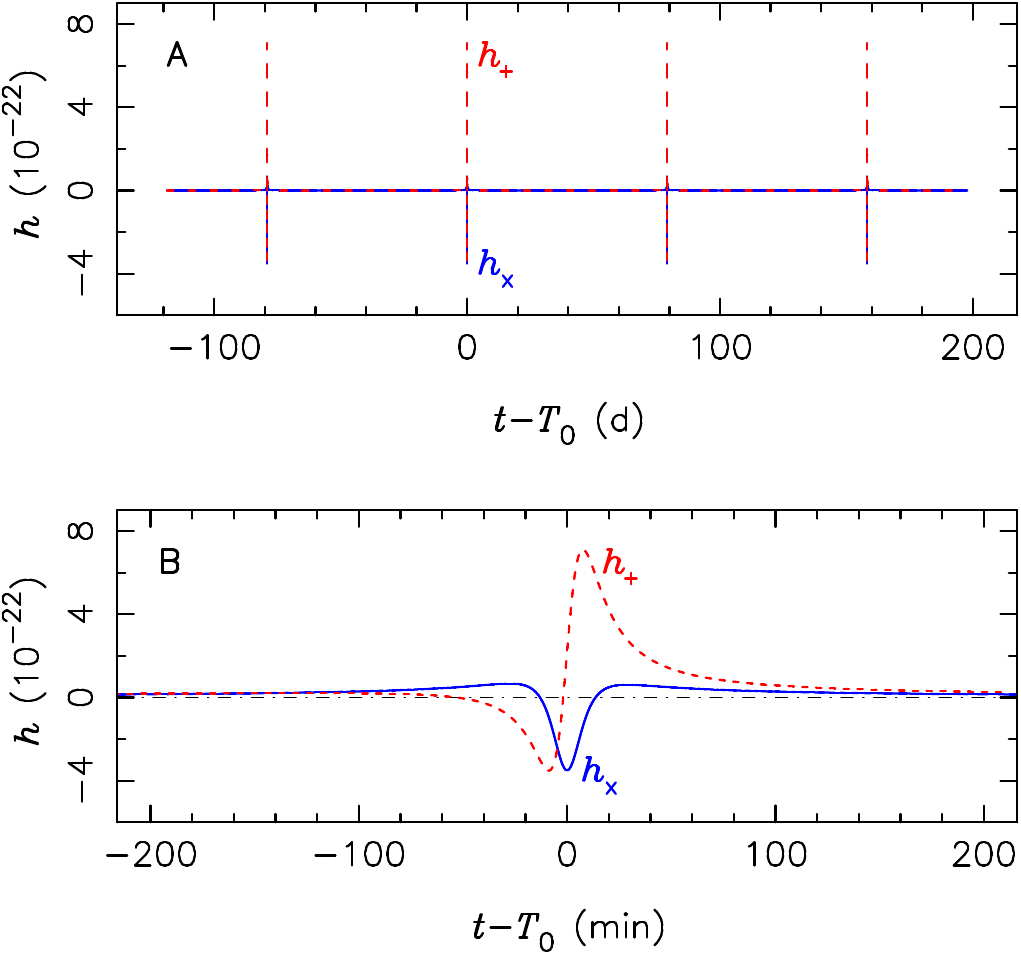}
    \caption{{\bf Strain amplitude of GW bursts from the PSR J0653+0443 DNS system.} Panel ({\bf A}) shows the strain amplitude of $h_+$ and $h_\times$ plotted as a function of time (in days), calculated by assuming a luminosity distance of 1.0~kpc and taking the parameters in Table~\ref{tab:timingPara}. With an extreme eccentricity of 0.992386, the GW emission becomes stronger in the short periastron passage, generating periodical bursts \cite{Poisson+2014grav.book.....P}. Panel ({\bf B}) presents the zoomed view (in minutes) for the detailed waveform around periastron ($t=T_0$).}
    \label{fig:hplusvstime}
\end{figure}

\subsection{GW emission and future evolution}

After the formation of PSR~J0653+0443, the evolution continues as the GW emission carries away the orbital energy and angular momentum of the system. 
The GW emission leads to the decrease of orbital parameters $a_{\rm pc}$, $P_{\rm orb}$ and $e$ following \cite{Peters+1964PhRv..136.1224P}:
\begin{equation}
\left\{
\begin{aligned}
    &\Dot{a}_{\rm pc}=-2\frac{L_{\rm GW}a^2_{\rm pc}}{Gm_{\rm p}m_{\rm c}}, \\
    &\Dot{P}_{\rm orb}=\frac{3}{2}\frac{P_{\rm orb}}{a_{\rm pc}}\Dot{a}_{\rm pc},\\
    &\Dot{e}=-\frac{304}{15}e\frac{G^3m_{\rm p}m_{\rm c}(m_{\rm p}+m_{\rm c})}{c^5a_{\rm pc}^4(1-e^2)^{\frac{5}{2}}}\left(1+\frac{121}{304}e^2\right).
\end{aligned}
\right.
\end{equation}
Here $L_{\rm GW}$ is the GW luminosity. The equation relating $a_{\rm pc}$ to $e$ is \cite{Peters+1964PhRv..136.1224P} 
\begin{equation}
    a_{\rm pc}=c_0\frac{e^{12/19}}{1-e^2}\left(1+\frac{121}{304}e^2\right)^{870/2299},
\end{equation}
where $c_0$ is a constant. The minimum separation between the pulsar and its companion is 
\begin{equation}
 a_{\rm pc}(1-e)=c_0\frac{e^{12/19}}{1+e}\left(1+\frac{121}{304}e^2\right)^{870/2299}\simeq 0.57c_0,\ \ \ \text{when}~ e\sim1.
\end{equation}
Therefore, during the decreases of the orbital period and semi-major axis, the minimum separation almost remains the same after the formation of PSR J0653+0443.

Currently, the fundamental GW frequency is $f_0=2/P_{\rm orb}=292.7$ nHz. 
The PSR~J0653+0443 DNS system has a high eccentricity of $e_0=0.9923869$, 
its GW emission is characterized as ``periodical bursts" (Fig.~\ref{fig:hplusvstime}). Bursting GW signals from highly eccentric binaries can provide information about the astrophysical sources, and they can be detected using matched filtering \cite{Loutrel+2020arXiv200911332L, Xuan+2025PhRvD.111d3018X}. The detailed waveform for the plus and cross polarizations ($h_+$ , $h_{\times}$) can be calculated by \cite{Poisson+2014grav.book.....P}
\begin{equation}
    \left\{
    \begin{aligned}
        h_+=&~-\frac{2G^2m_{\rm p}m_{\rm c}}{c^4a_{\rm pc}(1-e^2)d_{\rm L}}\left\{(1+\cos^2i)\left[\cos(2\theta+2\omega)+\frac{5}{4}e\cos(\theta+2\omega)+\frac{1}{4}e\cos(3\theta+2\omega)\right.\right.\\
        &~\left.\left.+\frac{1}{2}e^2\cos(2\omega)\right]+\frac{1}{2}\sin^2i(e\cos\phi+e^2)\right\},\\
        h_\times=&~-\frac{2G^2m_{\rm p}m_{\rm c}}{c^4a_{\rm pc}(1-e^2)d_{\rm L}}\cos i\left[2\sin(2\theta+2\omega)+\frac{5}{2}e\sin(\theta+2\omega)+\frac{1}{2}e\sin(3\theta+2\omega)+e^2\sin(2\omega))\right],
    \end{aligned}
    \right.
\end{equation}
here $d_{\rm L}$ is the luminosity distance, and we take $d_{\rm L}=1.0$~kpc according to the model-derived distances from the pulsar DM value.  
%
Taking the best-fitted parameters of the  PSR~J0653+0443 DNS system in Table~\ref{tab:timingPara}, the strain amplitude $h_+$ and $h_\times$ as a function of time is shown in Fig.~\ref{fig:hplusvstime}. 
\begin{figure}[t]
    \centering
    \includegraphics[width=0.6\columnwidth]{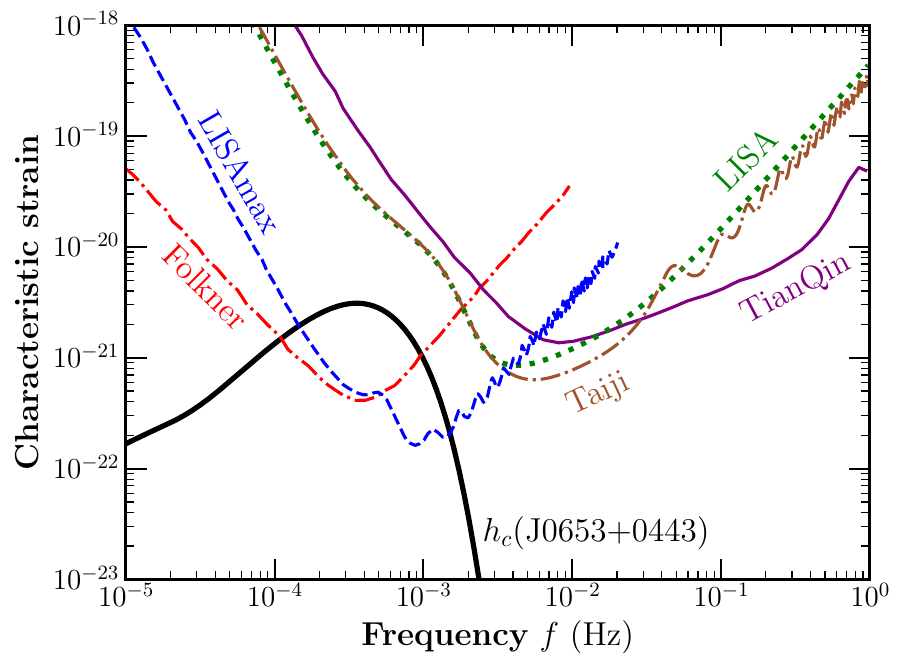}
    \caption{{\bf Characteristic GW strain of PSR~J0653+0443 and the sensitivity curves of LISA, TianQin and Taiji}. The characteristic strain for the spectral waveform $h_c$ of PSR~J0653+0443 is calculated by assuming a luminosity distance of $d_L=1.0$~kpc, compared to the noise amplitude of the LISA \cite{LISA:2017pwj}, Taiji \cite{Hu:2017mde}, TianQin \cite{TianQin:2015yph, Gong:2021gvw}, Folkner \cite{Mueller+2019BAAS}, and LISAmax \cite{2023CQGra..40s5022M}, observed for ten years.
    }
    \label{fig:sensitivity}
\end{figure}
The time-average luminosity $L_{\rm GW}$ is  \cite{Peters+1964PhRv..136.1224P}
\begin{equation}
L_{\rm GW}=\frac{32}{5}\frac{G^4m_{\rm p}^2m_{\rm c}^2(m_{\rm p}+m_{\rm c})}{c^5a_{\rm pc}^5(1-e^2)^{\frac{7}{2}}}\left(1+\frac{73}{24}e^2+\frac{37}{96}e^4\right) =  1.9\times10^{-4}~L_\odot, 
\end{equation}
here $L_\odot$ is the solar luminosity.

The GW characteristic strain amplitudes $h_{\rm c}(f)$ is defined as $h_{\rm  c}(f) = 2f\sqrt{|\tilde{h}_+(f)|^2+ |\tilde{h}_\times(f)|^2}$ at a given frequency $f$, where $\tilde{h}_+(f)$ and $\tilde{h}_\times(f)$ represent the Fourier transform of ${h}_+(t)$ and ${h}_\times(t)$, respectively \cite{Moore+2015CQGra..32a5014M}. 
The GWs emitted by the PSR~J0653+0443 DNS system are in the observational bands of future space-based GW detectors such as LISA \cite{LISA:2017pwj}, Taiji \cite{Hu:2017mde}, 
TianQin \cite{TianQin:2015yph, Gong:2021gvw}, 
Folkner \cite{Mueller+2019BAAS}, 
and LISAmax \cite{2023CQGra..40s5022M}.
%
The peak frequency $f_{\rm p0}$, at which the energy flux of GWs reaches its maximum \cite{Wen:2002km}, can be approximated by 
\begin{equation} \label{fpeq1}
f_{\rm p0}\approx \frac{(1+e_0)^{\gamma_{\rm GW}}}{(1-e_0^2)^{3/2}}f_0 =0.357 \text{ mHz},
\end{equation}
where $\gamma_{\rm GW}=1.1954$. The peak frequency $f_{\rm p0}$ is corresponding to about $2440$th harmonics of the orbital frequency. 
%
The characteristic strain $h_{\rm c}(f)$ of the PSR~J0653+0443 DNS system, however, is well below the noise amplitudes 
of the LISA \cite{LISA:2017pwj}, Taiji \cite{Hu:2017mde}, and TianQin \cite{TianQin:2015yph, Gong:2021gvw}, as shown in Fig.~\ref{fig:sensitivity}, 
%
%
%
%
%
too low to be detectable. 
For future space-based gravitational wave concepts such as Folkner \cite{Mueller+2019BAAS} and LISAmax \cite{2023CQGra..40s5022M}, the gravitational wave signal from this DNS system may be detectable over a ten year observation period, with a signal-to-noise ratio of $\sim 5.1$ and $8.4$, respectively, which may thus provide a direct test for the GW spectrum predicted for extremely eccentric binary systems. 

In the future, the GW emission will lead to the merger of the PSR J0653+0443 DNS system, producing a merger event like GW170817 \cite{Abbott+2017PhRvL.119p1101A} and GW190425 \cite{Abbott+2020ApJ...892L...3A}. The coalescence timescale $\tau_{\rm GW}$ is \cite{Peters+1964PhRv..136.1224P}
\begin{equation}
    \tau_{\rm GW}=\frac{12}{19}\frac{c_0^4}{\beta}\int_0^{e_0}\frac{e^\frac{29}{19}(1+\frac{121}{304}e^2)^\frac{1181}{2299}}{(1-e^2)^\frac{3}{2}}\mathrm{d}e,
\end{equation}
here $\beta = \frac{64G^3}{5c^5} m_{\rm p}m_{\rm c}(m_{\rm p}+m_{\rm c})$, $e_0$ is the current orbital eccentricity, and $e$ in this formula is the decreasing eccentricity of the future orbits. Using the best fitted parameters in Table~\ref{tab:timingPara}, we got $\tau_{\rm GW} = 2.54$~Gyr.

\end{document}